\documentclass[useAMS,usenatbib]{mnras}

\usepackage{times}
\usepackage{graphicx}
\usepackage{subfigure}
\usepackage{amsmath}

\newcommand{\gsim}{\mathrel{\hbox{\rlap{\lower.55ex \hbox {$\sim$}}
                   \kern-.3em \raise.4ex \hbox{$>$}}}}
\newcommand{\lsim}{\mathrel{\hbox{\rlap{\lower.55ex \hbox {$\sim$}}
                   \kern-.3em \raise.4ex \hbox{$<$}}}}

\title[Metallicity dependence of stellar properties ]{The statistical properties of stars and their dependence on metallicity}
\author[M. R. Bate]{Matthew R. Bate$^{1}$\thanks{E-mail:
M.R.Bate@exeter.ac.uk}\\
$^{1}$ School of Physics and Astronomy, University of Exeter, Stocker
Road, Exeter EX4 4QL  
}

\date{Accepted by MNRAS}
\begin{document}
\maketitle
\begin{abstract}
We report the statistical properties of stars and brown dwarfs obtained from four radiation hydrodynamical simulations of star cluster formation, the metallicities of which span a range from 1/100 to 3 times the solar value.  Unlike previous similar investigations of the effects of metallicity on stellar properties, these new calculations treat dust and gas temperatures separately and include a thermochemical model of the diffuse interstellar medium.
The more advanced treatment of the interstellar medium gives rise to very different gas and dust temperature distributions in the four calculations, with lower metallicities generally resulting in higher temperatures and a delay in the onset of star formation.  Despite this, once star formation begins, all four calculations produce stars at similar rates and many of the statistical properties of their stellar populations are difficult to distinguish from each other and from those of observed stellar systems.  We do find, however, that the greater cooling rates at high gas densities due to the lower opacities at low metallicities increase the fragmentation on small spatial scales (disc, filament, and core fragmentation).  This produces an anti-correlation between the close binary fraction of low-mass stars and metallicity similar to that which is observed, and an increase in the fraction of protostellar mergers at low metallicities. There are also indications that at lower metallicity close binaries may have lower mass ratios and the abundance of brown dwarfs to stars may increase slightly.  However, these latter two effects are quite weak and need to be confirmed with larger samples. 
\end{abstract}

\section{Introduction}
\label{introduction}

Within the past decade, it has become possible to perform radiation hydrodynamical calculations of star cluster formation that produce in excess of a hundred of stars and brown dwarfs from a single calculation.  Such numbers of stars and brown dwarfs allow meaningful comparisons to be made with observed stellar populations, for example, to determine whether the resulting stellar mass and multiplicity distributions are consistent with those of Galactic populations.  The first calculation to reproduce a wide variety of the observed statistical properties of stellar systems was that of \cite{Bate2012}, which produced more than 180 stars and brown dwarfs, including 40 multiple systems.  The stellar mass function produced by this calculation was in good agreement with the observed Galactic initial mass function (IMF), the multiplicity of the stellar systems was found to increase with primary mass with values in agreement with the results from field star surveys, and the properties of the multiple systems (e.g., distributions of mass ratios, separations, and orbital orientations) also reproduced many of the observed characteristics.  Subsequent calculations have also been able to produce realistic stellar populations \citep[e.g.][]{Bate2014,JonBat2018a}, some of which have included additional physical processes such as protostellar outflows \citep*{KruKleMcK2012} and magnetic fields \citep{Myersetal2013,Myers_etal2014,Krumholz_etal2016,Cunningham_etal2018}.

Now that we are able to perform hydrodynamical calculations of the formation of stellar groups and clusters with realistic properties, we have the potential to determine directly how the statistical properties of stellar systems depend on environment and initial conditions by performing a series of calculations.   \cite{Myersetal2011} and \cite{Bate2014} each performed a series of radiation hydrodynamical calculations in which the opacity of the gas was varied to mimic variations in the metallicity ranging from $0.05-1$ and $0.01-3$ times solar metallicity, respectively.  Both studies found no significant variation in the stellar mass distributions with opacity. Furthermore, \cite{Bate2014} found that the multiplicity fractions and properties of the multiple systems were also statistically indistinguishable between the different calculations.  \cite{Bate2014} did find a slight decrease in the ratio of stars to brown dwarfs at the lowest metallicity ($Z=0.01~{\rm Z}_\odot$), and a slight decrease in the separations of multiple systems with decreasing metallicity.  Both of these effects are consistent with the lower opacity leading to more rapid cooling at high densities and, therefore, an increase in small-scale fragmentation.  But the magnitude of these effects was small enough that they could have been due to random variation (the calculations each produced between 170 and 198 objects).

Another issue with these previous studies of the dependence of stellar properties on metallicity is that changing the metallicity of a molecular cloud affects much more than just its opacity \cite[see the introduction of][]{Bate2014}.  In particular, all of the above studies have assumed that the local gas and dust temperatures are identical.  This is a reasonable approximation at high densities ($n_{\rm H} \gsim 10^5~{\rm cm}^{-3}$) with solar or super-solar metallicity \citep{BurHol1983, Goldsmith2001, GloCla2012a}.  However, at lower densities or metallicities the gas and dust temperatures can be poorly coupled \citep{Omukai2000, TsuOmu2006, Dopckeetal2011, NozKozNom2012, ChiNozYos2013, Dopckeetal2013} and the gas temperature is typically higher than the dust temperature \citep[e.g.][]{GloCla2012c}.  Since fragmentation and gas accretion depend sensitively on the gas temperature, the star formation may be poorly modelled by only changing the opacity.

To improve the thermal modelling of star formation calculations, particularly at low densities and metallicities, \cite{BatKet2015} developed a new method that combines radiative transfer with a thermochemical model of the diffuse interstellar medium (ISM).  This method treats gas and dust temperatures separately, and includes prescriptions for a variety of heating and cooling mechanism that are important for low-density gas (heating from the interstellar radiation field, cosmic rays, and molecular hydrogen formation; cooling via atomic and molecular line emission).  The thermochemical model also includes simple chemical models for hydrogen (i.e., the fractions in atomic and molecular forms) and carbon (i.e., the fractions in the forms of C$^+$, neutral carbon, and CO).

In this paper, we report results from four radiation hydrodynamical calculations of star cluster formation that employ the new radiative transfer/diffuse ISM method of \cite{BatKet2015}.  The calculations are identical to each other, except for their metallicity which takes values of 1/100, 1/10, 1, and 3 times solar metallicity.  We follow the collapse of each of the molecular clouds to form a cluster of stars and then compare the properties of the stars and brown dwarfs to determine how sensitive their statistical properties are to variations in metallicity.  In Section \ref{sec:method} we provide summaries of the method and initial conditions. The results from the calculations are presented in Section \ref{sec:clouds}.  In Section \ref{sec:discussion}, we discuss the origins and properties of close multiple stellar systems in some detail, and compare our results with similar previously published calculations.  Finally, in Section \ref{conclusions} we provide our conclusions.

\vspace{-12pt}

\section{Method}
\label{sec:method}

The calculations were performed using the smoothed particle
hydrodynamics (SPH) code, {\tt sphNG}, based on the original 
version of \citeauthor{Benz1990} 
(\citeyear{Benz1990}; \citealt{Benzetal1990}), but substantially
modified using the methods described in \citet{BatBonPri1995}, \citet{PriMon2007},
\citet*{WhiBatMon2005}, \citet{WhiBat2006} and 
parallelised using both OpenMP and MPI.

Gravitational forces between particles and a particle's 
nearest neighbours are calculated using a binary tree.  
The smoothing lengths of particles varied in 
time and space and were set such that the smoothing
length of each particle 
$h = 1.2 (m/\rho)^{1/3}$ where $m$ and $\rho$ are the 
SPH particle's mass and density, respectively
\cite[see][for further details]{PriMon2007}.  The SPH equations were 
integrated using a second-order Runge-Kutta-Fehlberg 
integrator \citep{Fehlberg1969} with individual time steps for each particle
\citep{BatBonPri1995}.
To reduce numerical shear viscosity, the
\cite{MorMon1997} artificial viscosity was employed 
with $\alpha_{\rm_v}$ varying between 0.1 and 1 while $\beta_{\rm v}=2 \alpha_{\rm v}$
\citep[see also][]{PriMon2005}.

\begin{table*}
\begin{tabular}{lccccccccccc}\hline
Calculation & Initial Gas & Metallicity & No. Stars & No. Brown  & Mass of Stars \&  & Mean  & Mean & Median & Stellar\\
& Mass  &   & Formed & Dwarfs Formed & Brown Dwarfs & Mass & Log-Mass & Mass & Mergers \\
 & M$_\odot$ &  Z$_\odot$ & & & M$_\odot$ & M$_\odot$ &M$_\odot$ & M$_\odot$ \\ \hline
Metallicity 1/100 & 500 &  0.01 & $\geq 93$& $\leq 49$& 49.8 & $0.35\pm0.04$ & $0.15\pm0.02$ & 0.14 & 17 \\   
Metallicity 1/10 & 500 &  0.1 & $\geq 116$& $\leq 58$& 73.4 & $0.42\pm0.05$ & $0.17\pm0.02$ & 0.15 & 11 \\   
Solar Metallicity & 500 &  1.0 & $\geq 158$& $\leq 97$& 90.1 & $0.35\pm0.04$ & $0.16\pm0.01$ & 0.15 & 14\\  
Metallicity 3 & 500 & 3.0 & $\geq 177$& $\leq 81$& 92.0 & $0.36\pm0.03$ & $0.18\pm0.01$ & 0.17 & 6 \\  \hline 
\end{tabular}
\caption{\label{table1} The parameters and overall statistical results for each of the four radiation hydrodynamical calculations.  All calculations were run to 1.20~$t_{\rm ff}$.  All calculations employ sink particles with $r_{\rm acc}=0.5$~AU and no gravitational softening.  Brown dwarfs are defined as having final masses less than 0.075 M$_\odot$.  The numbers of stars (brown dwarfs) are lower (upper) limits because some brown dwarfs were still accreting when the calculations were stopped.  Changing the metallicity results in no significant difference in the mean and median masses of the stellar populations.  However, at the end of the calculations less gas is converted into stars in calculations with lower metallicities, and the average number of stellar mergers per star increases with decreasing metallicity.}
\end{table*}

\vspace{-6pt}

\subsection{Radiative transfer and the diffuse ISM model}
\label{hydro}

The calculations employed the combined radiative transfer and diffuse ISM model
that was developed by \cite{BatKet2015}.  For the details of the method, the reader
is directed to that paper.  Here we only briefly describe the main elements of the method.

The gas has an ideal gas equation of state for the gas pressure
$p= \rho T_{\rm gas} \cal{R}/\mu$, where 
$T_{\rm gas}$ is the gas temperature, $\mu$ is the mean molecular weight of the gas,
and $\cal{R}$ is the gas constant.  
The thermal evolution takes into account the translational,
rotational, and vibrational degrees of freedom of molecular hydrogen 
(assuming a 3:1 mix of ortho- and para-hydrogen; see
\citealt{Boleyetal2007}).  It also includes molecular
hydrogen dissociation, and the ionisations of hydrogen and helium.  
The hydrogen and helium mass fractions are $X=0.70$ and 
$Y=0.28$, respectively.  
For this composition, the mean molecular weight of the gas is initially $\mu = 2.38$ 
(the gas is taken to be entirely molecular initially).
The contribution of metals to the equation of state is neglected.

The thermal evolution combines the flux-limited diffusion radiative transfer method of 
\cite{WhiBatMon2005, WhiBat2006}, with a diffuse ISM model that is similar to 
that of \cite{GloCla2012c} but with a greatly simplified chemical model.
The gas, dust, and radiation fields all have separate temperatures.  The dust temperature
is set by assuming that the dust is in local thermodynamic equilibrium (LTE) with the total
radiation field (i.e., the combination of the local interstellar radiation field plus any protostellar 
radiation), but also accounts for the collisional exchange of thermal energy between the 
dust and the gas.  \cite{BatKet2015} implemented two different dust-gas collisional energy transfer
rates; here we use the rate given by \cite{HolMcK1989} and also used by \cite{GloCla2012b}.

For the gas, various heating and cooling processes are included.  Heating mechanisms 
include cosmic rays heating the gas by direct collision, 
heating of the gas indirectly through the photoelectric release of hot electrons 
from dust grains due to photons from the interstellar radiation field, 
and heating due to the formation of molecular hydrogen on dust grains.
Gas cooling mechanisms include electron recombination, atomic oxygen and carbon
fine-structure cooling, and molecular line cooling.  Because we do not have an 
explicit chemical model for oxygen, the abundance of atomic
oxygen is assumed to scale in proportional to $(1-{\rm CO/C})$ (i.e., the abundance of
atomic oxygen decreases as CO is formed).  

We employ simple chemical models to treat the evolution of hydrogen and carbon.
The abundances of C$+$, neutral carbon, CO, and the depletion of CO on to dust grains
are computed using the model of \cite{KetCas2008}.  For hydrogen, we evolve the
atomic and molecular hydrogen abundances using the same molecular hydrogen 
formation and dissociation rates as those used by \cite{Gloveretal2010}.

The interstellar radiation field (ISRF) is required to determine both the heating rate of the
dust grains and the photoelectric heating rate of the gas.  In both cases, 
the ISRF is attenuated due to dust extinction inside the molecular cloud.
To describe the ISRF we use the analytic form of \cite*{ZucWalGal2001}, modified to
include the `standard' UV component from \cite{Draine1978} in the energy range $h\nu=5-13.6$~eV.

The opacity of the matter is set in the same manner as in \cite{Bate2014}.  At low
temperatures when dust is present we use the solar
metallicity opacities of \cite{PolMcKChr1985}, 
and for other metallicities we assume that the opacity scales
linearly with the metallicity.  We note that dust properties themselves may 
change at different metallicities \cite[e.g.,][]{RemyRuyer2014}, but we do not attempt to take
this into account.  At higher temperatures we use the gas opacities of \cite{Fergusonetal2005}
with $X=0.70$ which cover heavy element abundances from $Z=0$ to $Z=0.1$. We take
the solar abundance to be ${\rm Z}_\odot=0.02$.  See Section 2.2 of \cite{Bate2014} for further
details.

\vspace{-6pt}

\subsection{Sink particles}
\label{sec:sinks}

The calculations followed the hydrodynamic collapse of each protostar through the first core
phase and into the second collapse (that begins at densities of
$\sim 10^{-7}$~g~cm$^{-3}$) due to molecular hydrogen dissociation \citep{Larson1969}.
However, due to the decreasing size of the time steps, sink particles
\citep{BatBonPri1995} were inserted when the density exceeded
$10^{-5}$~g~cm$^{-3}$.  This density is 
just two orders of magnitude before the
stellar core begins to form (density $\sim 10^{-3}$~g~cm$^{-3}$) and the associated free-fall time
is only one week.

A sink particle is formed by 
replacing the SPH gas particles contained within $r_{\rm acc}=0.5$ au 
of the densest gas particle in a region undergoing second collapse 
by a point mass with the same mass and momentum.  Any gas that 
later falls within this radius is accreted by the point mass 
if it is bound and its specific angular momentum is less than 
that required to form a circular orbit at radius $r_{\rm acc}$ 
from the sink particle.  Thus, gaseous discs around sink 
particles can only be resolved if they have radii $\gsim 1$ au.
Sink particles interact with the gas only via gravity and accretion.
There is no gravitational softening between sink particles.
The angular momentum accreted by a sink particle (its spin) is recorded but plays no further role in the calculation.
The sink particles do not contribute radiative feedback 
\citep[see][for detailed discussions of this limitation]{Bate2012,JonBat2018b}.
Sink particles are merged if they
pass within 0.03 au of each other (i.e., $\approx 6$~R$_\odot$).

\begin{figure*}
\centering
    \includegraphics[width=5cm]{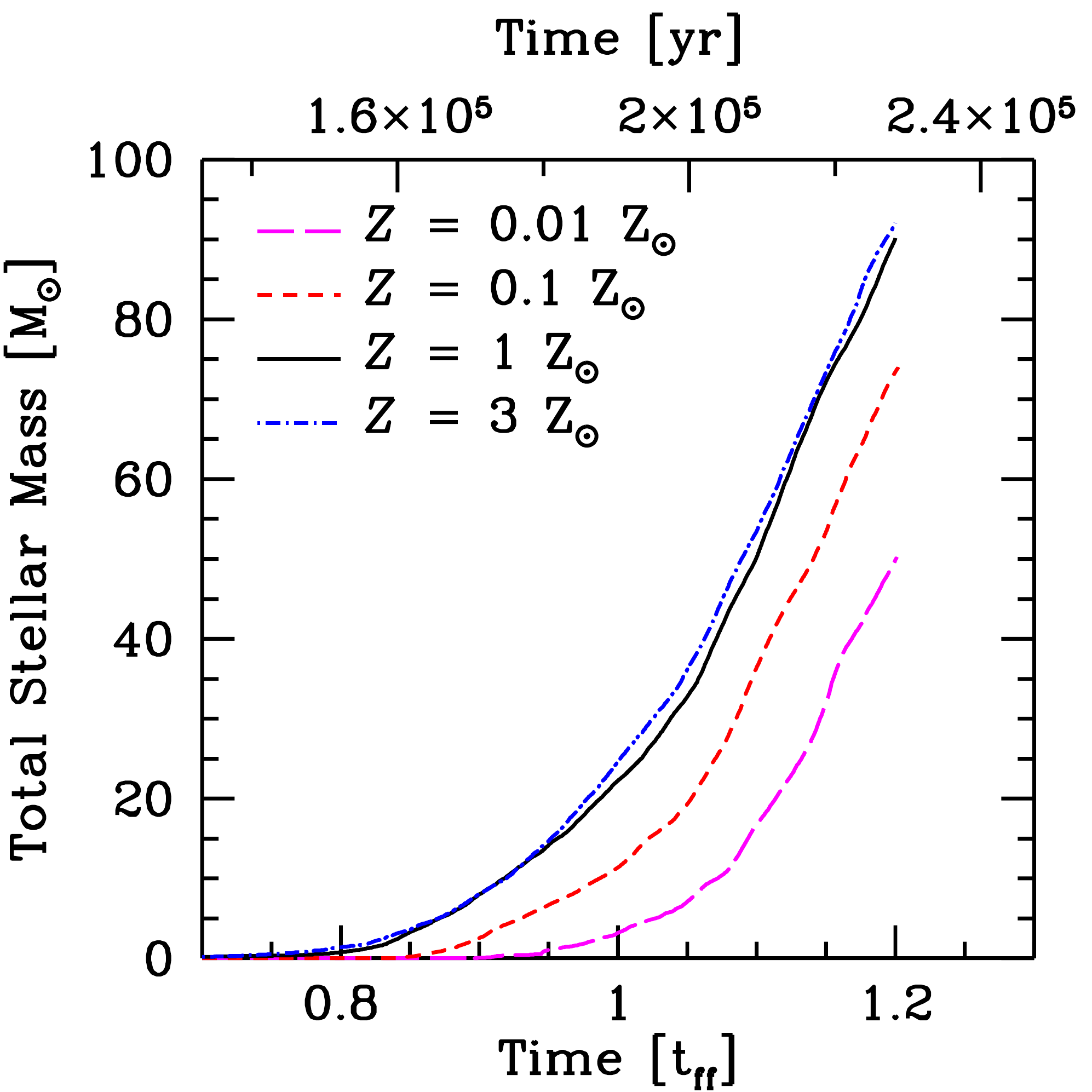} \hspace{0.3cm}
    \includegraphics[width=5cm]{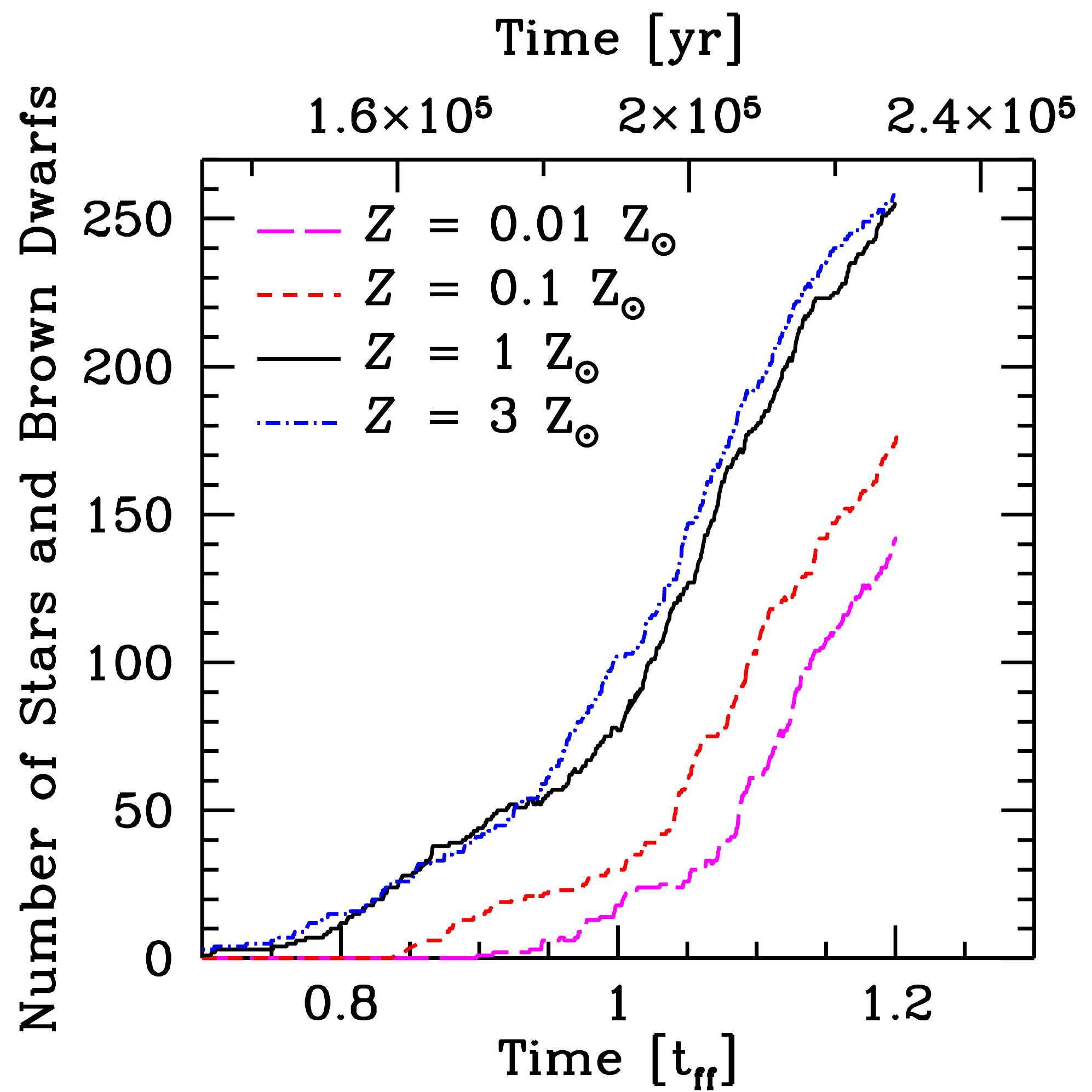} \hspace{0.3cm}
    \includegraphics[width=5cm]{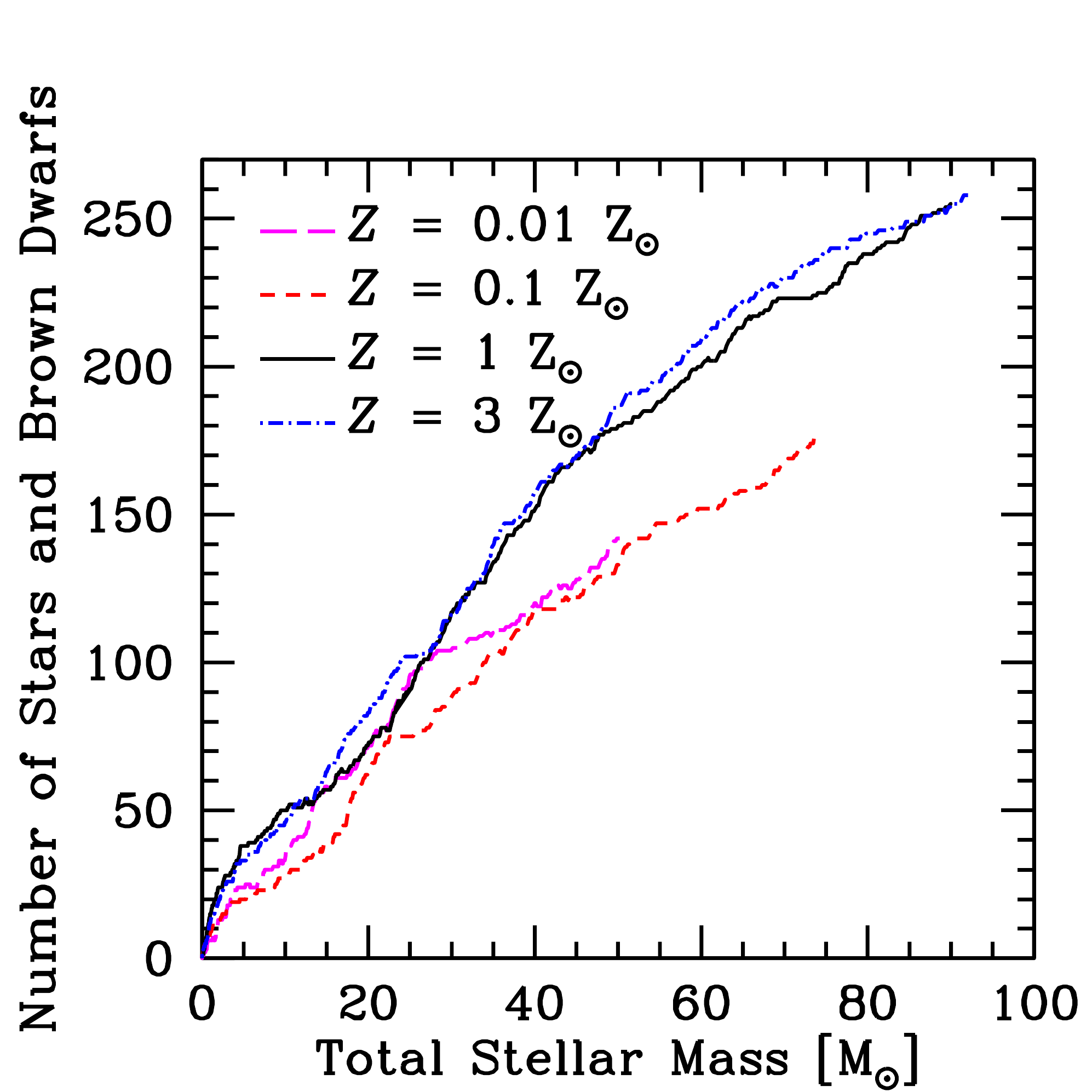}
\caption{The star formation rates obtained for each of the four radiation hydrodynamical calculations.  We plot: the total stellar mass (i.e., the mass contained in sink particles) versus time (left panel), the number of stars and brown dwarfs (i.e., the number of sink particles) versus time (centre panel), and the number of stars and brown dwarfs versus the total stellar mass (right panel).  The different line types are for metallicities of 1/100 (magenta, long dashed), 1/10 (red, short-dashed), 1 (black, solid), and 3 (blue, dot-dashed) times solar. Time is measured from the beginning of the calculation in terms of the free-fall time of the initial cloud (bottom) or years (top). The main difference is that the star formation is delayed in the two lowest metallicity calculations, due to the warmer gas temperatures, but after $t\approx 1.08$~$t_{\rm ff}$ the star formation rates are very similar for all of the calculations.  }
\label{massnumber}
\end{figure*}

\subsection{Initial conditions and resolution}
\label{initialconditions}

The initial density and velocity structure for each calculation are identical to those used by \cite{Bate2012} and \cite{Bate2014}.  For a full description see \cite{Bate2012}.  Briefly, the initial 
conditions for each calculation consisted of a spherical, uniform-density, molecular cloud containing 
500 M$_\odot$ of molecular gas, with a radius of 0.404 pc (83300 au) giving an initial density of 
$1.2\times 10^{-19}$~g~cm$^{-3}$ ($n_{\rm H} \approx 6\times 10^4$~cm$^{-3}$) and an initial free-fall time of the cloud of 
$t_{\rm ff}=6.0\times 10^{12}$~s or $1.90\times 10^5$ years.    
Although each cloud was uniform in density, we imposed an initial 
supersonic `turbulent' velocity field in the same manner
as \citet*{OstStoGam2001} and \cite*{BatBonBro2003}.
We generated a divergence-free random Gaussian velocity field with 
a power spectrum $P(k) \propto k^{-4}$, where $k$ is the wavenumber 
on a $128^3$ uniform grid and the velocities of the particles were interpolated from the grid.  
The velocity field was normalised so that the kinetic energy 
of the turbulence was equal to the magnitude of the gravitational potential 
energy of the cloud (the initial root-mean-square Mach number of the turbulence was ${\cal M}=13.7$ at 10~K).

In \cite{Bate2014}, the temperature of the matter was set to 10~K initially, while in the calculations presented here the gas and dust temperatures are set such that the dust is initially in thermal equilibrium with the local interstellar radiation field, and the gas is in thermal equilibrium such that heating from the interstellar radiation field and cosmic rays is balanced by cooling from atomic and molecular line emission and collisional coupling with the dust.  This produces clouds with dust temperatures that are warmest on the outside and coldest at the centre.  For the highest metallicity, the dust temperature ranges from $T_{\rm dust}=6.3-17$~K. For solar metallicity, $T_{\rm dust}=7.1-17$~K, for $Z=0.1~{\rm Z}_\odot$, $T_{\rm dust}=12-17$~K and for the lowest metallicity, $T_{\rm dust}=16-18$~K. The gas temperatures tend to vary less, with the bulk of the gas in each calculation beginning at $T_{\rm gas}=9.1-9.8$~K.

The calculations used $3.5 \times 10^7$ SPH particles to model the cloud.  This resolution is sufficient to resolve the local Jeans mass throughout the calculation, which is necessary to model fragmentation of collapsing molecular clouds correctly (\citealt{BatBur1997, Trueloveetal1997, Whitworth1998, Bossetal2000}; \citealt*{HubGooWhi2006}).  More recently, there has been much discussion in the literature about the resolution necessary to resolve fragmentation in isolated gravitationally unstable discs (\citealt{Nelson2006,MerBat2011,MerBat2012,HopChr2013,Rice_etal2014,YouCla2015,YouCla2016,LinKra2016}; \citealt*{TakTsuInu2016,BaeKlaKra2017,DenMayMer2017}; \citealt{Klee_etal2017}).  As yet, there is no consensus as to the resolution that is necessary and sufficient to capture fragmentation of such discs.  Moreover, the gravitationally unstable discs that form in the calculation discussed in this paper are usually accreting rapidly, rather than evolving in isolation.  Rapid accretion can be important for driving fragmentation \citep{Bonnell1994,BonBat1994a,Hennebelle_etal2004, KraMatKru2008, Kratter_etal2010}, adding another complication.  \cite{KraLod2016} provide a recent review of gravitational instabilities in circumstellar discs.  The fact that the criteria for disc fragmentation is not well understood should be kept in mind as a caveat.

\section{Results}
\label{sec:results}

\subsection{Cloud evolution}
\label{sec:clouds}

Each of the four radiation hydrodynamical calculations was evolved to $t=1.20~t_{\rm ff}$ (228,300~yrs).  By this time, each of them had converted different amounts of gas into stars, and produced different numbers of stars and brown dwarfs (sink particles with accretion radii of 0.5~AU).  The initial conditions and the statistical properties of the stars and brown dwarfs produced by each calculation are summarised in Table \ref{table1}.  There is a clear trend with metallicity in that the lower metallicity clouds produce fewer objects and less gas is converted to stars, although the numbers for the clouds with solar metallicity ($Z={\rm Z}_\odot$) and three times solar metallicity ($Z=3~{\rm Z}_\odot$) are similar.  However, the mean masses of the stars and brown dwarfs (linear or logarithmic) and the median masses are statistically indistinguishable.

Protostellar mergers occur in all of the calculations (see the last column of Table \ref{table1}).  In these calculations, mergers occur when two sink particles pass within $6~{\rm R}_\odot$.  This merger radius was chosen because low-mass protostars that accrete at high rates ($\dot{M}_* \sim 10^{-6} - 10^{-5}$~M$_\odot$~yr$^{-1}$) are thought to have radii of $2-3~{\rm R}_\odot$ \citep[e.g.][]{Larson1969,HosOmu2009}.  \cite{Bate2014} found that calculations with lower opacities gave more mergers. In Table \ref{table1}, the absolute number of stellar mergers does not display a consistent trend with metallicity.  However, it is important to note that fewer objects are produced with lower metallicity -- there is a consistent trend such that the mean number of mergers per protostar decreases steadily with increasing metallicity (from 1 merger for every 9 objects formed at $Z=0.01~{\rm Z}_\odot$ to 1 merger for every 44 objects at $Z=3~{\rm Z}_\odot$), agreeing with the trend from \cite{Bate2014}.

\begin{figure*}
\centering
    \includegraphics[width=17cm]{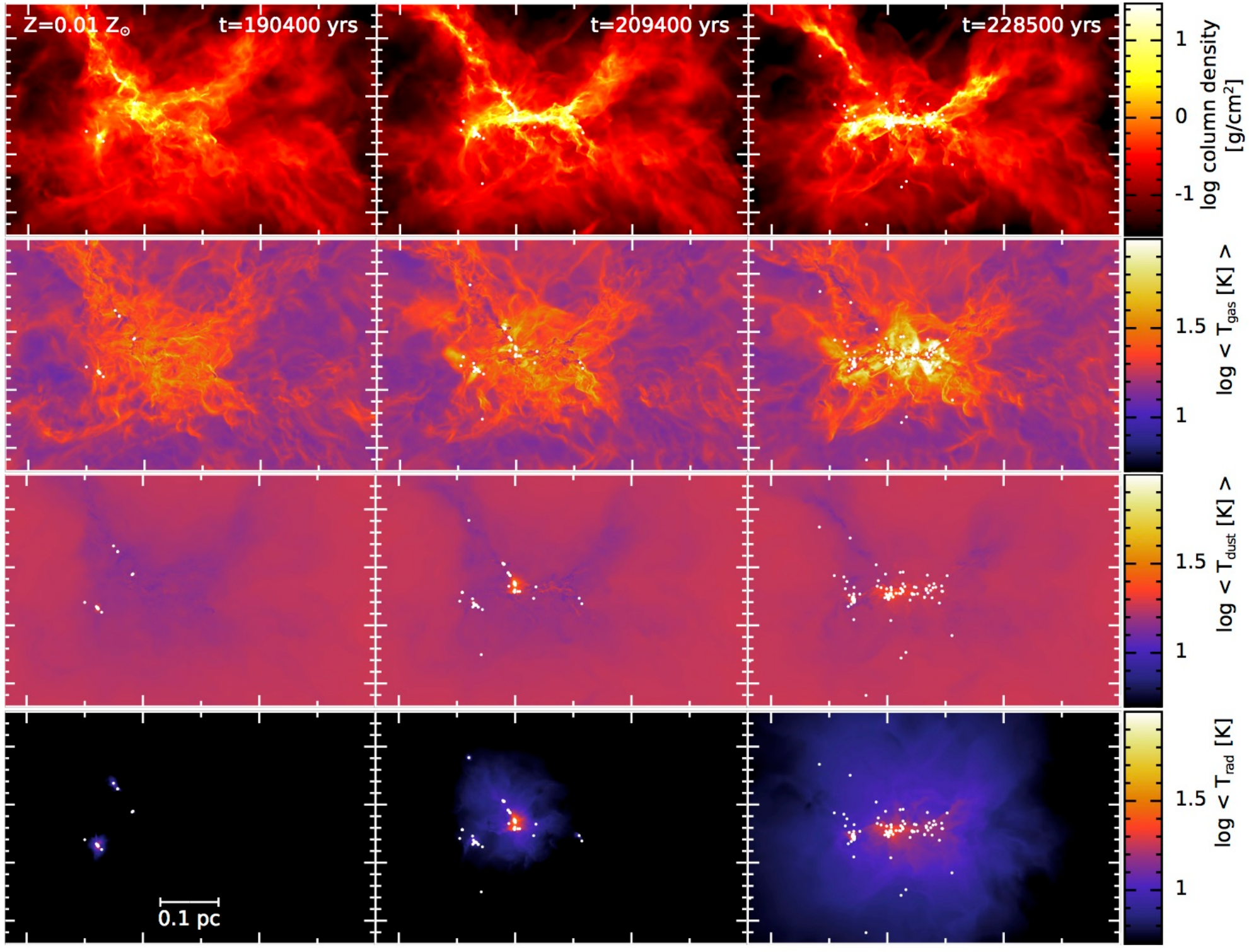} \vspace{0cm}
\caption{Column density and temperature snapshots at three different times ($t=1.00, 1.10, 1.20$~$t_{\rm ff}$ ) for the calculation with a metallicity of 1/100 times solar. From top to bottom, the rows give column density and the mass-weighted gas temperature, dust temperature, and protostellar radiation temperature.  The colour scales are logarithmic.  The column density scale covers $0.03-30$~g~cm$^{-3}$, and the temperature scales cover $5-100$K.  The stars and brown dwarfs are plotted using white circles. The gas temperatures in the dense parts of the cloud are much hotter than the dust temperature due to the low metallicity and poor gas-dust coupling. Animations of the evolution of the column density and gas and dust temperatures are provided in the Additional Supporting Information. }
\label{fig:DTZ001}
\end{figure*}

\begin{figure*}
\centering
    \includegraphics[width=17cm]{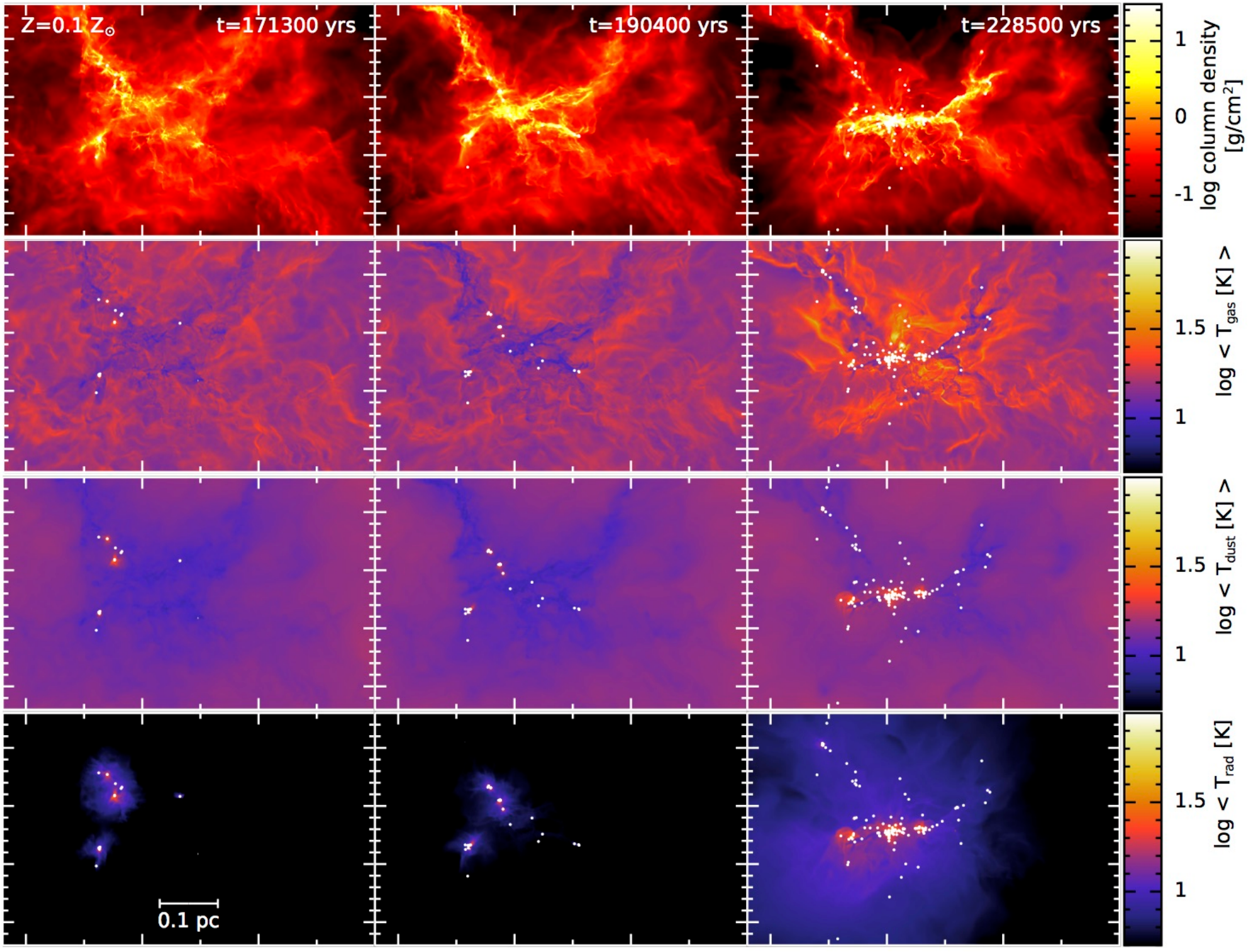} \vspace{0cm}
\caption{Column density and temperature snapshots at three different times ($t=0.90, 1.00, 1.20$~$t_{\rm ff}$ ) for the calculation with a metallicity of 1/10 times solar (note the times differ from those in Fig.~\ref{fig:DTZ001}). From top to bottom, the rows give column density and the mass-weighted gas temperature, dust temperature, and protostellar radiation temperature.  The colour scales are logarithmic.  The column density scale covers $0.03-30$~g~cm$^{-3}$, and the temperature scales cover $5-100$K.  The stars and brown dwarfs are plotted using white circles. The gas temperatures in the dense parts of the cloud tend to be hotter than the dust temperature, but less than in the 1/100 metallicity case.}
\label{fig:DTZ01}
\end{figure*}

\begin{figure*}
\centering
    \includegraphics[width=17cm]{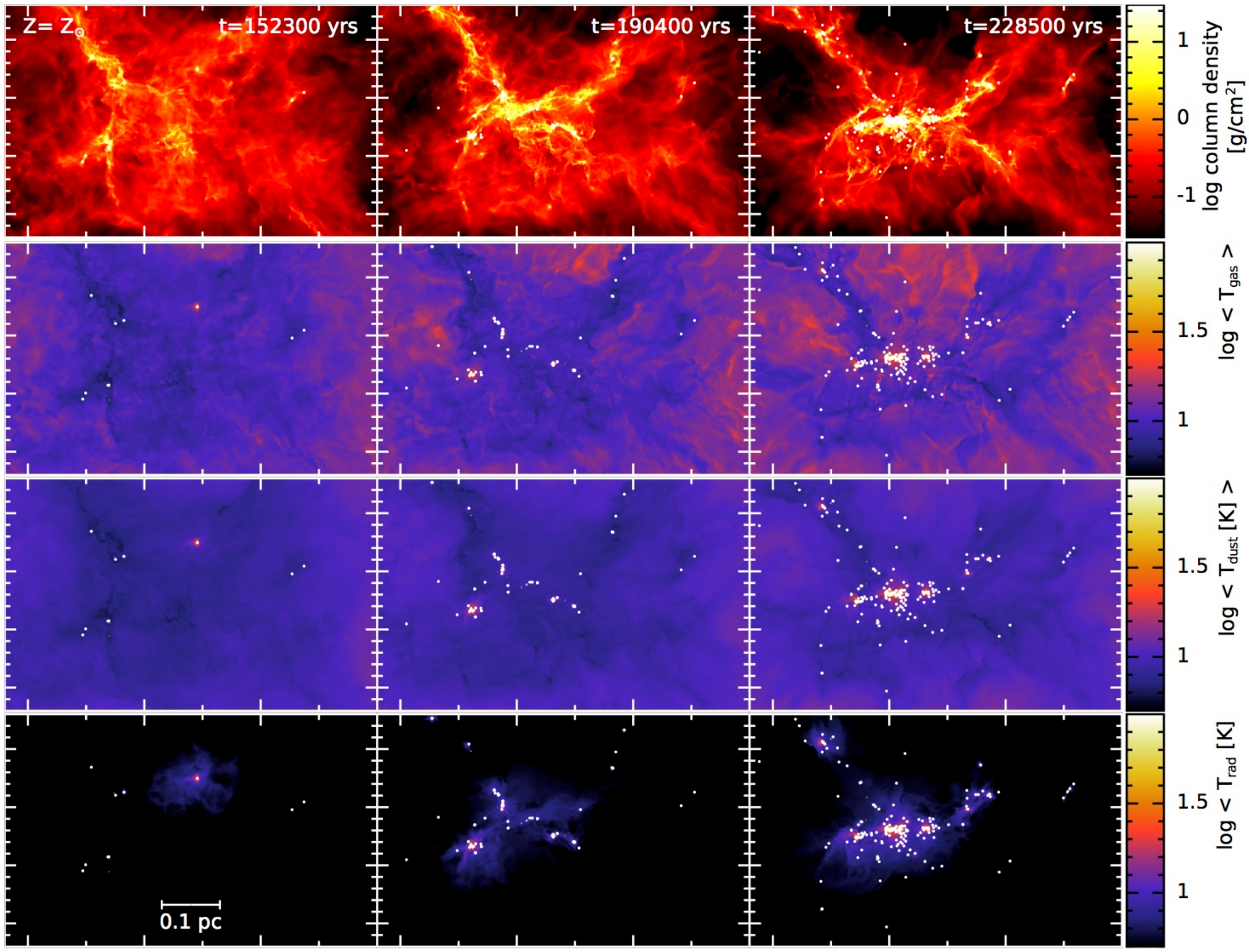} \vspace{0cm}
\caption{Column density and temperature snapshots at three different times ($t=0.80, 1.00, 1.20$~$t_{\rm ff}$ ) for the calculation with solar metallicity (note the times differ from those in Figs.~\ref{fig:DTZ001} and Fig.~\ref{fig:DTZ01}). From top to bottom, the rows give column density and the mass-weighted gas temperature, dust temperature, and protostellar radiation temperature.  The colour scales are logarithmic.  The column density scale covers $0.03-30$~g~cm$^{-3}$, and the temperature scales cover $5-100$K.  The stars and brown dwarfs are plotted using white circles. The gas and dust temperatures are generally lower than in the low metallicity calculations.}
\label{fig:DTZ1}
\end{figure*}

\begin{figure*}
\centering
    \includegraphics[width=17cm]{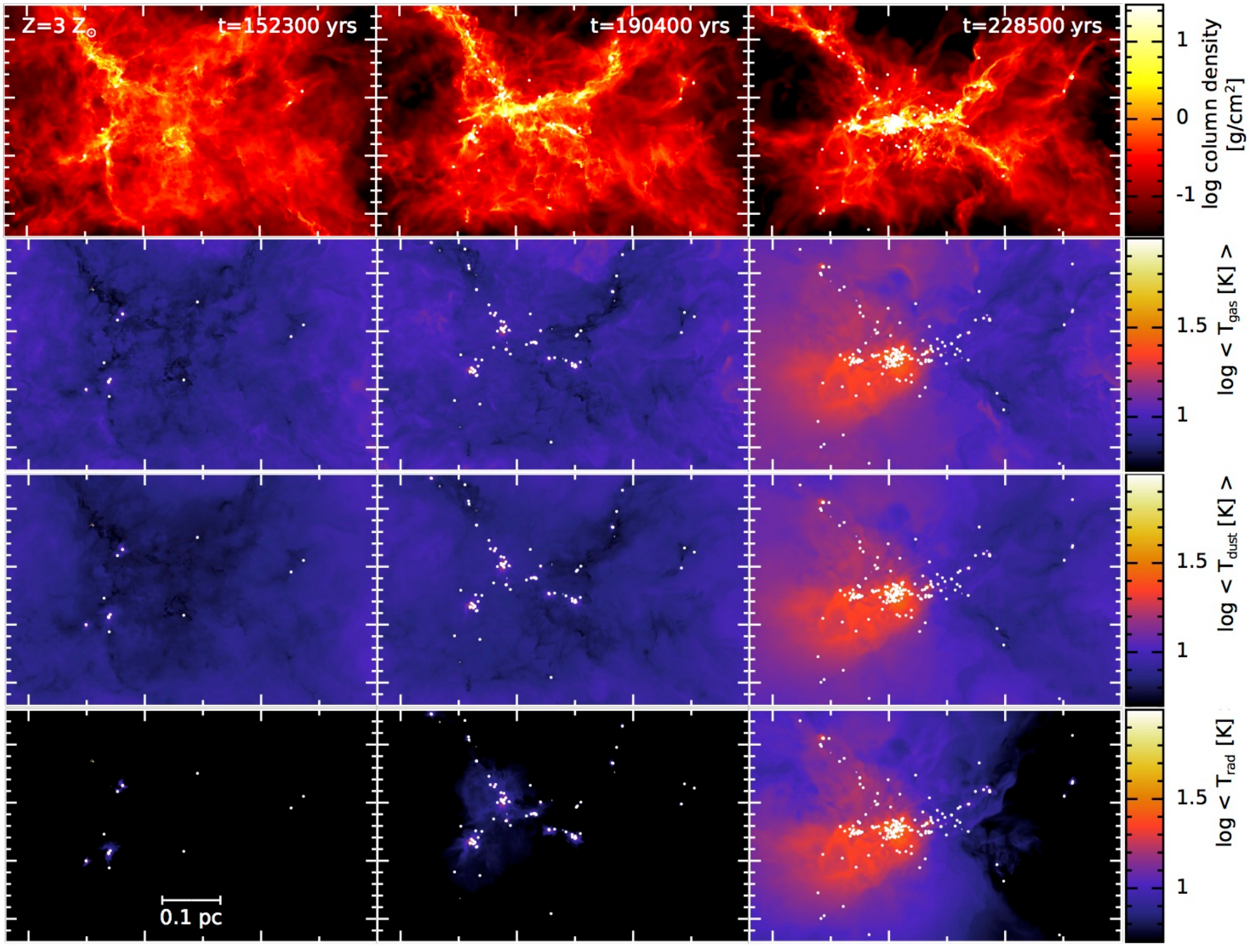} \vspace{0cm}
\caption{Column density and temperature snapshots at three different times ($t=0.80, 1.00, 1.20$~$t_{\rm ff}$ ) for the calculation with 3 times solar metallicity (the same times as Fig.~\ref{fig:DTZ1}, but different from those in Figs.~\ref{fig:DTZ001} and Fig.~\ref{fig:DTZ01}). From top to bottom, the rows give column density and the mass-weighted gas temperature, dust temperature, and protostellar radiation temperature.  The colour scales are logarithmic.  The column density scale covers $0.03-30$~g~cm$^{-3}$, and the temperature scales cover $5-100$K.  The stars and brown dwarfs are plotted using white circles. The gas and dust temperatures are well coupled, and the protostellar radiation is more effective at heating the gas and dust as the cloud becomes more optically-thick.}
\label{fig:DTZ3}
\end{figure*}

Fig.~\ref{massnumber} shows how the star formation rates evolve with time, both in terms of the total mass and the number of stars and brown dwarfs.  It is clear from these graphs that the trends of lower total stellar mass and smaller numbers of objects with lower metallicities at the end of the calculations arise primarily because the star formation is delayed with lower metallicity.  After $t=1.08~t_{\rm ff}$ (205,500~yrs), the star formation rates (the slopes of the lines in the left two panels) are indistinguishable.  But there is a delay in the star formation getting started, particularly with sub-solar metallicities.  In terms of reaching the same total stellar mass, there is a delay of $\approx 0.06~t_{\rm ff}$ (11,400~yrs) between the $Z=0.1~{\rm Z}_\odot$ and higher metallicity calculations, while the delay is doubled for the $Z=0.01~{\rm Z}_\odot$ calculation.  By the end of the calculations, almost 1/5 of the total mass has been converted into stars in the solar and super-solar metallicity calculations, but with the lowest metallicity only 1/10 of the total mass has been converted to stars.

\cite*{LeeChaMur2015} and \cite{MurCha2015} have recently developed an analytic model of the star formation rate in molecular clouds.  They find that in clouds in which the star formation occurs in a globally collapsing region, the total stellar mass grows with time approximately as $\propto t^2$.  Such a functional form agrees with the results presented in the left panel of Fig.~\ref{massnumber}.

In the right panel of Fig.~\ref{massnumber}, we plot the number of stars and brown dwarfs versus their total mass.  If these curves were straight lines lying on top of each other, they would show that the mean stellar masses were always the same.  The end points of each calculation almost lie on a linear relation, hence the indistinguishable mean masses in Table \ref{table1}.  However, we note that the trajectories differ.  The two high metallicity calculations follow similar paths, and the two low metallicity calculations follow similar paths (the latter with more mass associated with a given number of objects).  Whether this difference is significant is unclear -- the slopes of all the lines are similar over the last $\approx 30$~M$_\odot$ of star formation, so there may only be a transient difference.

The reason for the delay of the star formation at low metallicities is that the metal poor gas is hotter.  In Figs.~\ref{fig:DTZ001} to \ref{fig:DTZ3} we provide snapshots of the column densities and mass-weighed temperatures from the four calculations.  The times have been chosen to cover the period of star formation.  They cover the same time period for the two highest metallicity calculations, but shorter periods for the low metallicity calculations because of their delayed star formation.  In each figure, the second, third, and fourth rows give the separate gas, dust, and radiation temperature distributions, respectively.  The radiation temperature is that of the radiation field generated by the star formation itself; the interstellar radiation field that also heats the cloud is treated separately.  We also provide animations of the evolution of the column density and the gas and dust temperatures in the online Additional Supporting Information.

There is a clear progression of the temperatures with metallicity.  The gas temperatures are generally hotter for lower metallicity \citep[e.g.][]{Omukai2000,GloCla2012c}.  The typical gas temperatures in the dense gas before star formation has progressed very far are $T_{\rm gas} \approx 20-50$~K for $Z=0.01~{\rm Z}_\odot$, $10-20$~K for $Z=0.1~{\rm Z}_\odot$, $5-15$~K for $Z={\rm Z}_\odot$, and $5-10$~K for $Z=3~{\rm Z}_\odot$.  The gas is hotter at lower metallicities primarily because it cannot cool as efficiently.  Both the reduced amount of dust, that dominates the cooling at high densities, and the reduction of the atomic and molecular abundances (that dominate the cooling at low densities) contribute.  When shocks and other compressive motions in the clouds generate thermal energy, this cannot be radiated away as easily at low metallicity and, thus, the gas is substantially hotter.  

A secondary effect is that at lower metallicities the inner parts of the clouds are more exposed to the ISRF, which is the same for all four calculations.  This can be seen in the dust temperatures on large scales (that primarily reflect the low density parts of the cloud where the dust and gas are poorly thermally coupled).  At the lowest metallicity, most of the dust has temperatures of $T_{\rm dust} \approx 17$~K as it is essentially exposed to the unattenuated ISRF.  At $Z=0.1~{\rm Z}_\odot$, the dust temperatures range cover $T_{\rm dust} \approx 10-17$~K as the ISRF is significantly attenuated by dust extinction in the denser parts of the cloud.  For solar and super-solar metallicities the extinction is even stronger, resulting in dust temperatures of $T_{\rm dust} \approx 8-12$~K for $Z={\rm Z}_\odot$, and $T_{\rm dust} \approx 5-10$~K for $Z=3~{\rm Z}_\odot$.

As the star formation ramps up, the radiation fields generated by the gas collapsing onto the protostars become stronger.  These primarily heat the gas and dust with the highest densities.  The effect is more obvious at higher metallicities because, as we have already seen, at low metallicity much of the gas and dust is already quite warm.  In the $Z=3~{\rm Z}_\odot$ calculation, the opacities are so high that the inner parts of the cloud are becoming optically thick even to infrared radiation, meaning that the protostellar radiation takes time to propagate out of the cloud and leading to substantially larger temperatures on scales of tens of thousands of AU near the end of the calculation than would be reached without the feedback.

The higher temperatures at lower metallicities give higher gas pressures.  These help to support the clouds against gravity and delay the collapse of the low metallicity clouds, as seen in Fig.~\ref{massnumber}.  The effect of the higher pressures is also seen in the column density snapshots in Figs.~\ref{fig:DTZ001} to \ref{fig:DTZ3} (top rows);  the gas distributions are noticeably smoother at lower metallicities than at high metallicities.  Fig.~\ref{fig:DTZ3} also shows that at the highest metallicity, the gas and dust temperatures are well coupled throughout the bulk of the cloud.

\begin{figure*}
\centering 
    \includegraphics[height=5.0cm]{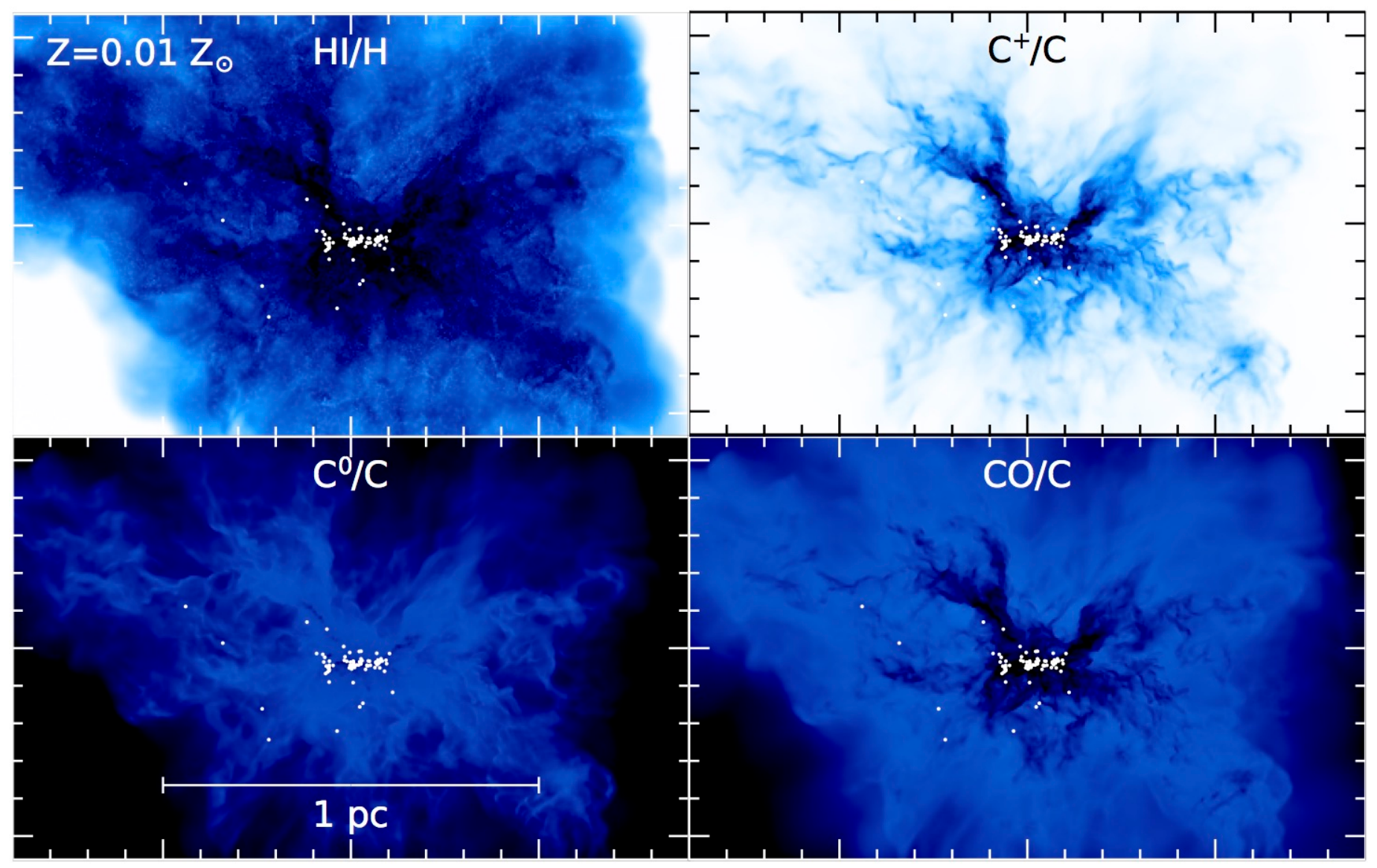} \vspace{0cm} \hspace{0.2cm}
    \includegraphics[height=5.0cm]{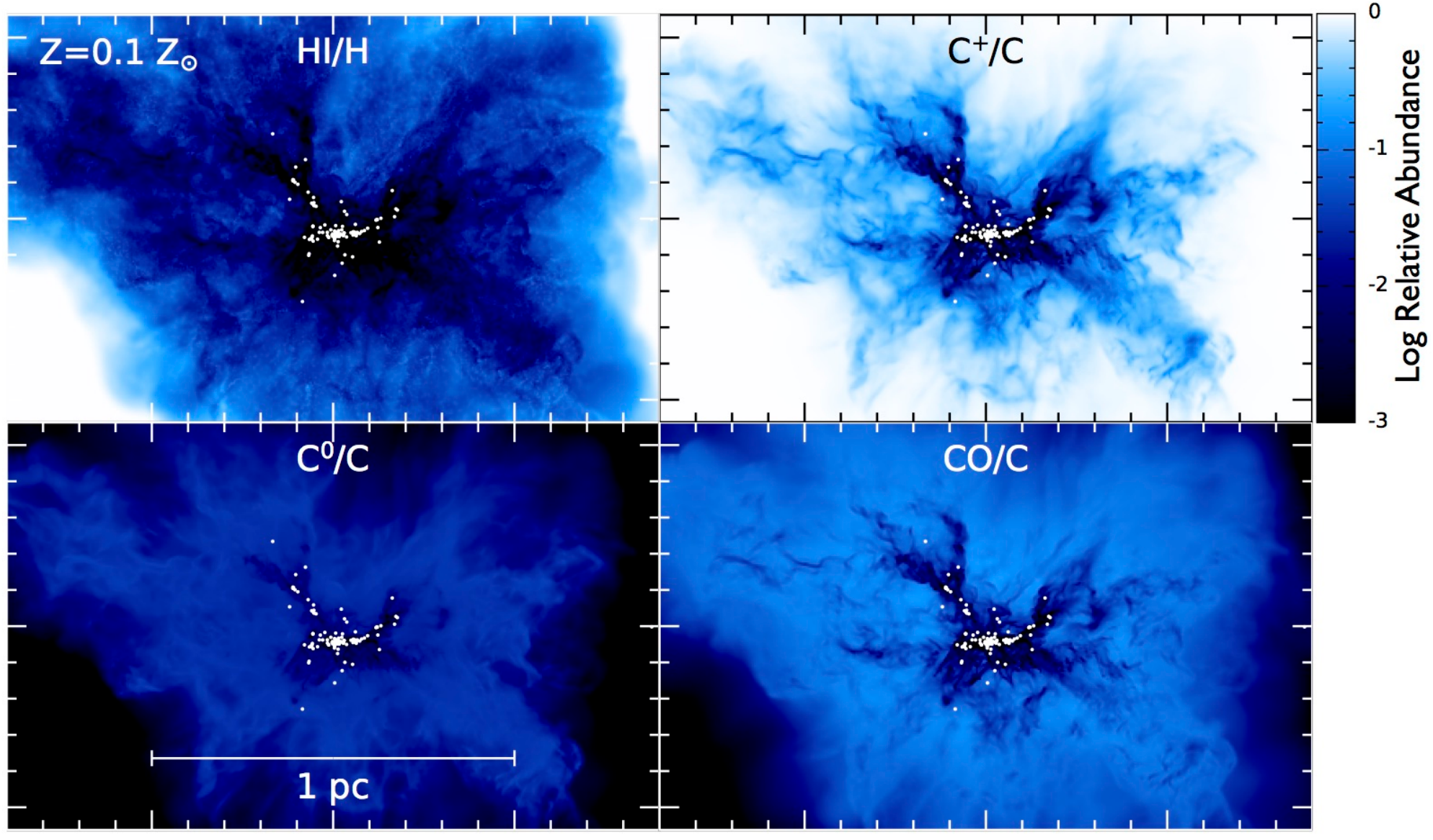} \vspace{0cm} \hspace{0.02cm}
    \includegraphics[height=5.0cm]{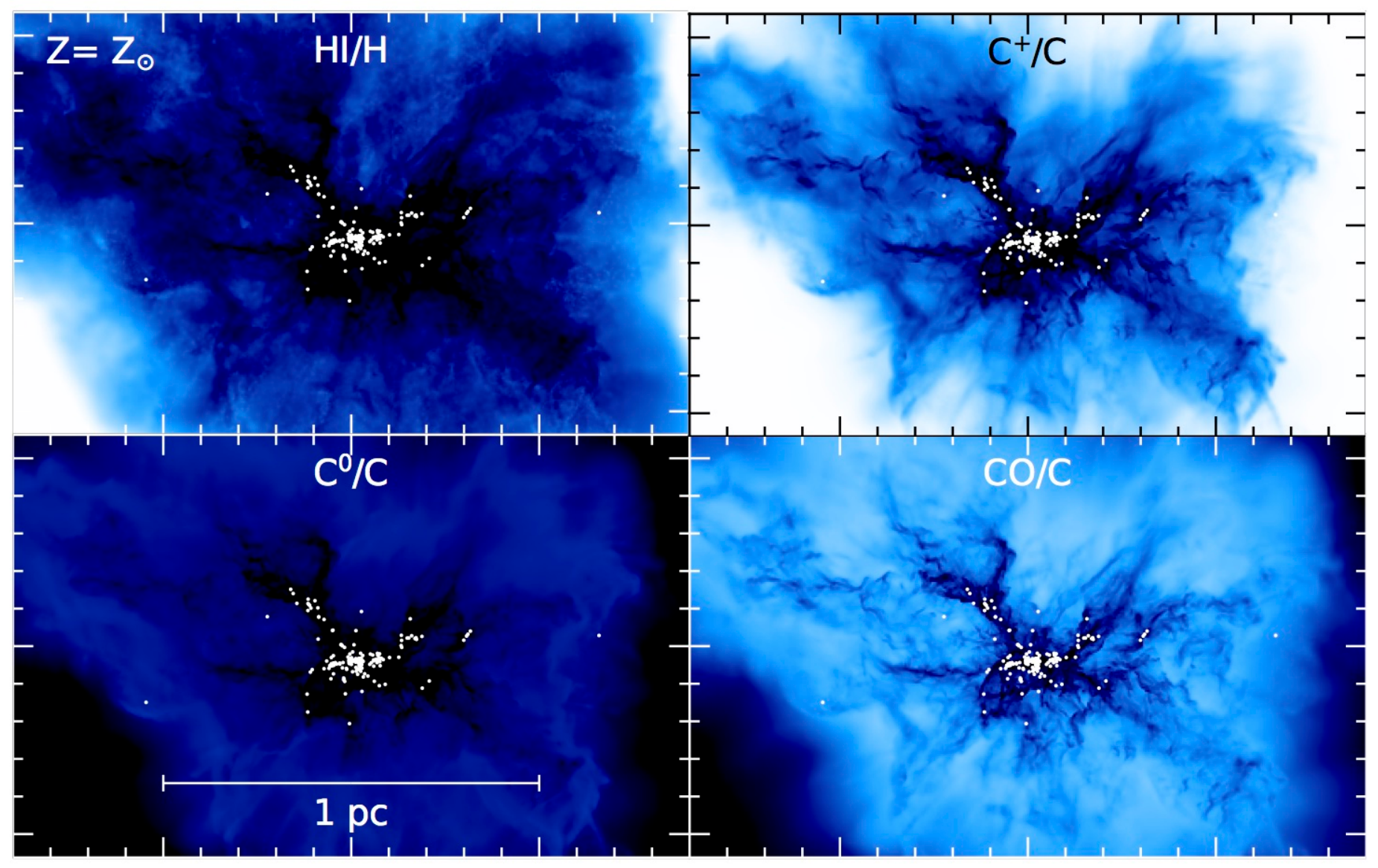} \vspace{0cm} \hspace{0.2cm}
    \includegraphics[height=5.0cm]{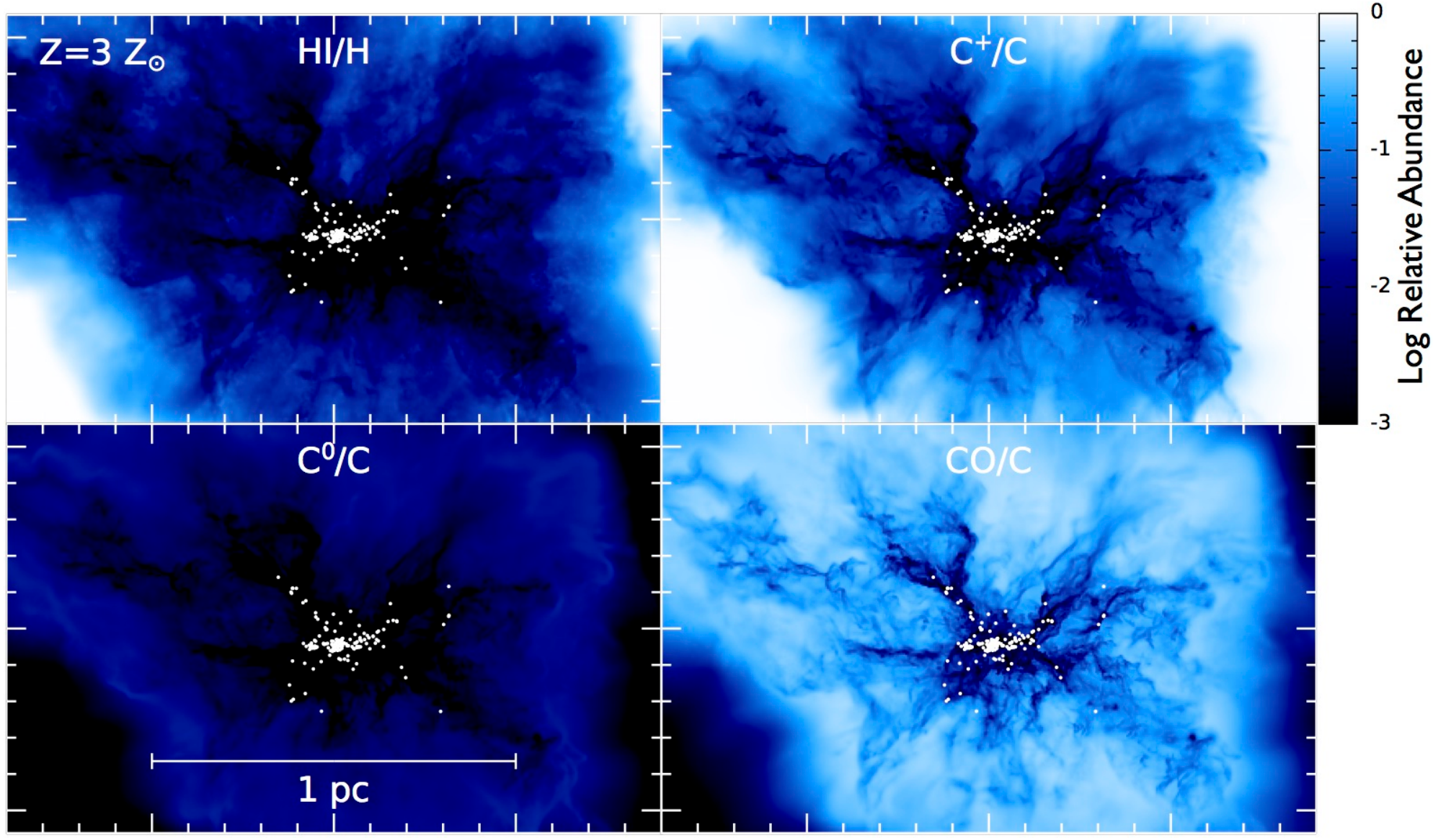} \vspace{0cm}
\caption{Snapshots of the mass-weighted abundances of atomic hydrogen, ionised carbon, neutral carbon, and CO at the end of each of the four calculations with differing metallicity ($t=1.20$~$t_{\rm ff}$). There are four groups of panels, one for each metallicity of 1/100, 1/10, 1, and 3 times solar. For each group of four panels, the top left panel gives the relative abundance HI/H, the top-right panel the relative abundance C$^+$/C, the lower-left panel the relative abundance C$^0$/C, and the lower-right panel the relative abundance of gas phase CO to the total carbon abundance.  The stars and brown dwarfs are plotted using white dots. The colour scale of relative abundances is logarithmic ranging from $10^{-3}$ to unity.  Generally, the outer parts of the clouds contain atomic hydrogen and C$^+$.  The inner parts of the cloud are primarily composed of H$_2$ and carbon is in the form of CO.  At the highest densities, CO freezes out onto grains (hence the drop in gas-phase abundance).  At lower metallicities, there is a greater fraction of atomic hydrogen and more of the carbon is in C$^+$ because there is less extinction of the ISRF by dust and more self-shielding of H$_2$.}
\label{fig:chemistrysnaps}
\end{figure*}

\begin{figure*}
\centering
    \includegraphics[height=3.5cm]{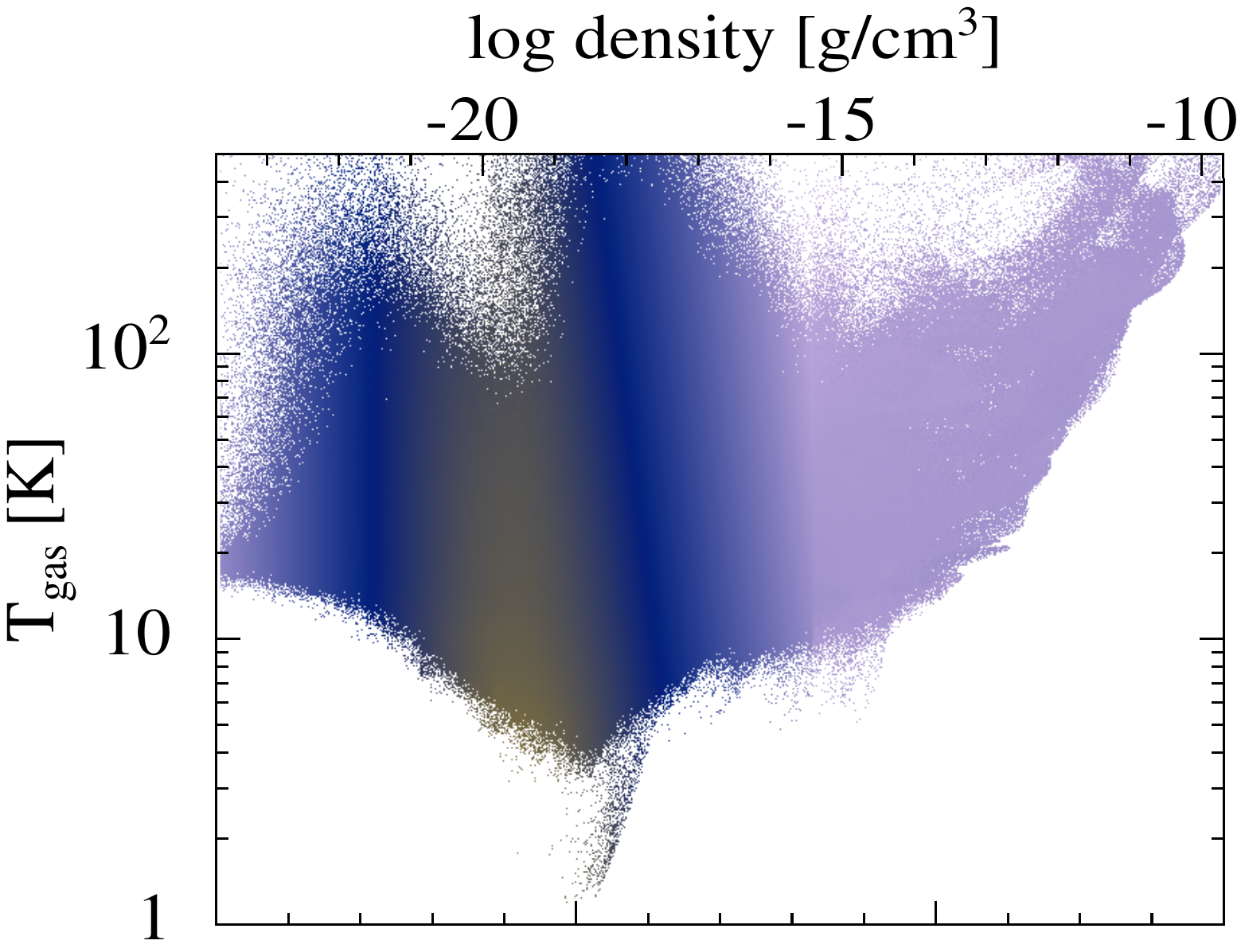} \vspace{0cm} 
    \includegraphics[height=3.5cm]{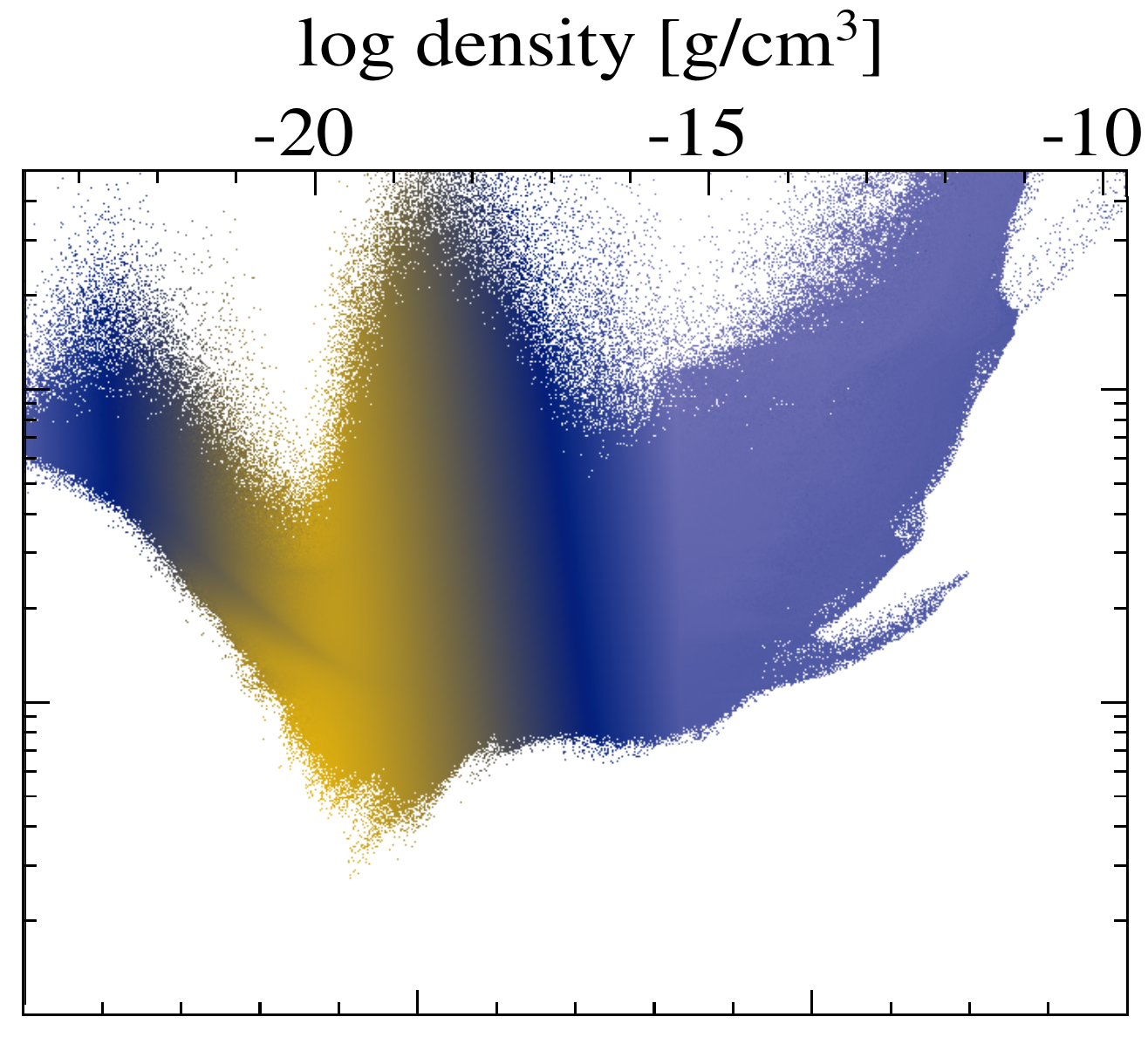} \vspace{0cm} 
    \includegraphics[height=3.5cm]{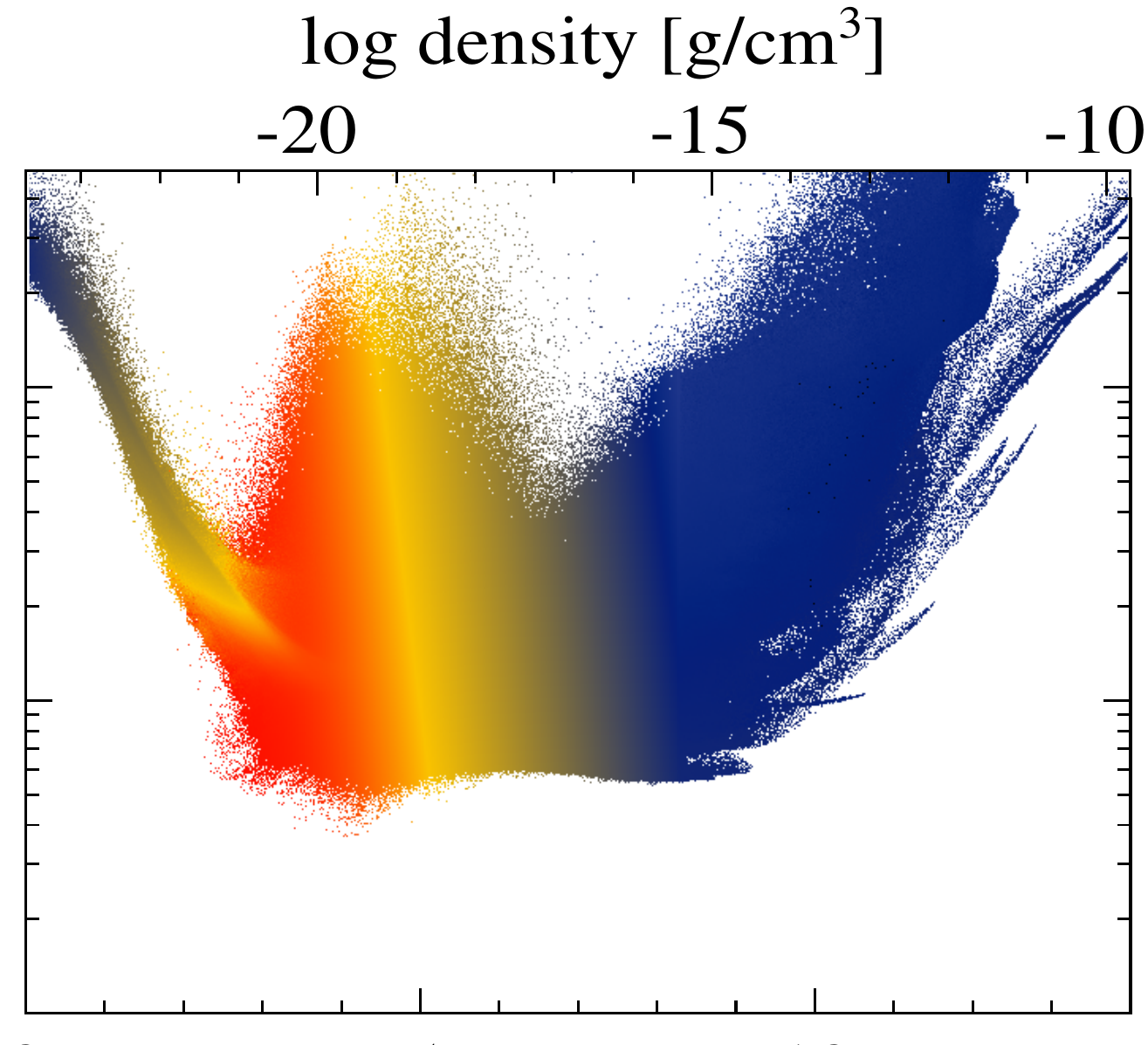} \vspace{0cm} 
    \includegraphics[height=3.5cm]{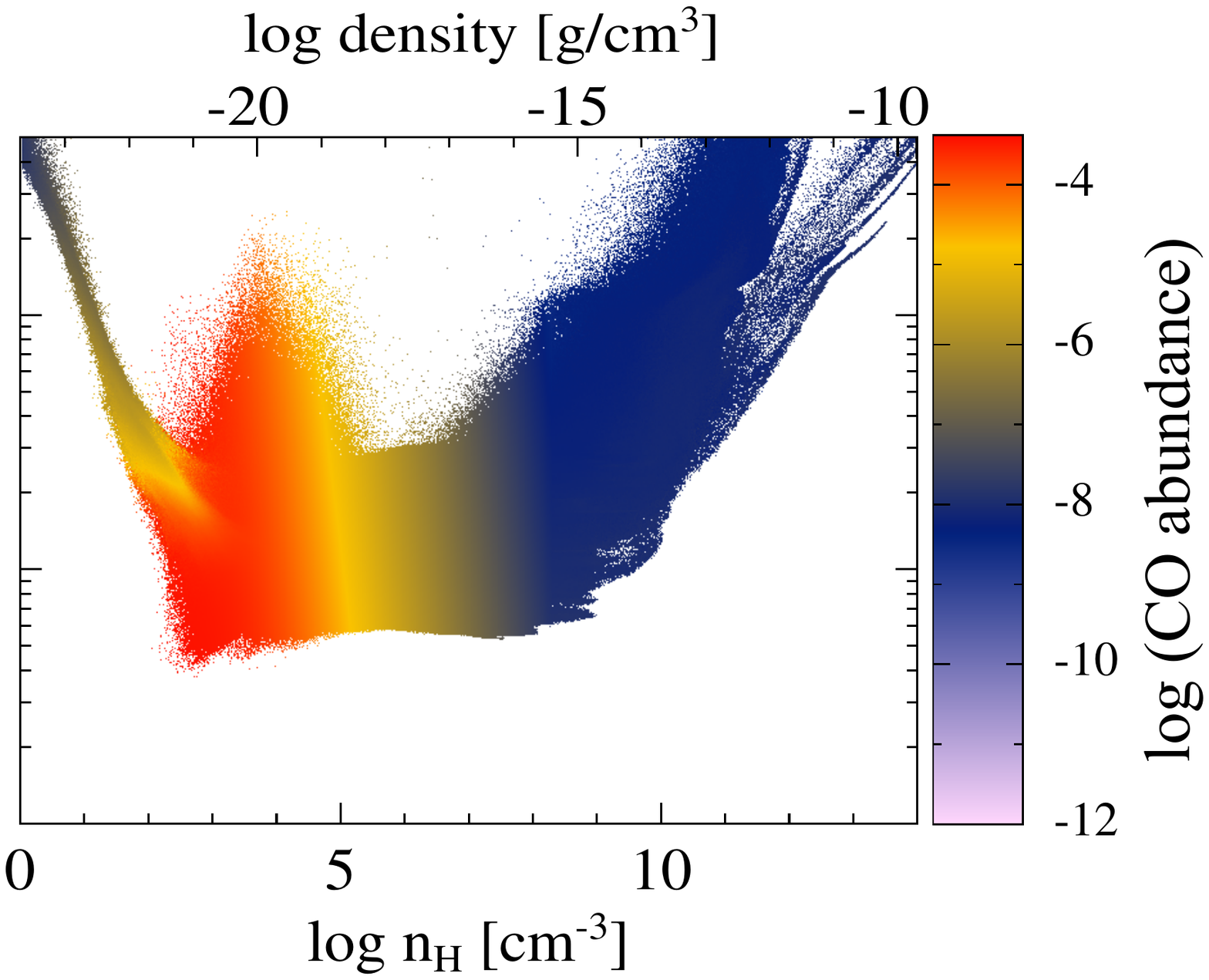} \vspace{0cm}  \hspace{0.01cm}
    \includegraphics[height=3.5cm]{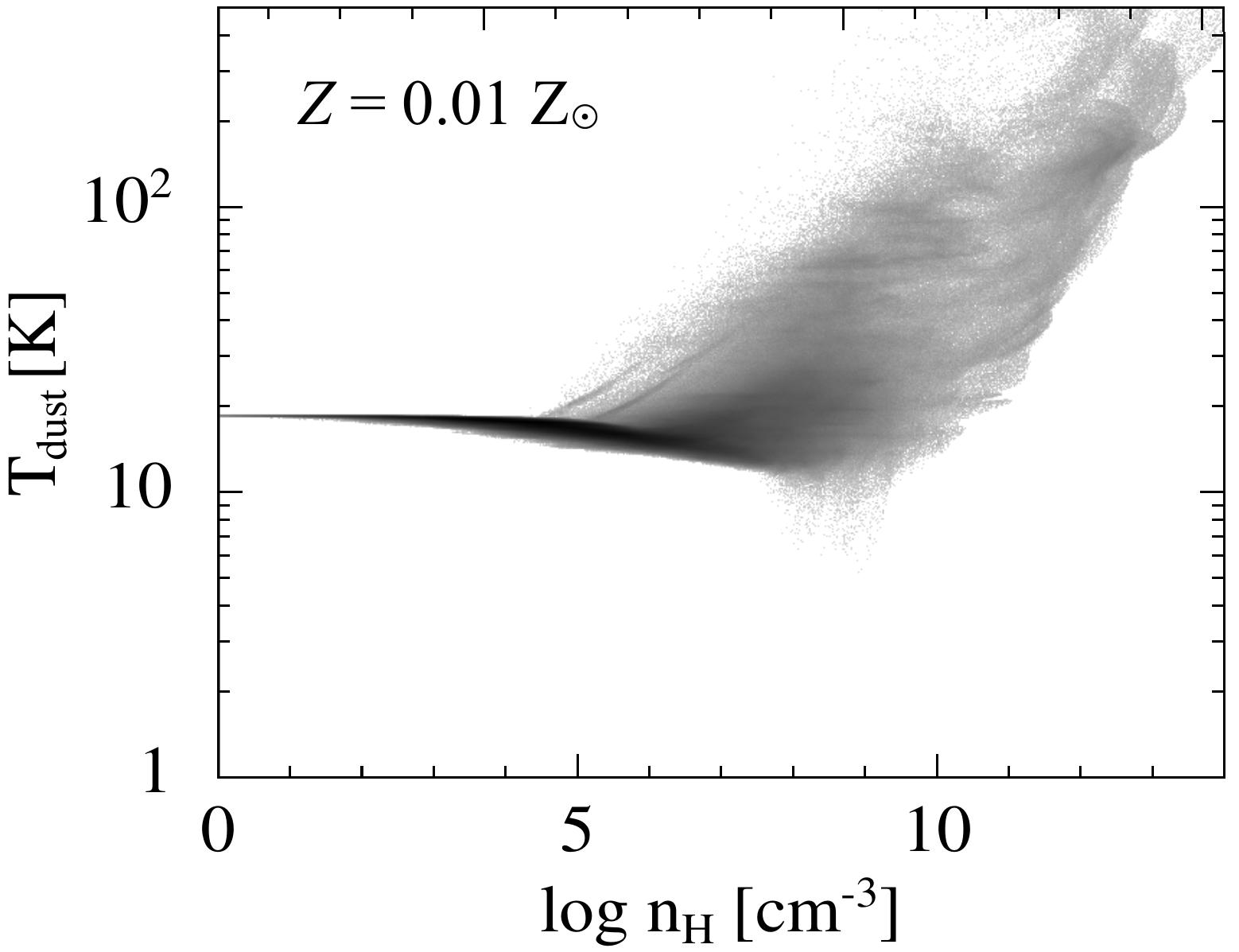} \vspace{0cm} 
    \includegraphics[height=3.5cm]{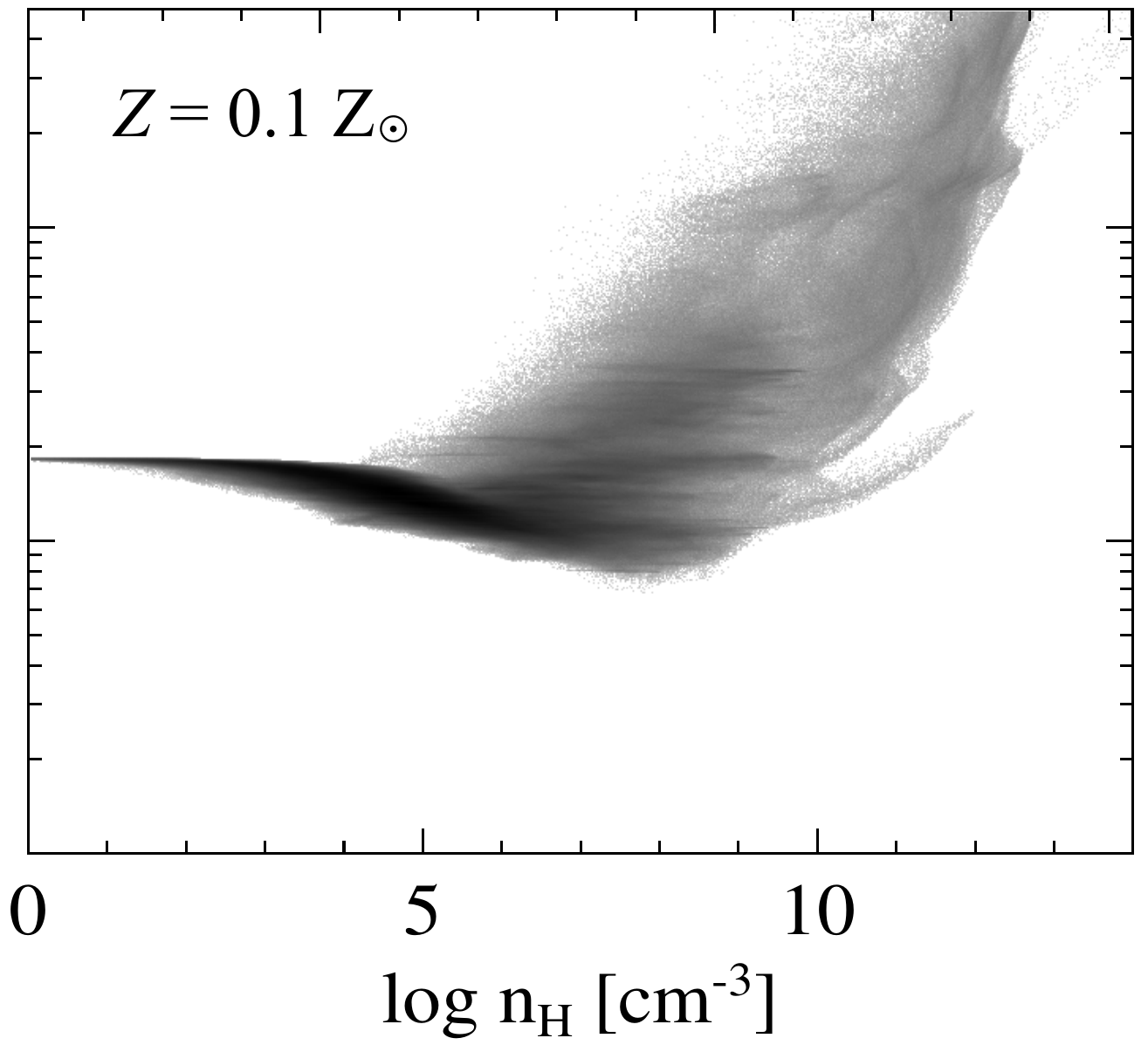} \vspace{0cm}
    \includegraphics[height=3.5cm]{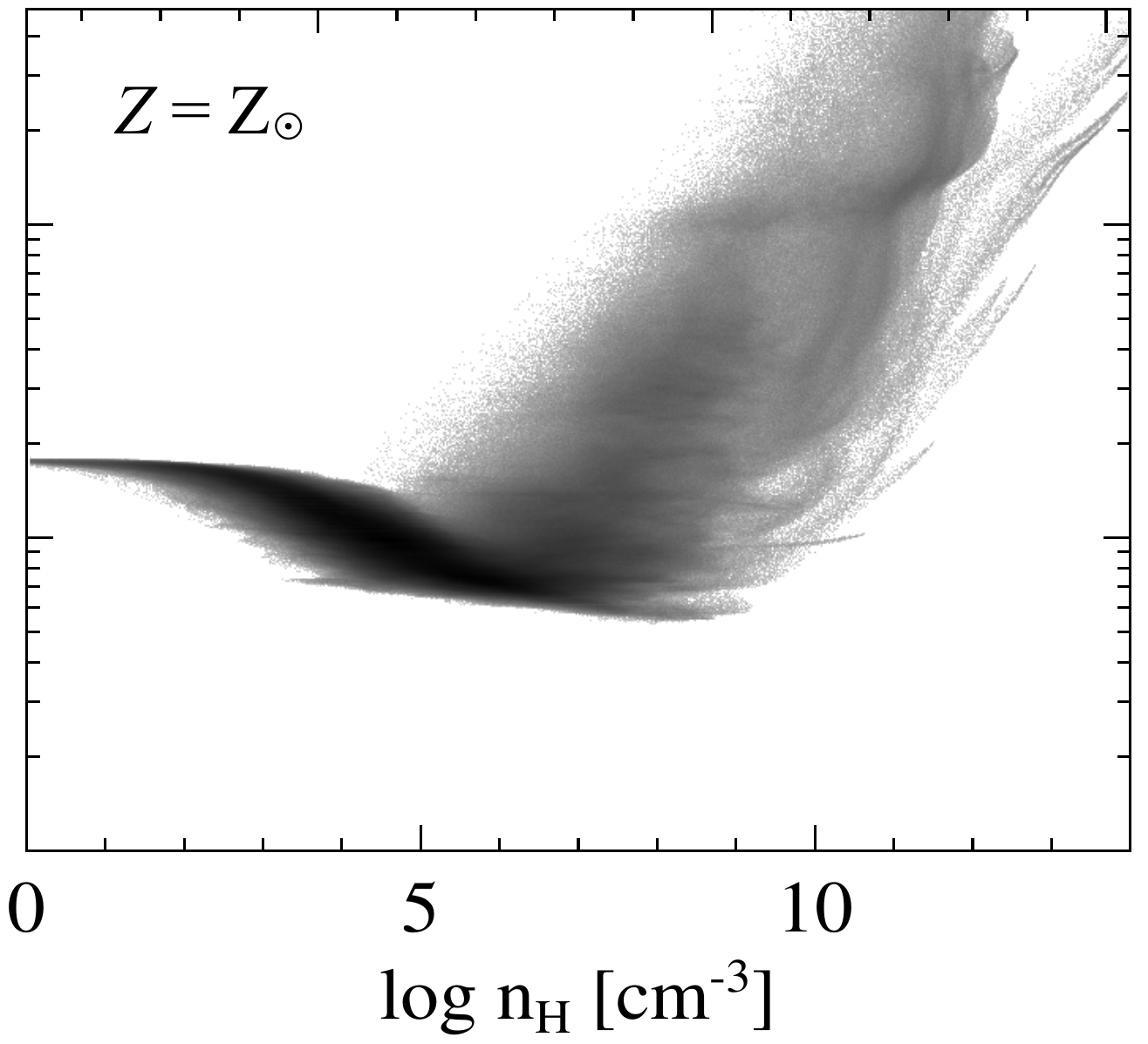} \vspace{0cm}
    \includegraphics[height=3.5cm]{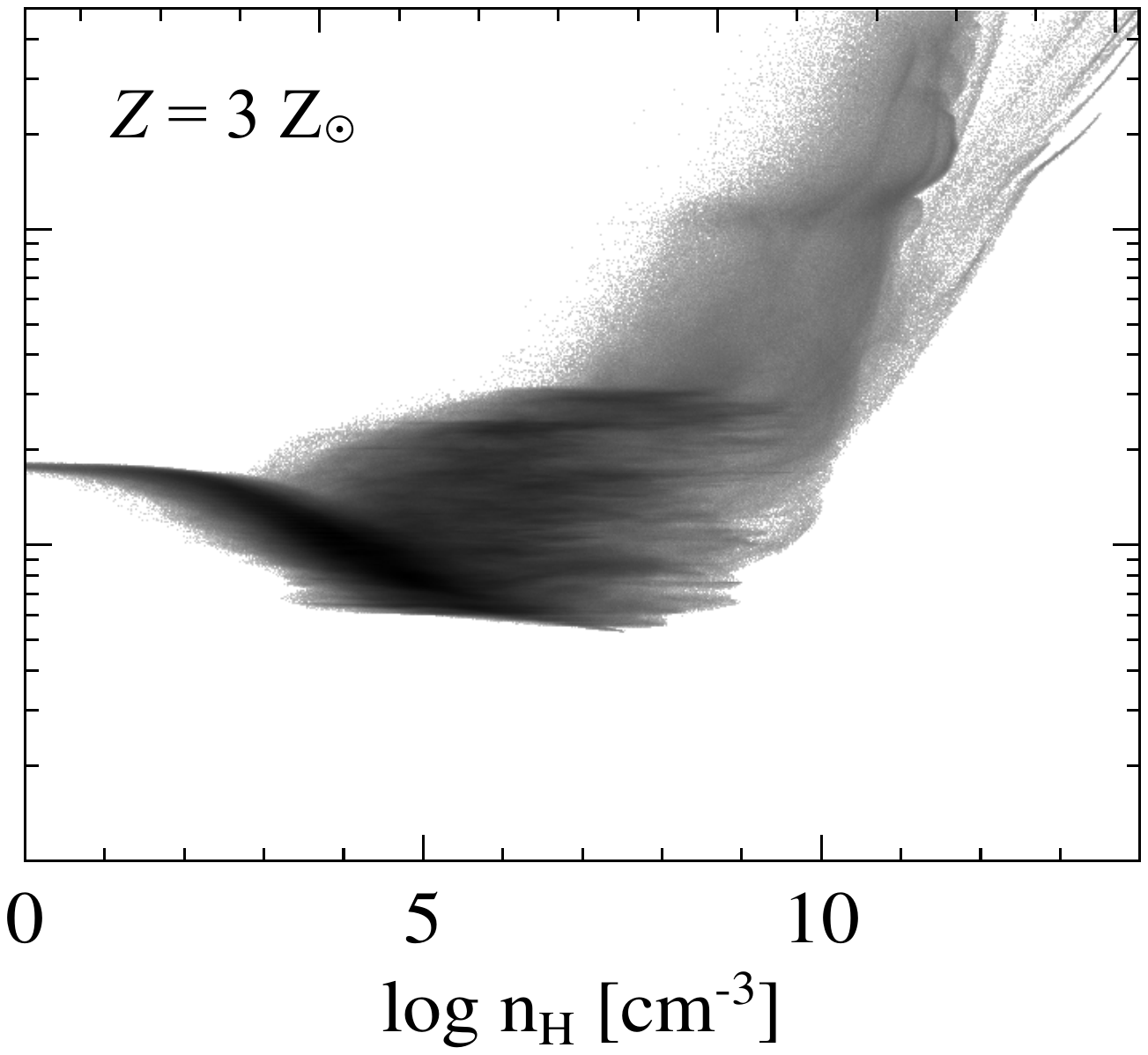} \vspace{0cm} 
\caption{Phase diagrams of temperature vs gas density at the end of each of the calculations ($t=1.20$~$t_{\rm ff}$). The upper row of panels gives the gas temperature, while the lower row gives the dust temperature. The colour scale in the upper panels gives the mean abundance of gas phase CO relative to hydrogen at that temperature and density. Note that the calculations include a prescription for the freeze out of CO onto dust grains, but they do not treat thermal desorption of CO from dust grains at $T_{\rm dust}\gsim 20$~K.  In the dust temperature plots, the grey scale is proportional to the logarithm of the number of fluid elements (i.e., darker regions contain more SPH particles).  The low-density gas ($n_{\rm H}\lsim 10^{10}~{\rm cm}^{-3}$) tends to be warmer at low metallicity than with high metallicity due to the poor cooling, but the high-density gas ($n_{\rm H}\gsim 10^{10}~{\rm cm}^{-3}$) tends to be cooler due to the reduced optical depths. The density above which the gas and dust temperatures become well-coupled increases with decreasing metallicity, from $n_{\rm H}\approx 10^5~{\rm cm}^{-3}$ at $Z=3~{\rm Z}_\odot$ to $n_{\rm H}\approx 10^9~{\rm cm}^{-3}$ at $Z=0.01~{\rm Z}_\odot$.}
\label{fig:phase}
\end{figure*}

\subsection{Chemistry}

The method used to perform these calculations is unique in that it is the first to combine a model of the diffuse ISM with radiative transfer to treat the radiation generated from the collapsing gas.  As part of the treatment of the diffuse ISM, the calculations provide some chemical information about the gas.  In particular, the hydrogen may be atomic, molecular, or a mixture of the two forms, and there is a model for the chemical form of carbon.  The form of hydrogen is important because molecular hydrogen formation can provide heating, while carbon is one of the primary coolants.  We model whether carbon is in the form of ionised carbon, C$^+$, neutral carbon, C$^0$, or molecular in the form of CO.  There is also a model for the freezing out of CO onto dust grains \citep[see][]{BatKet2015}.

In Fig.~\ref{fig:chemistrysnaps} we provide snapshots of the chemical make up of the clouds at the end of each of the four calculations.  The snapshots depict large scales because the chemistry of these species varies most in the low-density gas and in the outer parts of the clouds.  Note that the carbon abundances in the figure are scaled relative to the total carbon abundance in the particular calculation (i.e. the absolute carbon abundance is 100 times lower in the $Z=0.01~{\rm Z}_\odot$ calculation than in the solar metallicity calculation).

The first point to note is that the clouds modelled in this paper have high initial densities ($n_{\rm H} \approx 6\times 10^4$~cm$^{-3}$).  Thus, the hydrogen deep within the clouds is almost entirely molecular.  The atomic hydrogen abundance is only high in the outer parts of the clouds where it is exposed to the ISRF.  At low metallicity, the atomic hydrogen abundance within the cloud is higher since there is less extinction of the ISRF by dust and less self-shielding provided by the molecular hydrogen, but even with $Z=0.01~{\rm Z}_\odot$ the hydrogen is 97.7 percent molecular at the end of the calculation.

Similarly, in the outer parts of the clouds, carbon is almost entirely in the form of C$^+$ because it is exposed to the ISRF.  C$^+$ is a strong atomic line coolant and, along with atomic oxygen, provides much of the cooling at low-densities.  Deep within the clouds, the main gas phase form of carbon is CO, which is also a strong coolant.  At intermediate depths carbon is found in neutral atomic form, but it is never very abundant.  There is a similar dependence of the C$^+$/C relative abundance on metallicity as form HI/H except that the C$^+$/C ratio is always higher deeper into the cloud than for HI/H.  At low metallicity, there is less dust extinction, so the ISRF penetrates further into the cloud and there is a greater fraction of C$^+$.  The reverse is the case at super-solar metallicity.

Fig.~\ref{fig:phase} provides temperature-density phase diagrams at the end of each of the calculations. The upper panels use the gas temperature and also provide the gas-phase CO abundance relative to hydrogen using the colour scale.  Note that the chemistry model treats the freeze out of CO on to dust grains and desorption of CO by cosmic rays, but it does not treat the thermal desorption that occurs at dust temperatures above 20~K (e.g., in protostellar discs).  We neglect this because we only treat the chemistry in order to provide realistic gas temperatures and the primary coolant at the densities above which CO freeze out becomes important is usually the dust rather than the CO.
The lower panels of Fig.~\ref{fig:phase} use the dust temperature and the grey scale is proportional to the logarithm of the amount of dust at each temperature and density.  

In the gas temperature-density phase diagrams, we see that there is a large dispersion of the gas temperature at a given density; a barotropic equation of state would be a poor approximation in most of the parameter space. In general, the low-density gas ($n_{\rm H}\lsim 10^{10}~{\rm cm}^{-3}$) tends to be warmer at low metallicity than with high metallicity due to the poor cooling.  On the other hand, the high-density gas ($n_{\rm H}\gsim 10^{10}~{\rm cm}^{-3}$) tends to be cooler at low metallicity than high metallicity due to the reduced optical depths. Comparing the upper and lower panels of Fig.~\ref{fig:phase}, the density above which the gas and dust temperatures become well-coupled increases with decreasing metallicity, from $n_{\rm H}\approx 10^5~{\rm cm}^{-3}$ at $Z=3~{\rm Z}_\odot$ to $n_{\rm H}\approx 10^9~{\rm cm}^{-3}$ at $Z=0.01~{\rm Z}_\odot$.  Since the initial density of the clouds is $n_{\rm H} \approx 6\times 10^4$~cm$^{-3}$, this means that the dust and gas are thermally well coupled throughout the bulk of the cloud in the $Z=3~{\rm Z}_\odot$ calculation, as already noted from Fig.~\ref{fig:DTZ3}.

\subsection{The statistical properties of the stellar populations}

In this section, we compare the statistical properties of the stars and brown dwarfs that are formed in each of the four calculations with different metallicities.  As in papers that have discussed previous similar calculations \citep{Bate2009a,Bate2012,Bate2014}, we study the distributions of stellar masses, the multiplicity of the populations, the distributions of separations of multiple stellar systems, and the mass ratio distributions of binary systems.  However, for the purpose of relative brevity, we do not discuss other statistics such as the eccentricity distributions of multiple systems, the orbits of triple systems, the relative orientations of sink particle spins in multiple systems, the accretion histories or kinematics of the stars, or the distributions of closest encounters.  We do note, however, that we find no evidence that these other properties vary with metallicity.

\begin{figure*}
\centering \vspace{-0.5cm}
\mbox {    \includegraphics[width=8.cm]{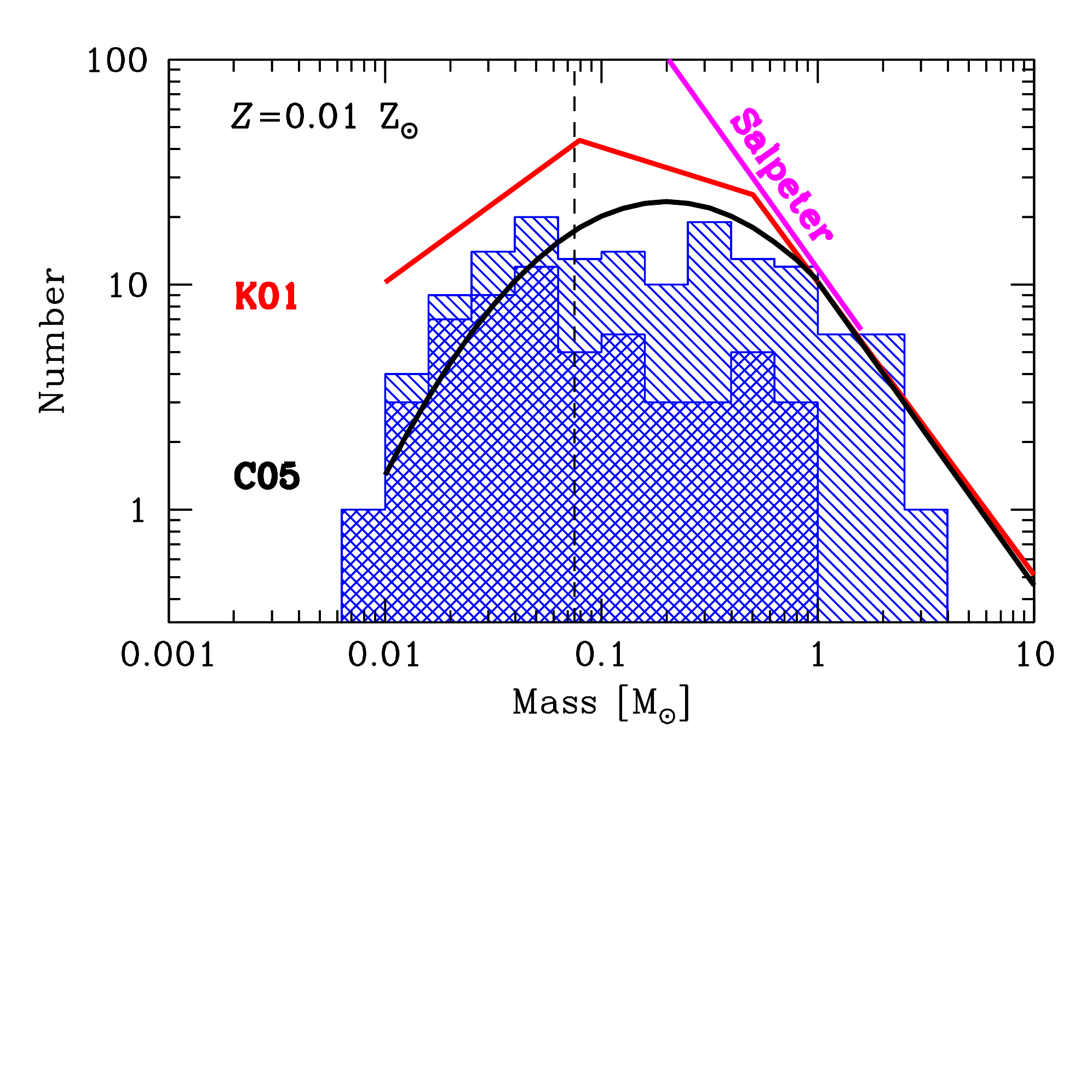}
    \includegraphics[width=8.cm]{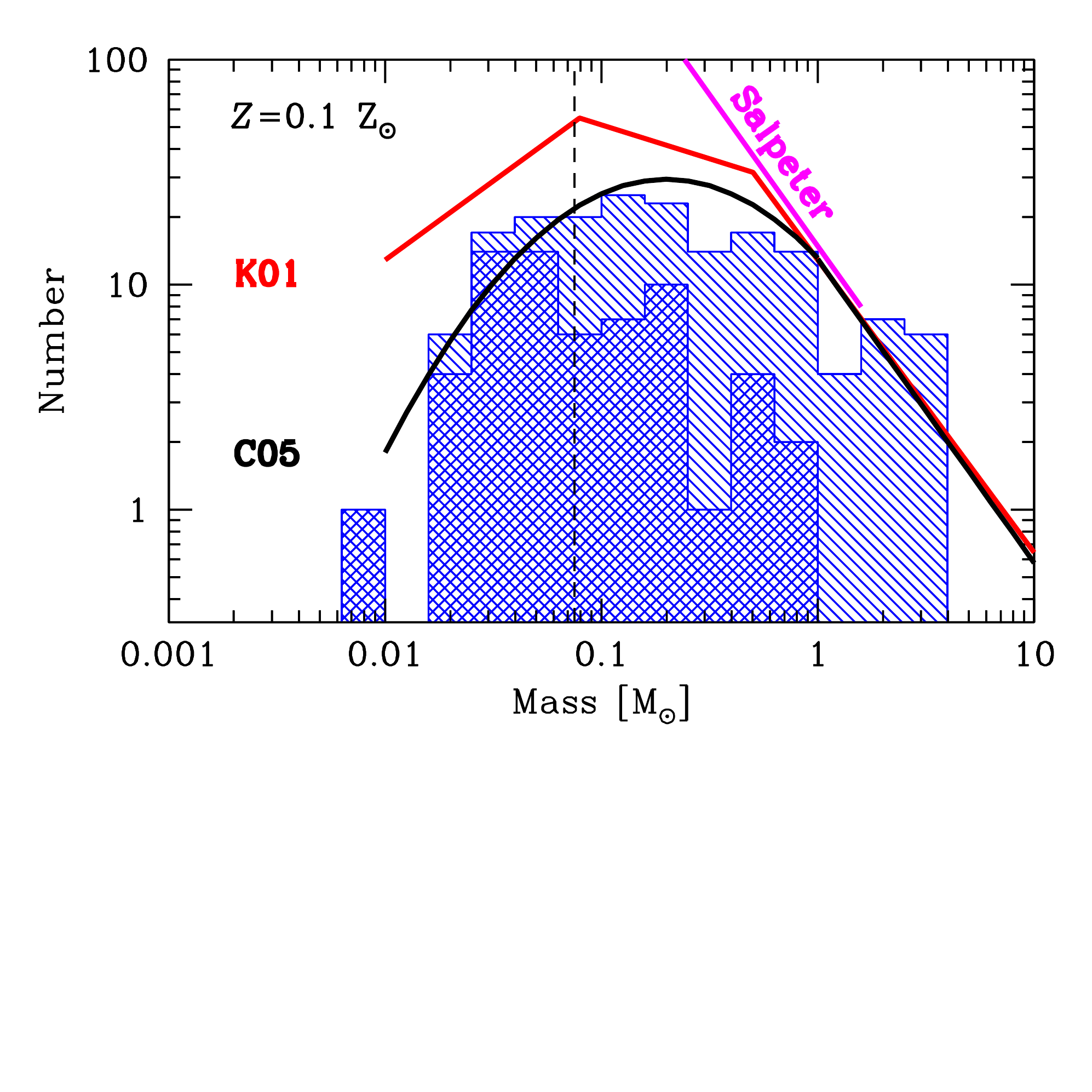} }\vspace{-2.8cm}
    \includegraphics[width=8.cm]{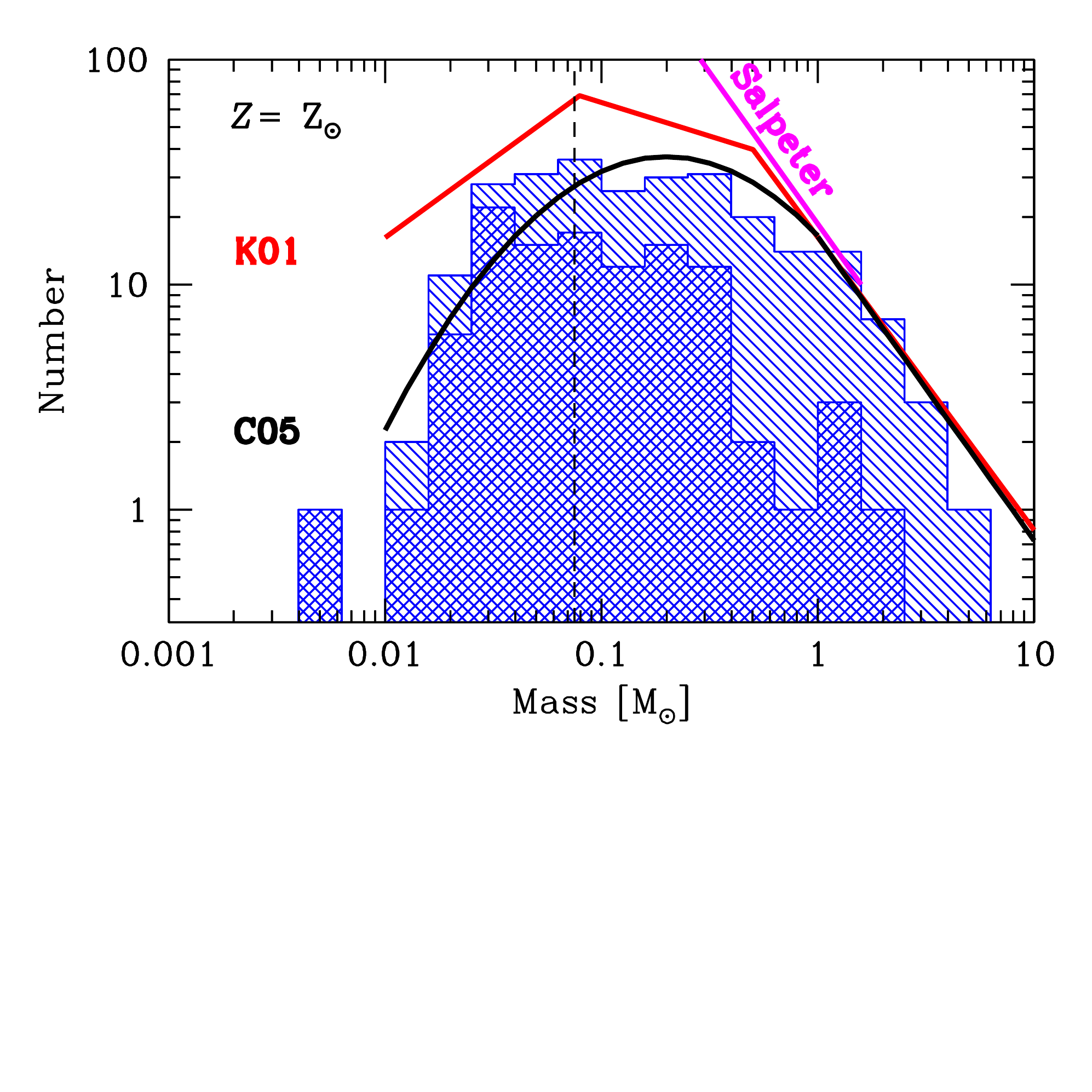}
    \includegraphics[width=8.cm]{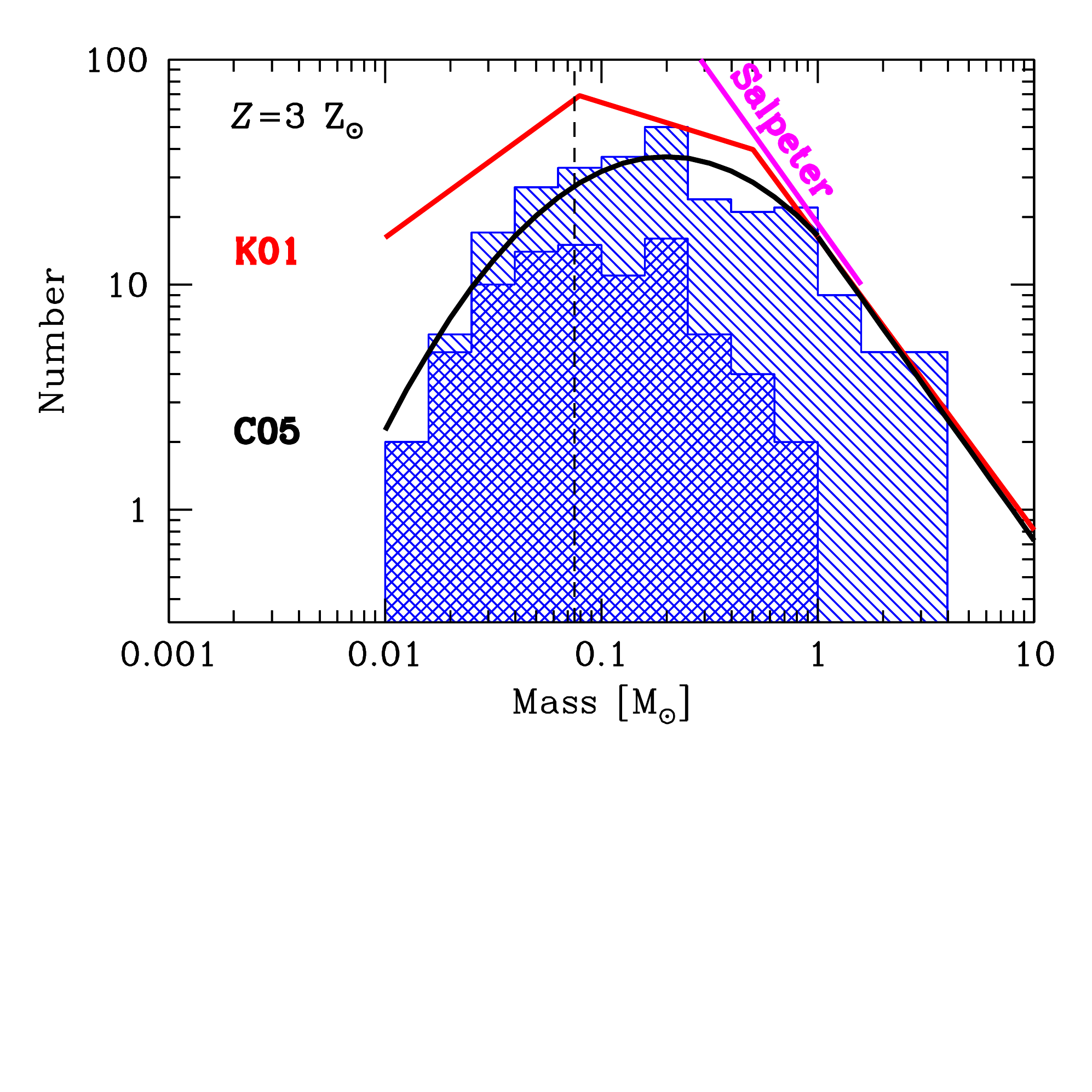}
    \vspace{-2.8cm}
\caption{Histograms giving the initial mass functions of the stars and brown dwarfs produced by the four radiation hydrodynamical calculations, each at $t=1.20t_{\rm ff}$.  The double hatched histograms are used to denote those objects that have stopped accreting (defined as accreting at a rate of less than $10^{-7}$~M$_\odot$~yr$^{-1}$), while those objects that are still accreting are plotted using single hatching.  Each of the mass functions are in reasonable agreement with the Chabrier (2005) fit to the observed IMF for individual objects.  Two other parameterisations of the IMF are also plotted: Salpeter (1955) and Kroupa (2001). Despite the metallicity varying by a factor of up to 300 between the calculations, the IMFs are statistically indistinguishable. }
\label{fig:IMF}
\end{figure*}

\begin{figure}
\centering
    \includegraphics[width=8.0cm]{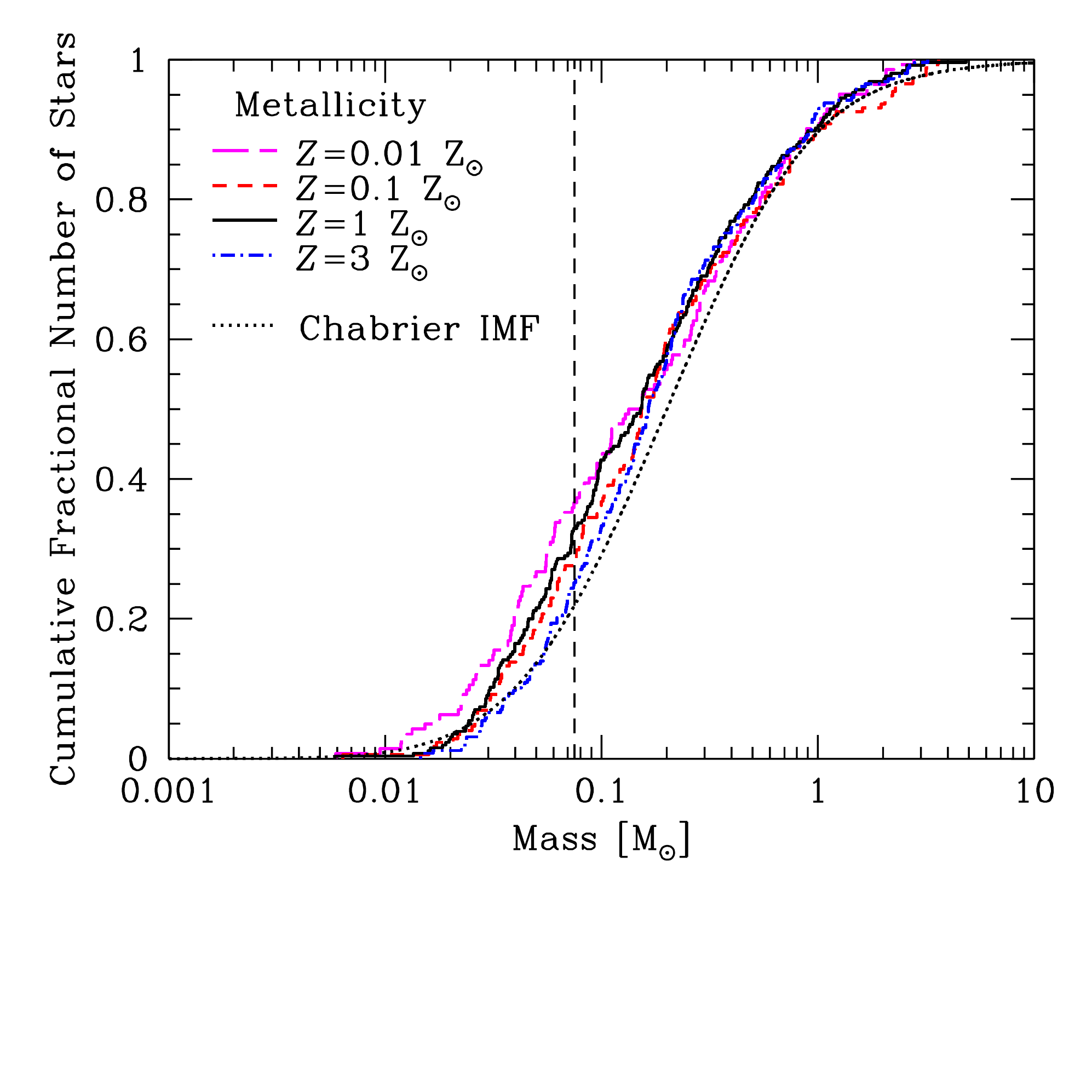}\vspace{-1.8cm}
\caption{The cumulative stellar mass distributions produced by the four radiation hydrodynamical calculations with metallicities of $Z=0.01~{\rm Z}_\odot$ (magenta long-dashed line), $Z=0.1~{\rm Z}_\odot$ (red dashed line), $Z={\rm Z}_\odot$ (black solid line), and $Z=3~{\rm Z}_\odot$ (blue dot-dashed line).  We also plot the Chabrier (2005) IMF (black dotted line).  The vertical dashed line marks the stellar/brown dwarf boundary.  The form of the stellar mass distribution does not vary significantly with different opacities: Kolmogorov-Smirnov tests show that even the two most different distributions ($Z=0.01~{\rm Z}_\odot$ and $Z=3~{\rm Z}_\odot$) have a 4\% probability of being drawn from the same underlying distribution (equivalent to a $\approx 2.1 \sigma$ difference).  However, it may be that the fraction of brown dwarfs increases slightly with decreasing metallicity.}
\label{fig:cumIMF}
\end{figure}

\subsubsection{The initial mass function}

In Fig.~\ref{fig:IMF}, we compare the differential IMFs at the end of each of the four radiation hydrodynamical calculations with different metallicities. We compare them with parameterisations of the observed Galactic IMF, given by \cite{Chabrier2005}, \cite{Kroupa2001}, and \cite{Salpeter1955}.  There is no clear difference between the mass distributions, suggesting that the IMFs produced by the calculations do not depend strongly on metallicity.

In Fig.~\ref{fig:cumIMF}, we compare the cumulative IMFs at the end of all four calculations.  We also plot the parameterisation of the observed Galactic IMF of \cite{Chabrier2005}.  Again, there is no strong difference between the stellar mass distributions.  There is an indication that the fraction of brown dwarfs may vary with metallicity.  In the most metal-poor calculation, 1/3 of the objects are brown dwarfs, while in the most metal-rich calculation only 1/4 of the objects are brown dwarfs.  However, the magnitude of the difference is small, so it is difficult to know whether it is statistically significant or not.  Performing Kolmogorov-Smirnov tests on each pair of distributions shows that, formally, they are consistent with random sampling from a single underlying distribution.  The two most different distributions are those from the most metal-poor and the most metal-rich calculations (i.e. $Z=0.01~{\rm Z}_\odot$ and $Z=3~{\rm Z}_\odot$).  But these have a 4\% probability of being drawn from the same underlying distribution (equivalent to a $\approx 2.1\sigma$ difference).  Larger numbers of objects would be required to definitively demonstrate such a metallicity dependence.

Each of the distributions is in reasonable agreement with the \cite{Chabrier2005} IMF, but formally all but the $Z=0.1~{\rm Z}_\odot$ mass distribution are statistically different.  Kolmogorov-Smirnov tests that compare the numerical distributions to \citeauthor{Chabrier2005}'s parameterisation give probabilities of $8\times 10^{-4}$ ($Z=0.01~{\rm Z}_\odot$), 0.03 ($Z=0.1~{\rm Z}_\odot$), $1\times 10^{-4}$ ($Z={\rm Z}_\odot$), and 0.004 ($Z=3~{\rm Z}_\odot$) of the numerical distributions being randomly drawn from the \cite{Chabrier2005} IMF.  The differences essentially arise from the fact that the median masses of all of the numerical distributions are slightly lower than the value of 0.20~M$_\odot$ use by \citeauthor{Chabrier2005} (see Table \ref{table1}).  Of course, this does not taken into account the observational uncertainty in the Galactic median mass.

Finally, we note that the calculations produce protostellar mass functions (PMFs) rather than IMFs \citep{FleSta1994a,FleSta1994b,McKOff2010} because some of the objects are still accreting when the calculations are stopped and the star formation has finished.  We compare our mass functions to the IMF simply because PMFs cannot be determined observationally.  \cite{Bate2012} showed that the form of the distribution of stellar masses in a calculation similar to those performed here (but not treating the diffuse interstellar medium or having separate gas and dust temperatures) did not change significantly with time during his calculation.  This is also true of the calculations presented here, although as seen from the right panel of Fig.~\ref{massnumber}, at any given time the mean mass may vary by up to $\approx 25$\%.  Although the maximum stellar mass and the total number of stars both increase with time, the form of the mass functions remains similar due to the production of new protostars.

\begin{figure*}
\centering \vspace{-0.2cm}
    \includegraphics[width=8.cm]{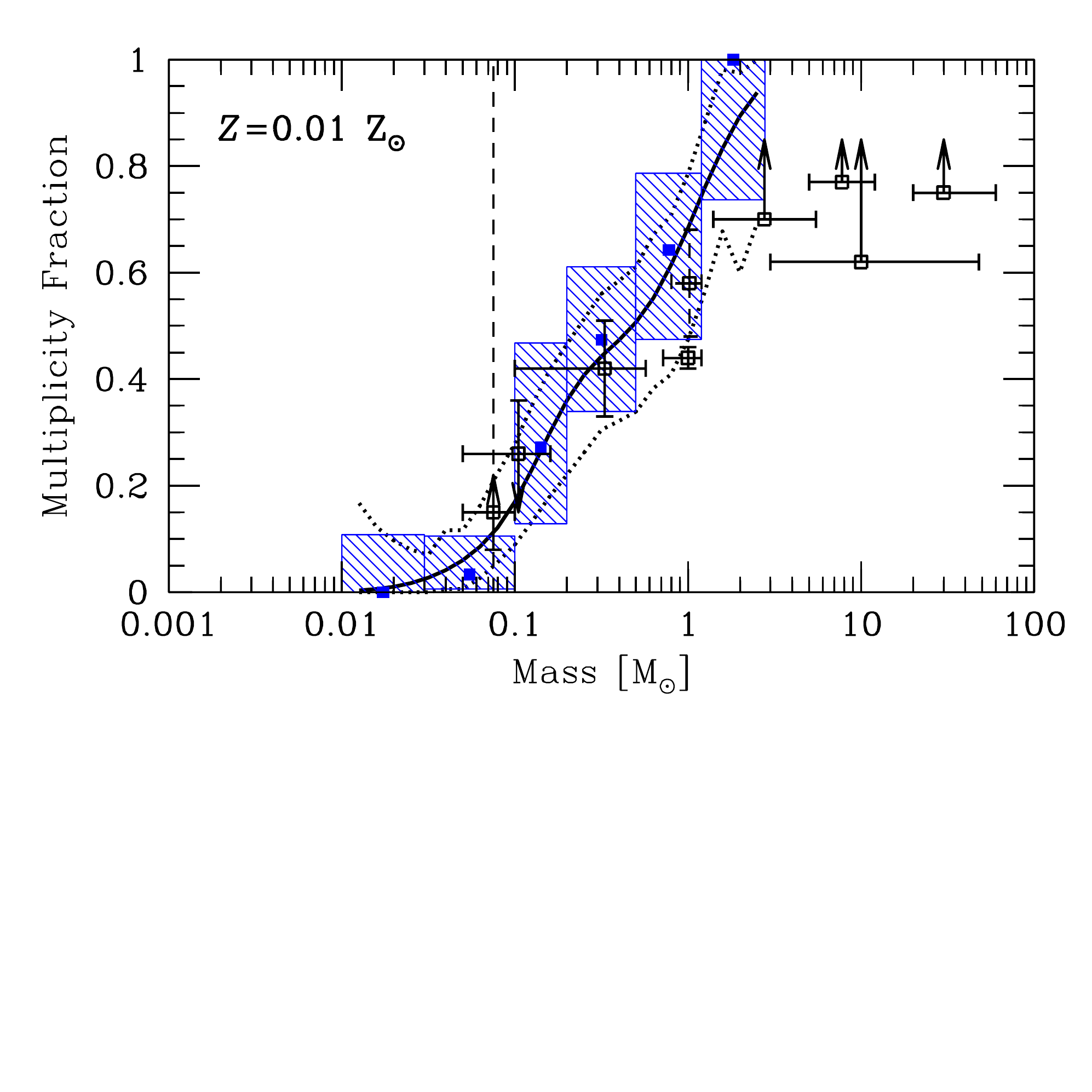}
    \includegraphics[width=8.cm]{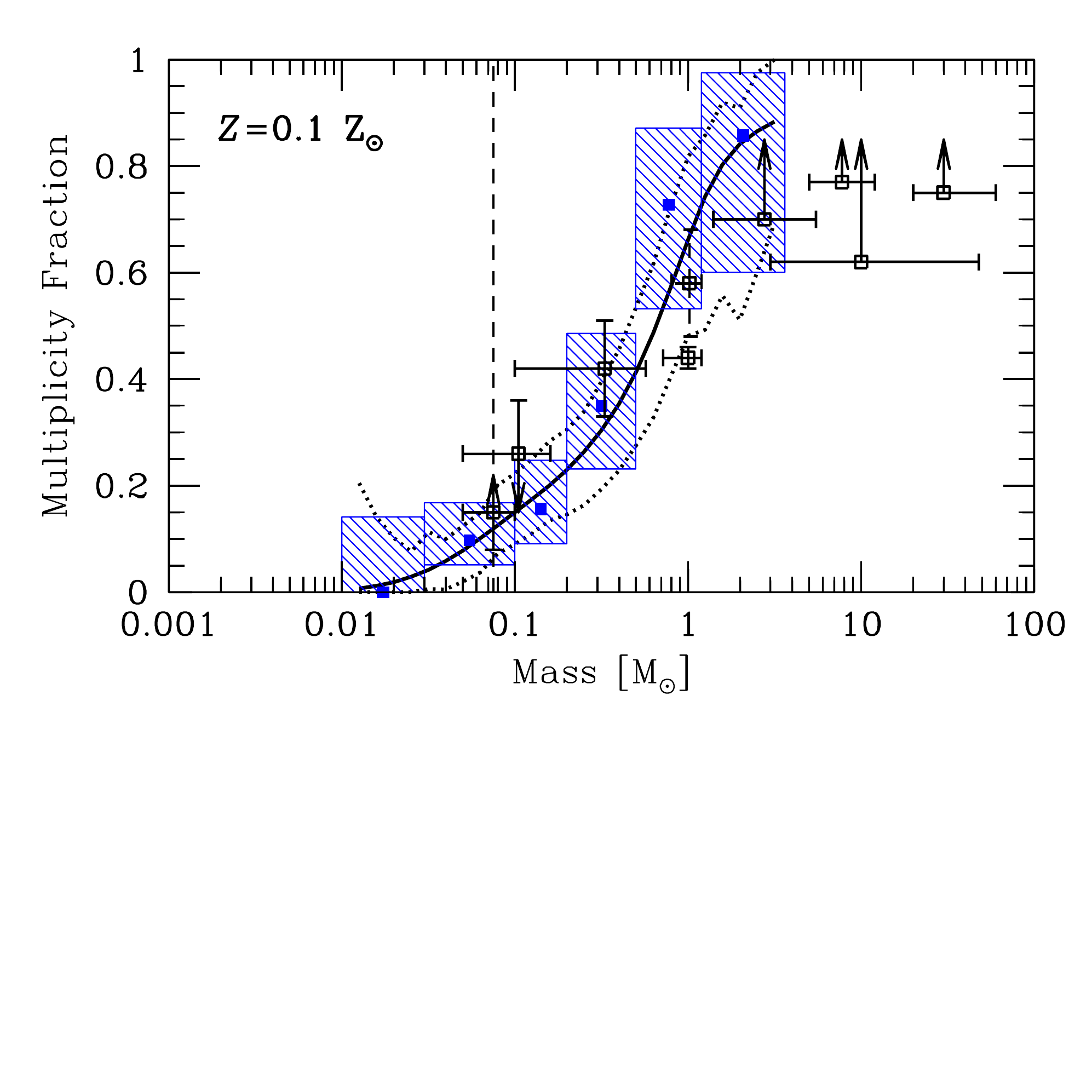}\vspace{-3cm}
    \includegraphics[width=8.cm]{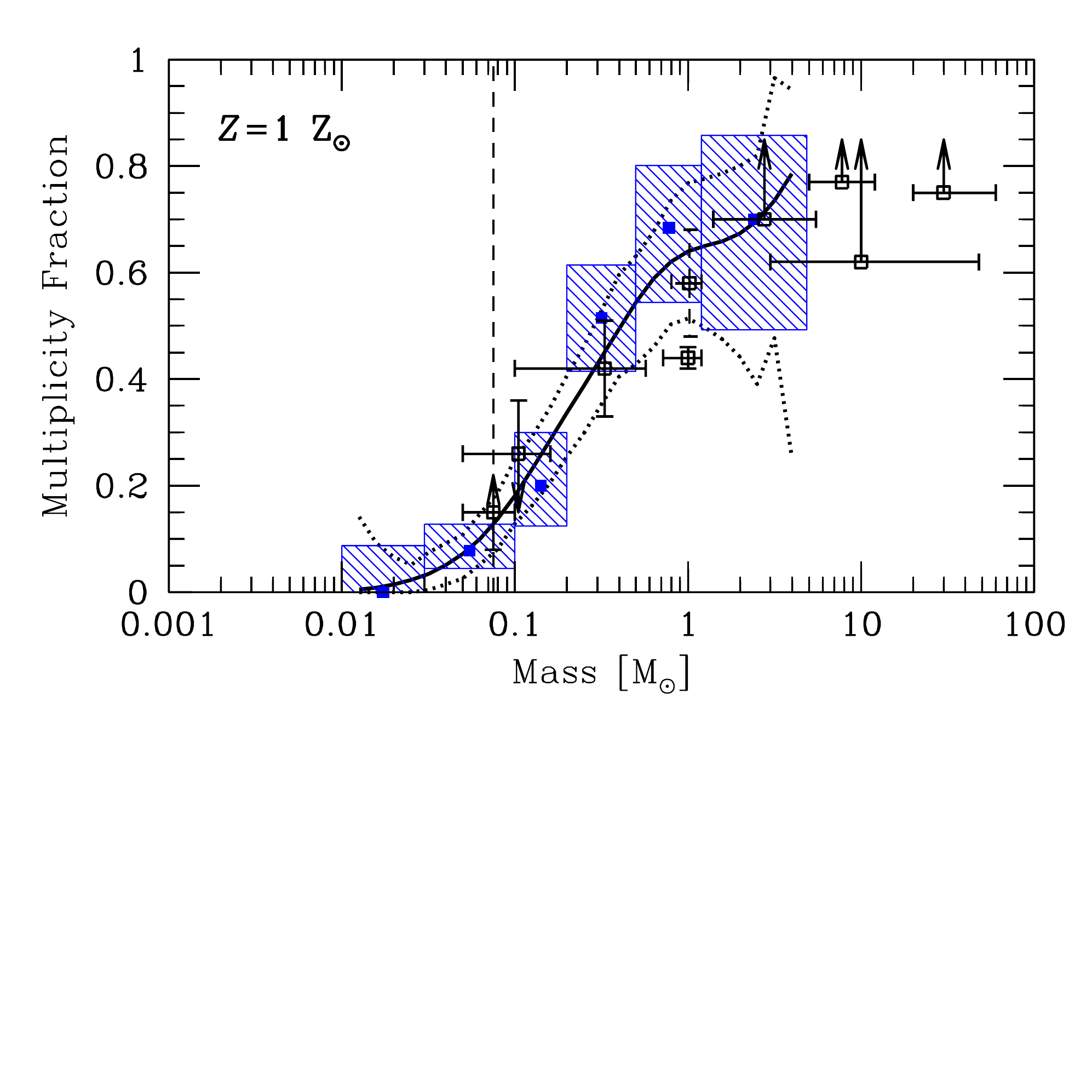}
    \includegraphics[width=8.cm]{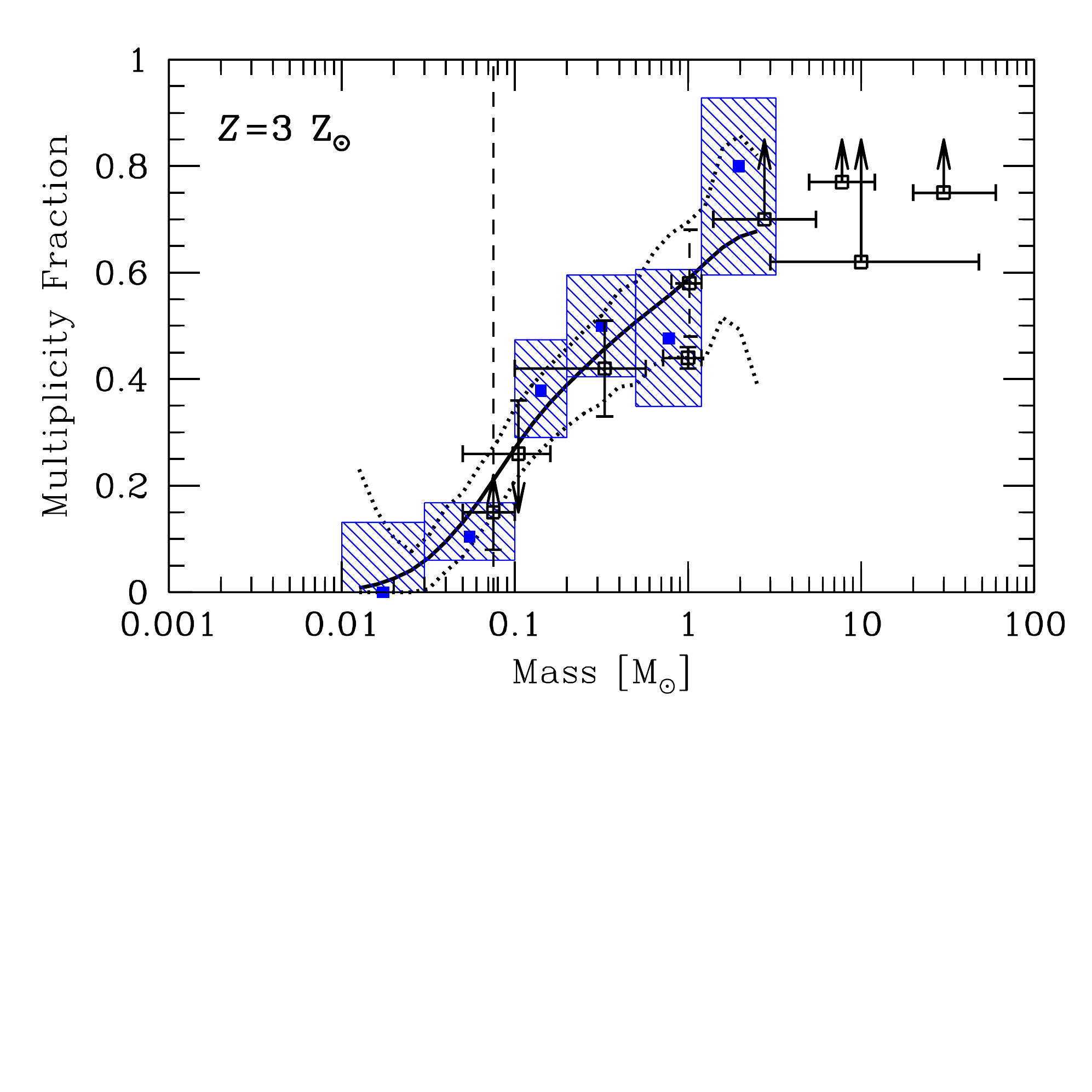}\vspace{-2.8cm}
\caption{Multiplicity fraction as a function of primary mass at the end of each of the calculations with different metallicities.  The blue filled squares surrounded by shaded regions give the results from the calculations with their $1\sigma$ statistical uncertainties.  The thick solid lines give the continuous multiplicity fractions from the calculations computed using a sliding log-normal average and the dotted lines give the approximate $1\sigma$ confidence intervals around the solid line. The open black squares with error bars and/or upper/lower limits give the observed multiplicity fractions from the surveys of \citet{Closeetal2003}, \citet{BasRei2006}, \citet{FisMar1992}, \citet{Raghavanetal2010}, \citet{DuqMay1991}, \citet{Kouwenhovenetal2007}, \citet{Rizzutoetal2013}, \citet{Preibischetal1999} and \citet{Masonetal1998}, from left to right.  Note that the error bars of the \citet{DuqMay1991} results have been plotted using dashed lines since this survey has been superseded by \citet{Raghavanetal2010}.  The observed trend of increasing multiplicity with primary mass is reproduced by all calculations.  Since multiplicity is a steep function of primary mass it is important that similar mass ranges are used when comparing simulations with observations. }
\label{multiplicity}
\end{figure*}

\subsubsection{Multiplicity as a function of primary mass}

The formation mechanisms of multiple systems and the evolution of their properties (e.g., separations) has been discussed in some detail by \cite{Bate2012} and will not be repeated here.  Our main purpose here is to determine whether or not a change of the metallicity affects stellar properties significantly.

As in \cite{Bate2009a}, \cite{Bate2012}, and subsequent papers, to quantify the fraction of stars and brown dwarfs that are in multiple systems, we use the multiplicity fraction, $mf$, defined as a function of stellar mass as
\begin{equation}
mf = \frac{B+T+Q}{S+B+T+Q},
\label{eq:mf}
\end{equation}
where $S$ is the number of single stars within a given mass range and, $B$, $T$, and $Q$ are the numbers of binary, triple, and quadruple systems, respectively, for which the primary has a mass in the same mass range.  As discussed by \cite{HubWhi2005} and \cite{Bate2009a}, this measure of multiplicity is relatively insensitive to both observational incompleteness (e.g., if a binary is found to be a triple it is unchanged) and further dynamical evolution (e.g., if an unstable quadruple system decays the numerator only changes if it decays into two binaries).

We use the same method for identifying multiple systems as that used by \cite{Bate2009a} and \cite{Bate2012}.  As in those papers, we identify binary, triple, and quadruple stellar systems, but we ignore higher-order multiples (e.g., a quintuple system consisting of a triple and a binary orbiting one another is counted as one triple and one binary).  We choose quadruple systems as a convenient point to stop as it is likely that most higher order systems are not stable and would decay in the long term.

In Table \ref{tablemult}, we provide the numbers of single and multiple star and brown dwarf systems produced by each calculation.  \cite{Bate2014} provided electronic ASCII tables of the properties of each of the stars, brown dwarfs, and multiple systems produced by the calculations discussed in that paper, and we do the same here.  We provide tables that list the masses, formation times, and final accretion rates of the stars and brown dwarfs (see Table \ref{tablestars} for an example). These tables are given file names such as {\tt Table3\_Stars\_Metal001.txt} for the $Z=0.01~{\rm Z}_\odot$ calculation.  We also provide tables that list the properties of each multiple system (see Table \ref{tablemultprop} for an example). These tables are given file names such as {\tt Table4\_Multiples\_Metal3.txt} for the $Z=3~{\rm Z}_\odot$ calculation.

The overall multiplicities for stars and brown dwarfs of all masses from each of the calculations are 29, 24, 27, and 33 per cent for the calculations with metallicities of 1/100, 1/10, 1, and 3 times solar, respectively.  The typical $1\sigma$ uncertainties are $\pm 5$ per cent.  Therefore, there is no evidence for a significant dependence of the overall multiplicity on metallicity.

\begin{figure*}
\centering
   \includegraphics[width=4.2cm]{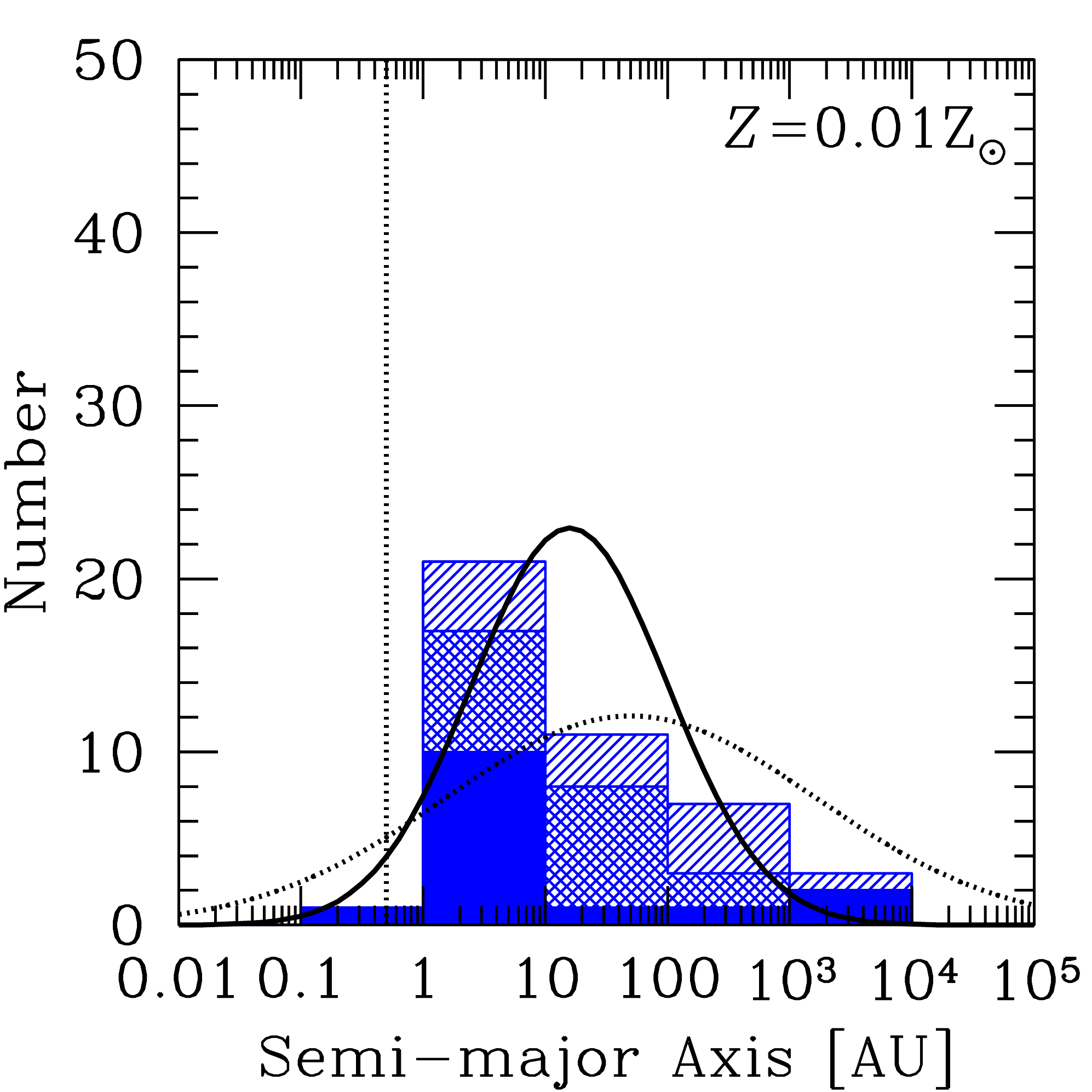} 
    \includegraphics[width=4.2cm]{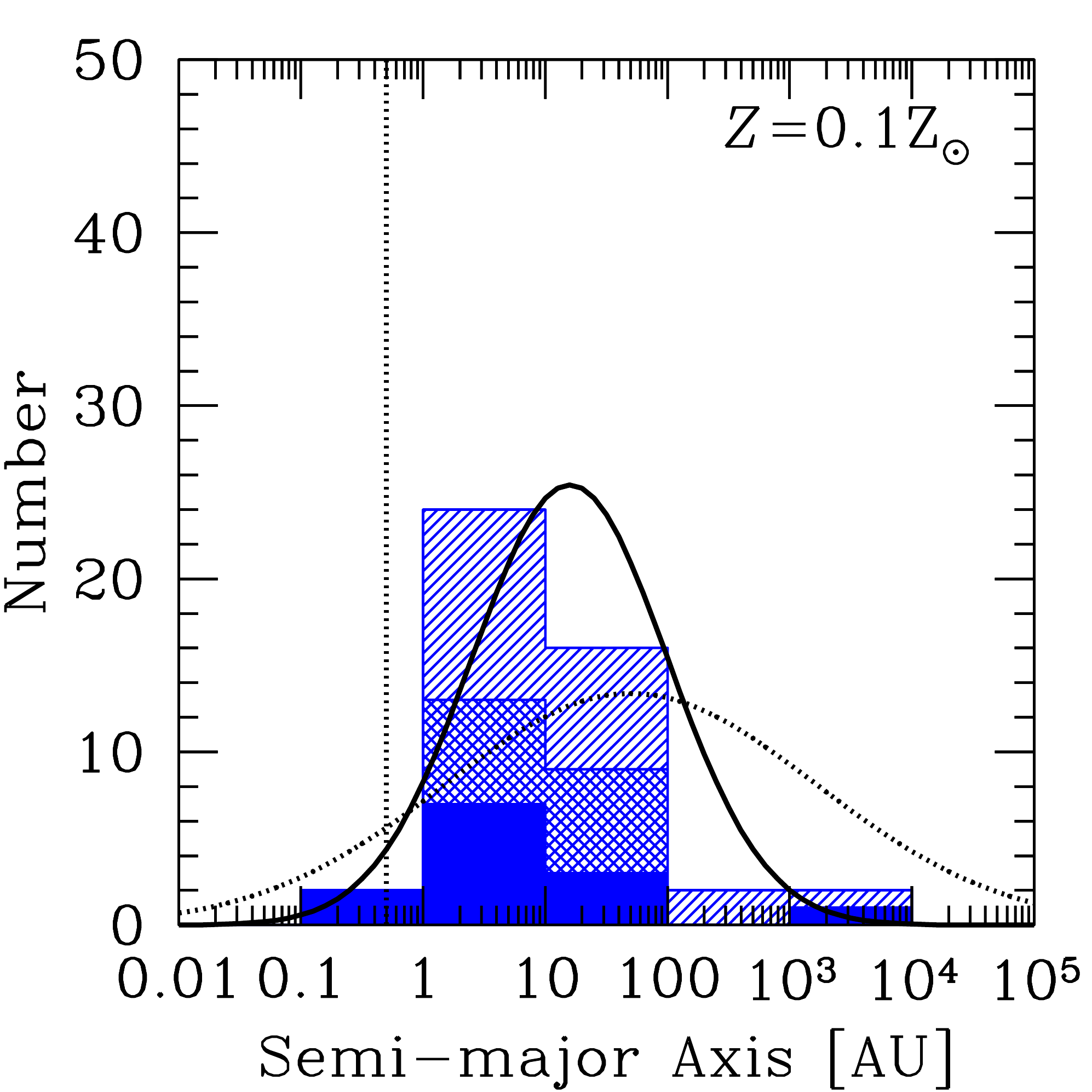}
    \includegraphics[width=4.2cm]{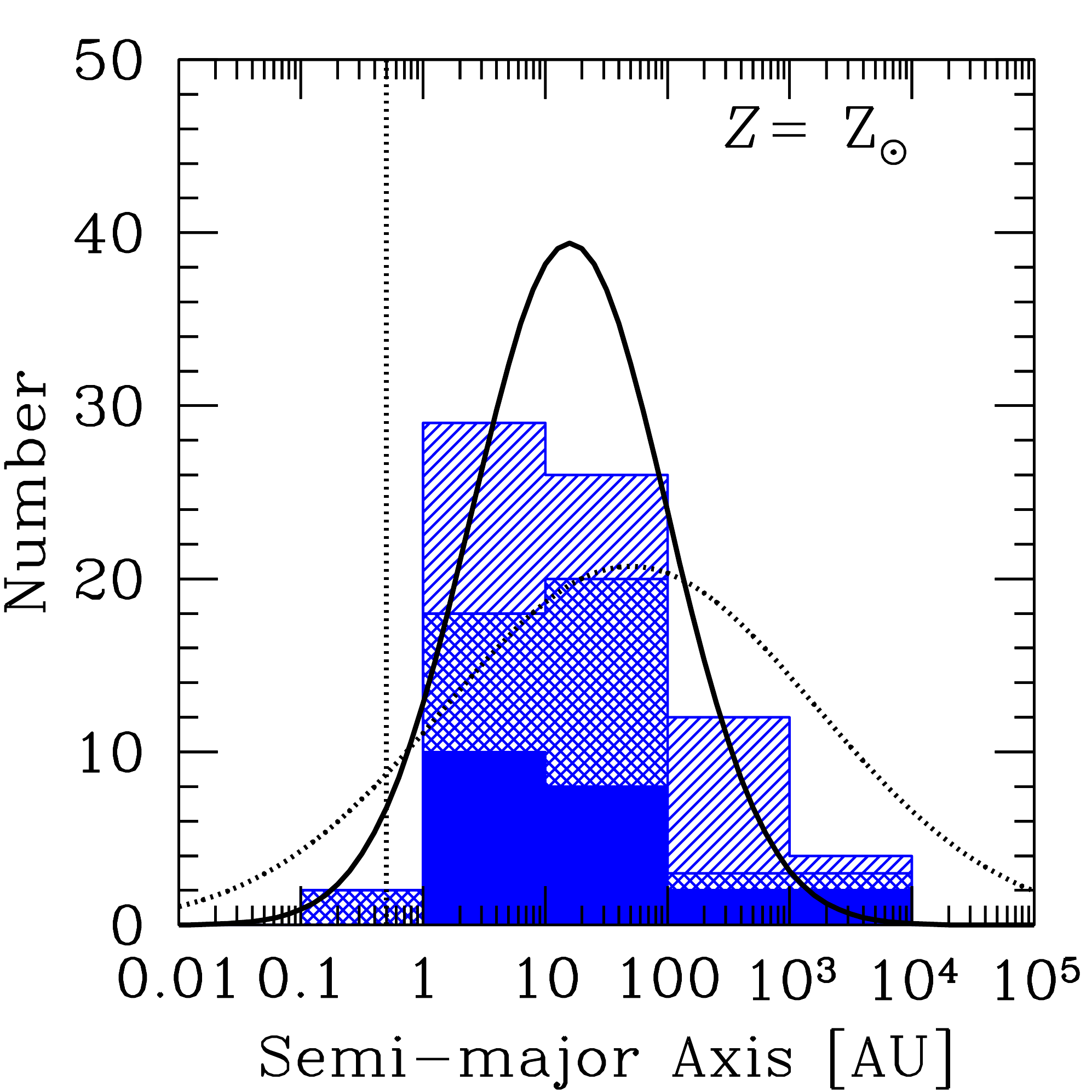} 
    \includegraphics[width=4.2cm]{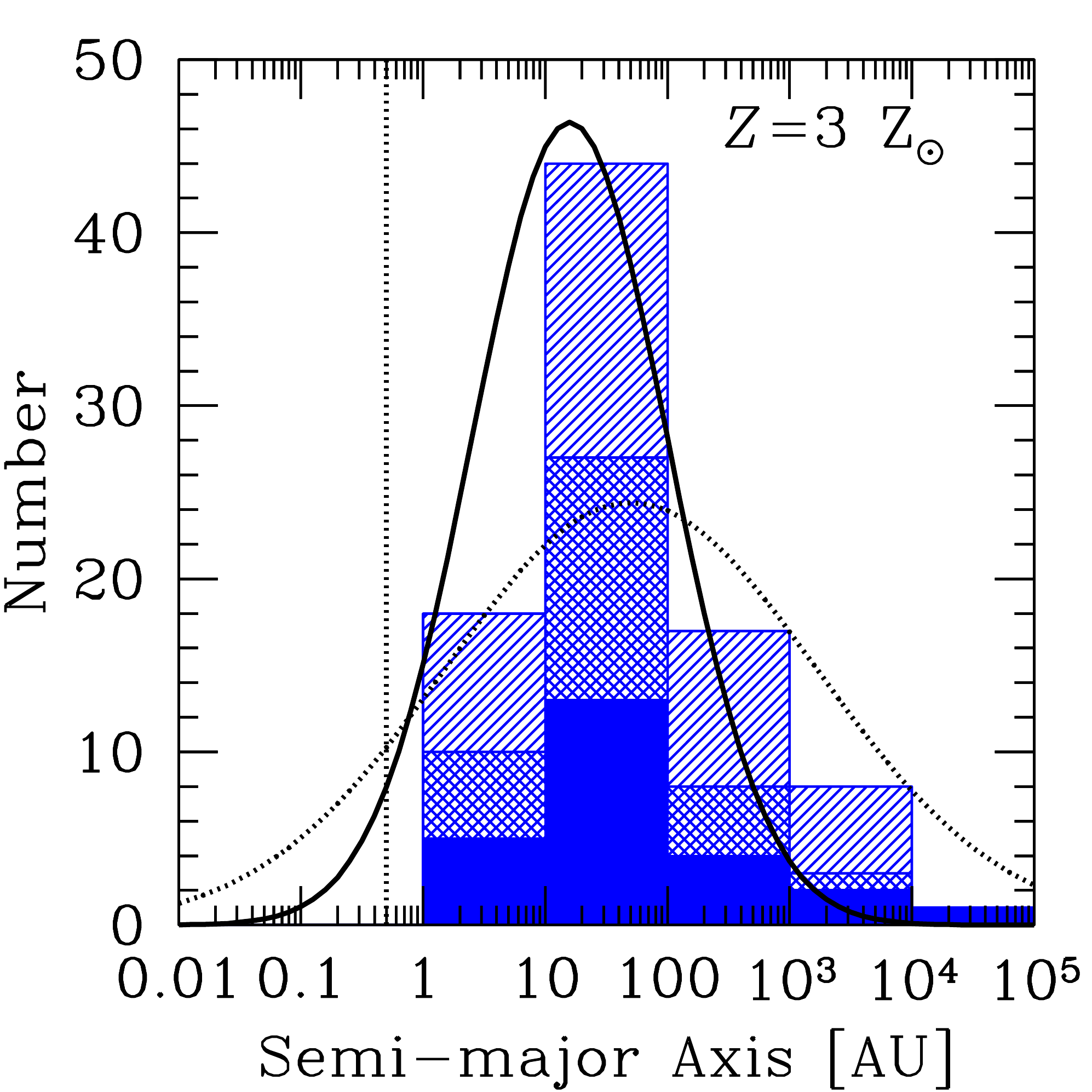}
\caption{The distributions of separations (semi-major axes) of multiple systems with stellar primaries ($M_*>0.1$~M$_\odot$) produced by the calculations with different metallicities.  The solid, double hatched, and single hatched histograms give the orbital separations of binaries, triples, and quadruples, respectively (each triple contributes two separations, each quadruple contributes three separations).  The solid curve gives the M-dwarf separation distribution (scaled to match the area) from the M-dwarf survey of \citet{Jansonetal2012}, and the dotted curve gives the separation distribution for solar-type primaries of \citet{Raghavanetal2010}. 
Note that since most of the simulated systems are low-mass, the distributions are expected to match the \citeauthor{Jansonetal2012} distribution better than that of \citeauthor{Raghavanetal2010}  The vertical dotted line gives the resolution limit of the calculations as determined by the accretion radii of the sink particles (0.5 AU).}
\label{separation_dist}
\end{figure*}

In Fig.~\ref{multiplicity}, for each of the four simulations we compare the multiplicity fraction of the stars and brown dwarfs as functions of stellar mass with the values obtained from various observational surveys (see the figure caption).  The results from the calculations have been plotted in two ways.  First, using the numbers given in Table \ref{tablemult} we compute the multiplicities in six mass ranges: low-mass brown dwarfs (masses $<0.03$~M$_\odot$), very-low-mass (VLM) objects excluding the low-mass brown dwarfs (masses $0.03-0.10$ M$_\odot$), low-mass M-dwarfs (masses $0.10-0.20$ M$_\odot$), high-mass M-dwarfs (masses $0.20-0.50$ M$_\odot$), K-dwarfs and solar-type stars (masses $0.50-1.20$ M$_\odot$), and intermediate mass stars (masses $>1.2$ M$_\odot$). The filled blue squares give the multiplicity fractions in these mass ranges, while the surrounding blue hatched regions give the range in stellar masses over which the fraction is calculated and the $1\sigma$ (68\%) uncertainty on the multiplicity fraction calculated using Poisson statistics.  Second, a thick solid line gives the continuous multiplicity fraction computed using a sliding log-normal-weighted average from the results from each simulation.  The width of the log-normal average is half a decade in stellar mass.  The dotted lines give the approximate $1\sigma$ (68\%) uncertainty on the sliding log-normal average.

All of the calculations produce multiplicity fractions that increase strongly with increasing primary mass, in qualitative agreement with observed stellar systems.  The actual values of the multiplicities of M dwarfs and high-mass brown dwarfs are in good agreement with the observed multiplicities of field stars.  For solar-type stars, all except the $Z=3~{\rm Z}_\odot$ calculation seem to over-produce multiple systems.  However, many of these systems are high-order multiple systems (see Table \ref{tablemult}) and some may well break up before they would become field stars: it is well known that the multiplicity of young stars tends to be higher than for field stars \citep{Leinertetal1993,GheNeuMat1993,Richichietal1994,Simonetal1995,Ghezetal1997,Duchene1999,Ducheneetal2007}.  There is no evidence for a strong dependence of the multiplicities on metallicity.

\begin{figure}
\centering
    \includegraphics[width=8cm]{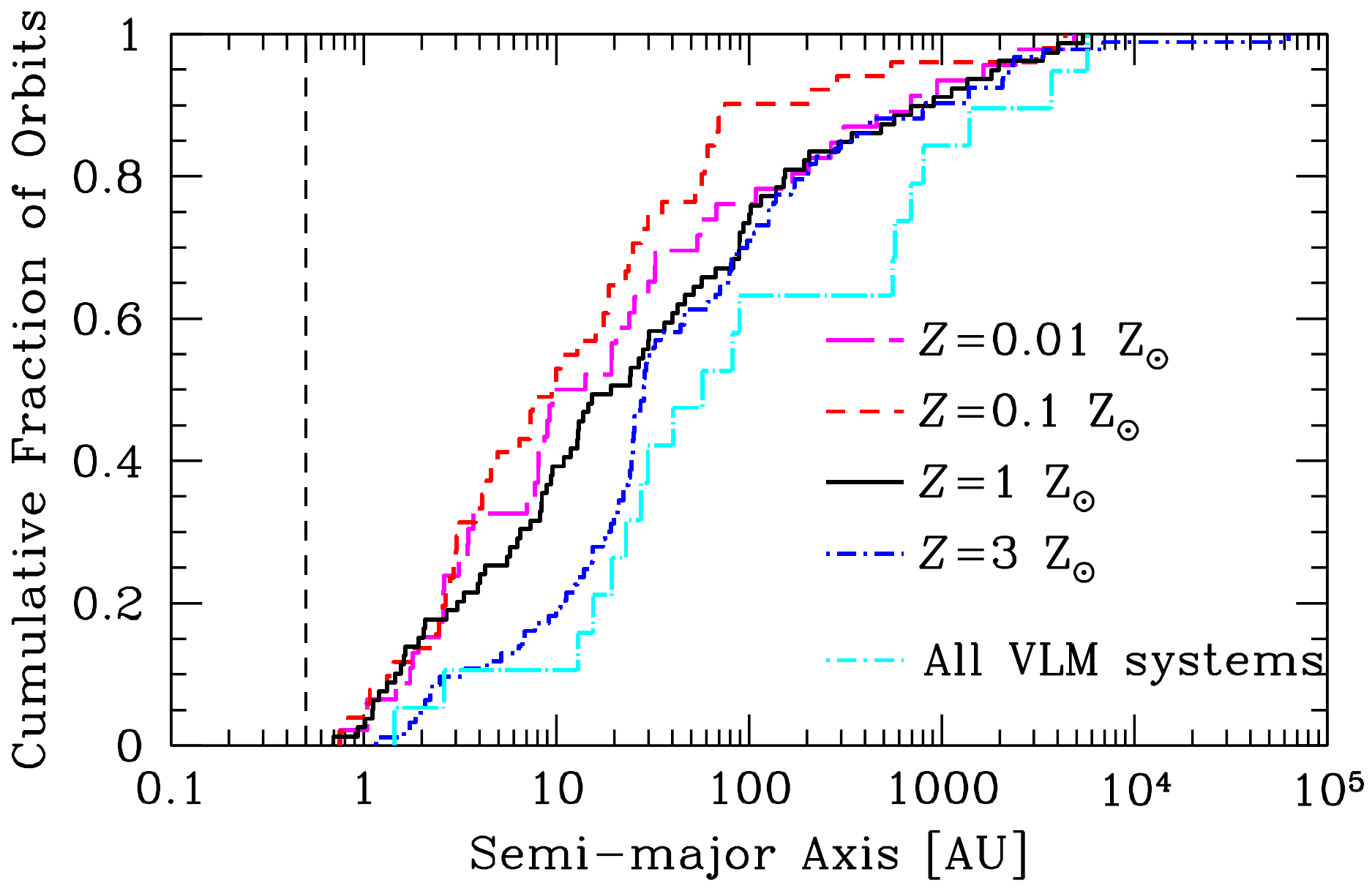}
\caption{The cumulative semi-major axis (separation) distributions of the multiple systems produced by the four calculations with different metallicities.  All orbits are included in the plot (i.e. two separations for triple systems, and three separations for quadruple systems), and the lines for different metallicities include both stellar and brown dwarf systems.  In addition, we give the separation distribution of all VLM multiple systems (all metallicities).  The vertical dashed line marks the resolution limit of the calculations as determined by the accretion radii of the sink particles.  Performing Kolmogorov-Smirnov tests on pairs of distributions shows that the distributions for the different metallicities are statistically indistinguishable, except for the highest metallicity which has a significant deficit of systems with separations $\lsim 20$~AU.  Note that, if anything, the VLM systems are wider than the stellar systems. }
\label{cumsep_comp}
\end{figure}

\subsubsection{Separation distributions of multiple systems}
\label{sec:separations}

Observationally, the mean and median separations of binaries are believed to depend on the mass of the primary \cite[see the review of][]{DucKra2013}.  For solar-type binaries, the mean separation (in the logarithm of separation) is observed to be $\approx 40$~AU \citep{DuqMay1991,Raghavanetal2010}.  M-dwarf binaries are found to be more tightly bound with a mean separation of $\approx 16$~AU \citep{FisMar1992,Jansonetal2012}.  VLM binaries (primary masses $M_1<0.1$~M$_\odot$) have a mean separation $\lsim 4$~AU \citep{Closeetal2003,Closeetal2007,Siegleretal2005}, with few VLM binaries have separations greater than 20~AU, especially in the field \citep{Allenetal2007}.  At higher masses, \cite{DeRosaetal2014} find the peak of the separation distribution for A-type stars lies at $\approx 400$~AU, although their survey only covered separations larger than $\approx 30$~AU.  Since there are approximately as many spectroscopic companions \citep{Abt1965,CarPri2007} as their are wider companions \citep{DeRosaetal2014}, this raises the prospect that the separation distribution of A-type stars is bimodal \citep{DucKra2013}.  O-type stars may also have a bimodal companion separation distribution \citep{Masonetal1998}, with the separation distribution of spectroscopic systems peaking at very short periods \citep{Sanaetal2012,Sana_etal2013,Sana2017}.

In the calculations performed for this paper, binaries as close as 0.03~AU (6~R$_\odot$) can be modelled before they are assumed to merge.  However, the sink particles have accretion radii of 0.5~AU, so gas is not modelled on scales smaller than this.  The lack of gas to provide dissipation on small scales inhibits the formation of very close systems \citep*{BatBonBro2002b}.

\begin{figure*}
\centering
    \includegraphics[width=4cm]{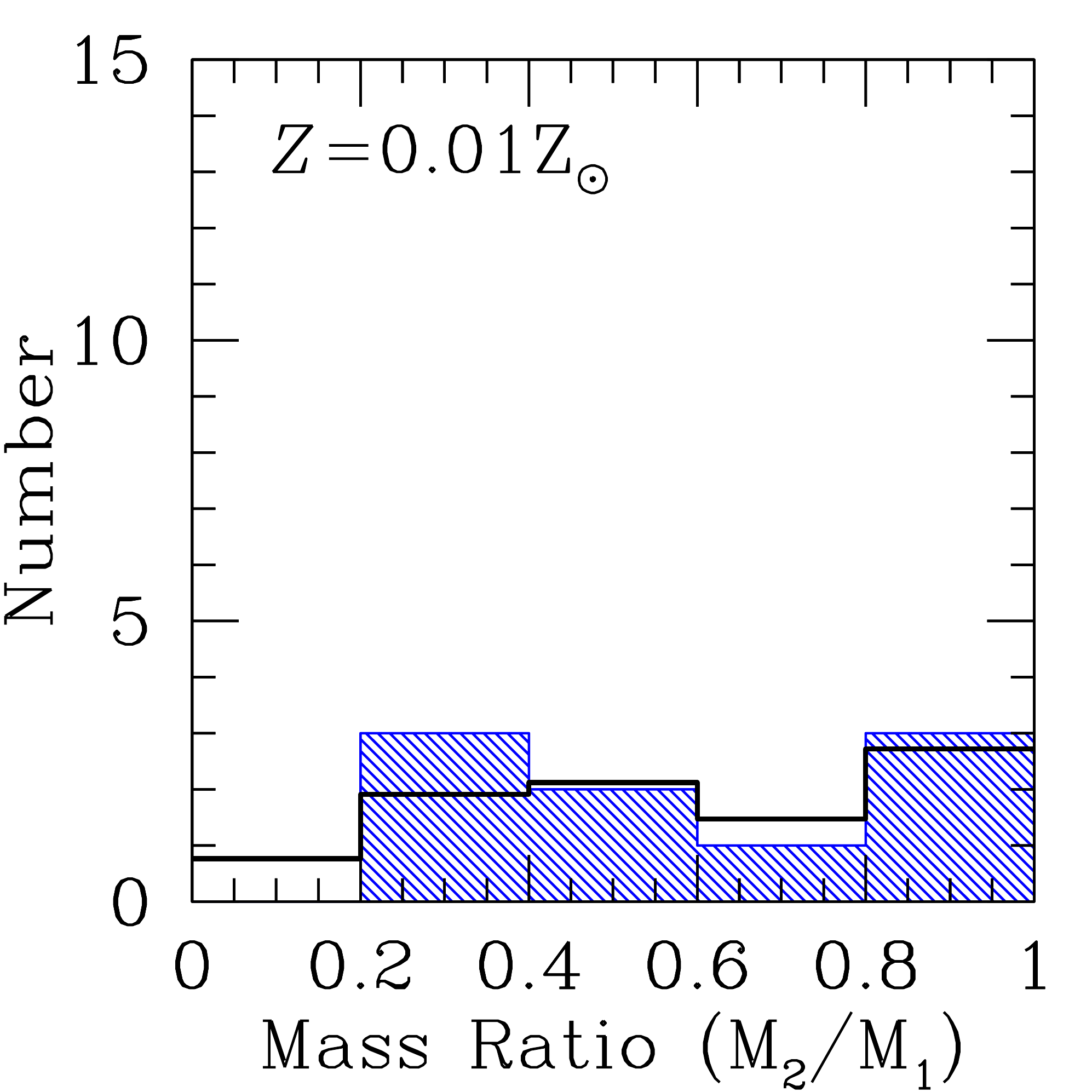}
    \includegraphics[width=4cm]{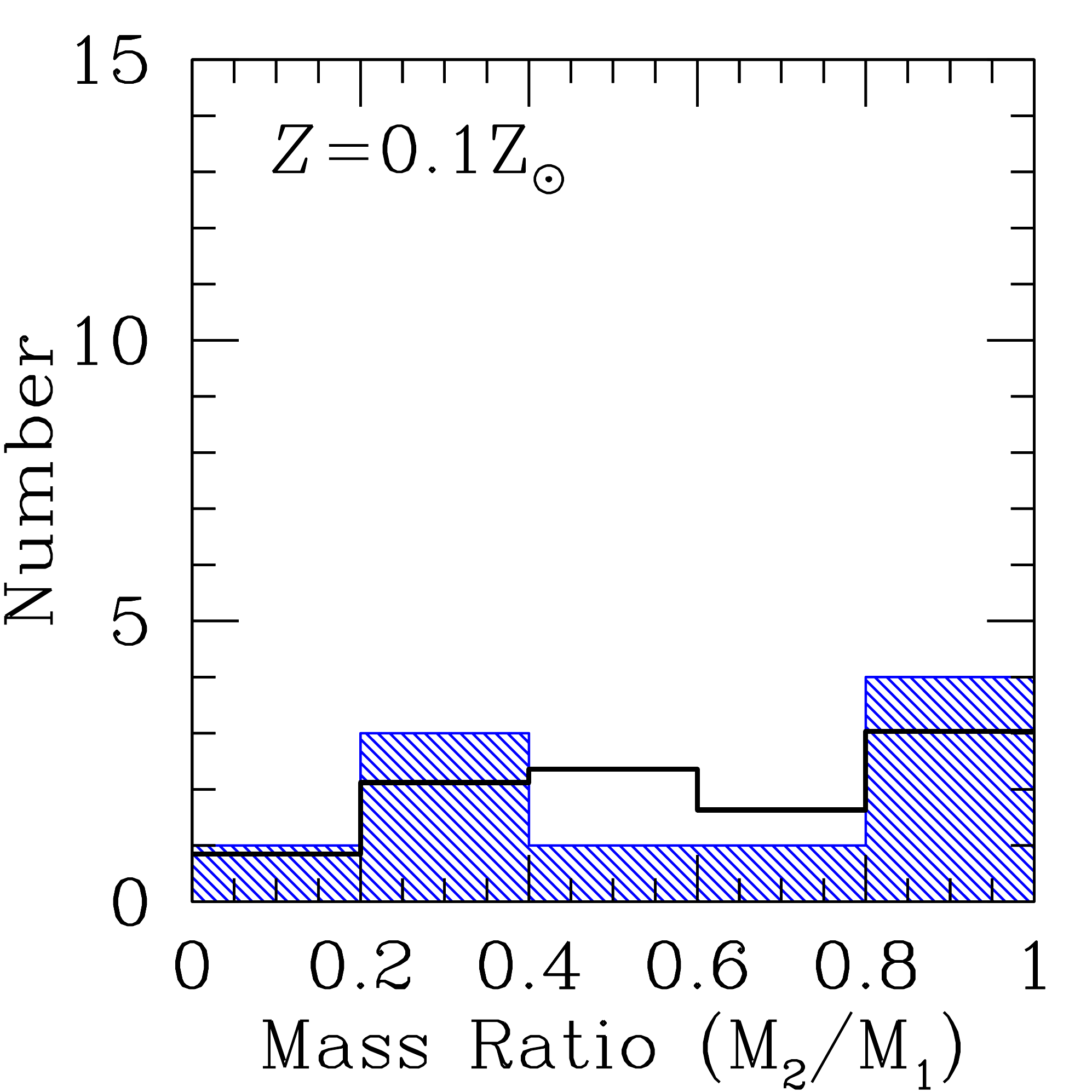}
    \includegraphics[width=4cm]{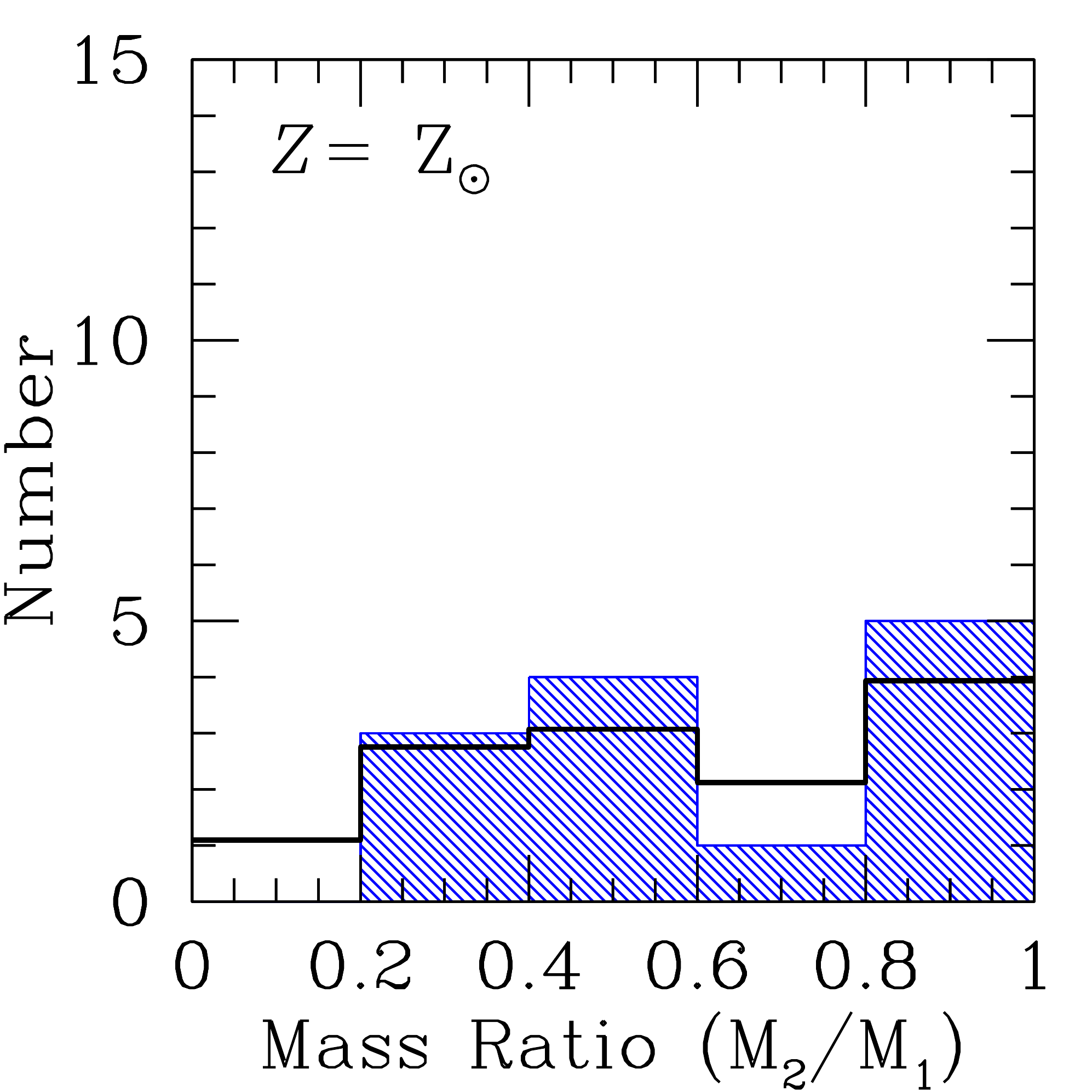}
    \includegraphics[width=4cm]{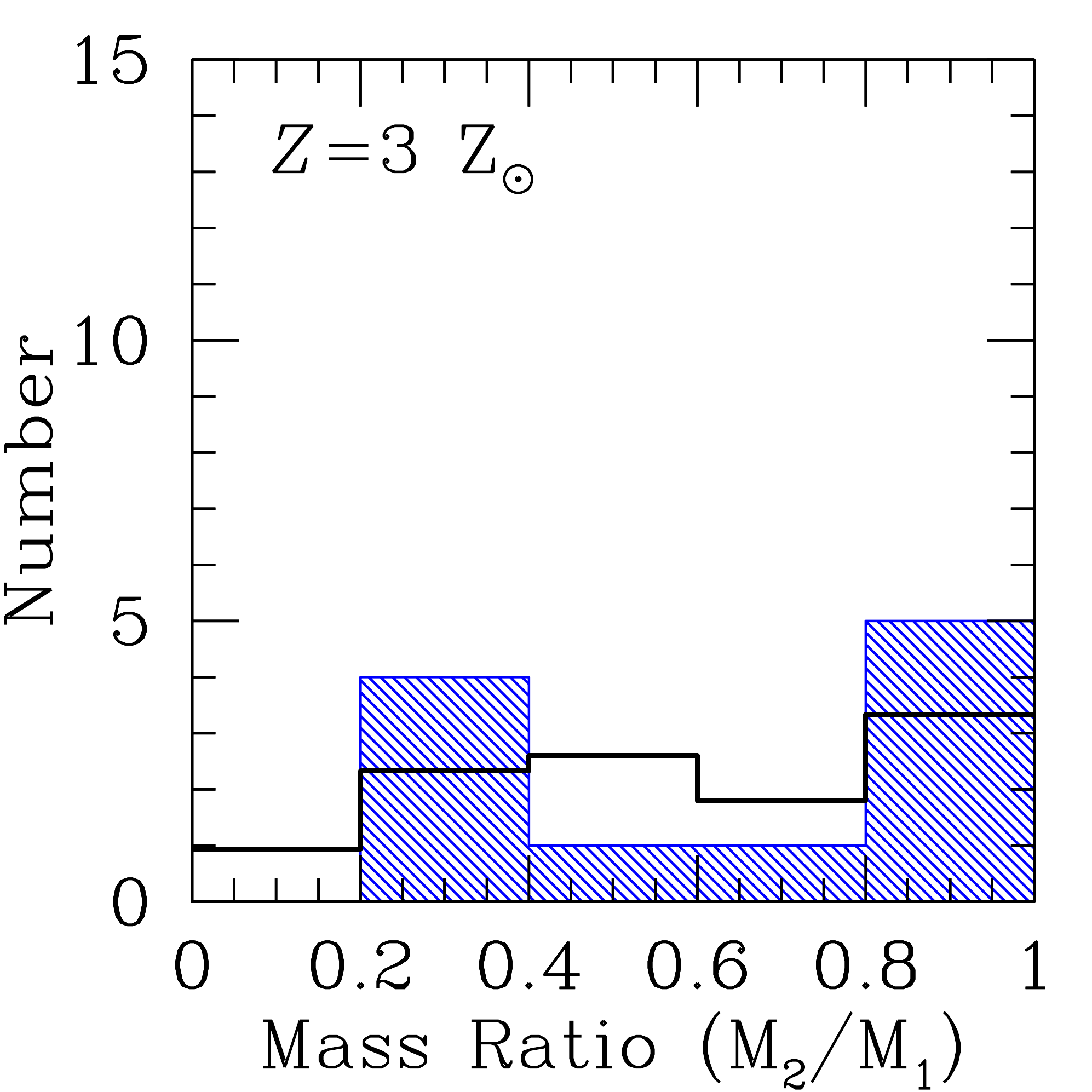}
    \includegraphics[width=4cm]{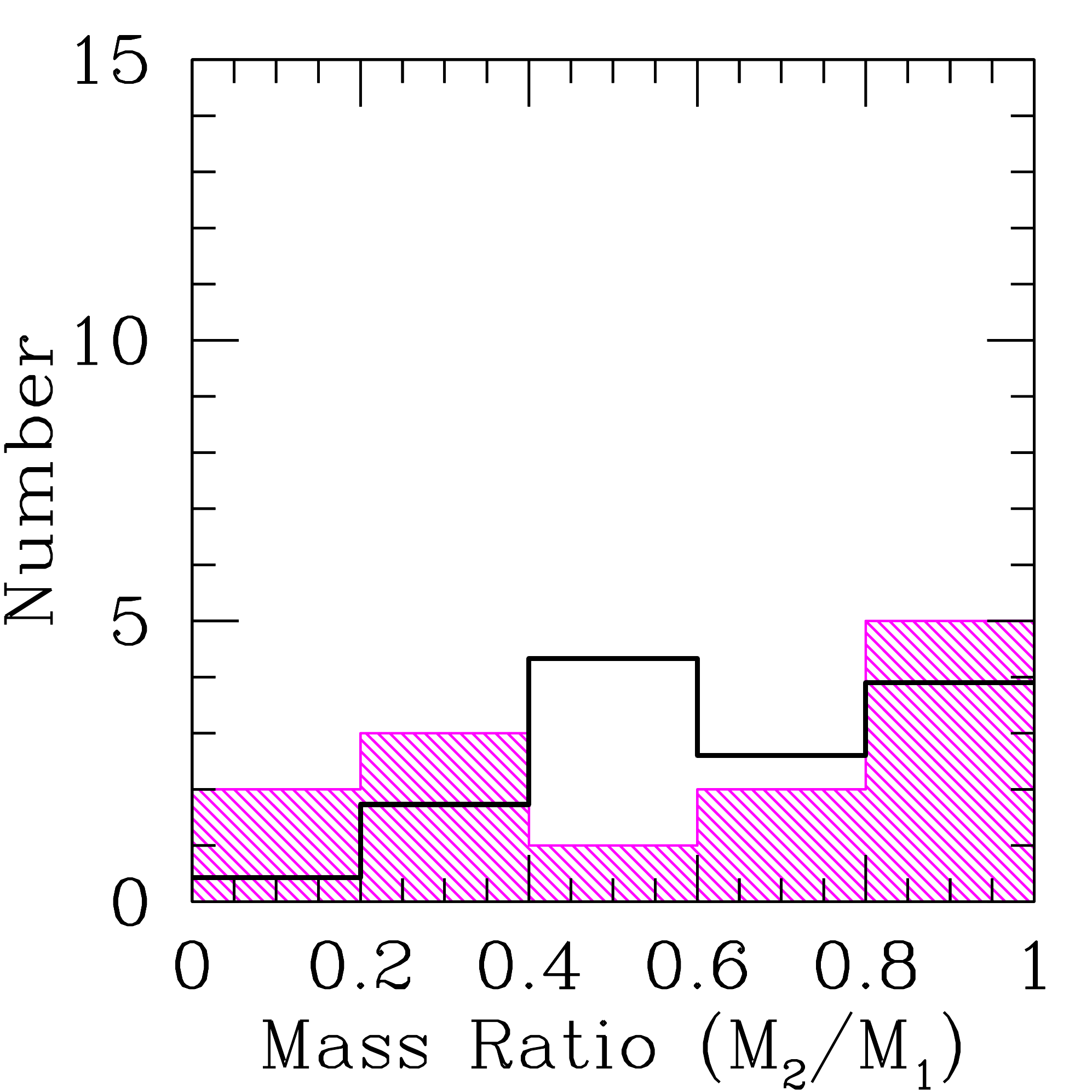}
    \includegraphics[width=4cm]{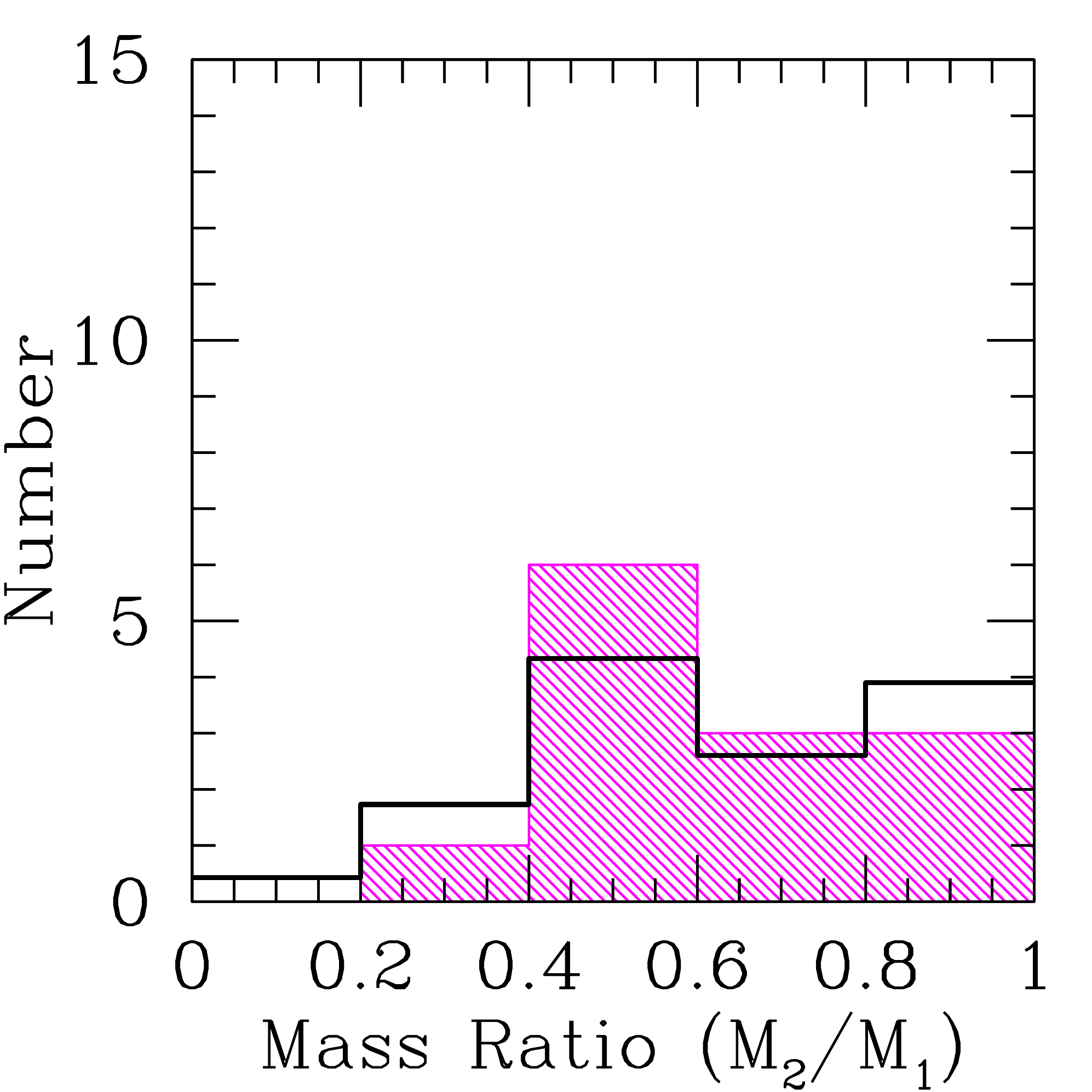}
    \includegraphics[width=4cm]{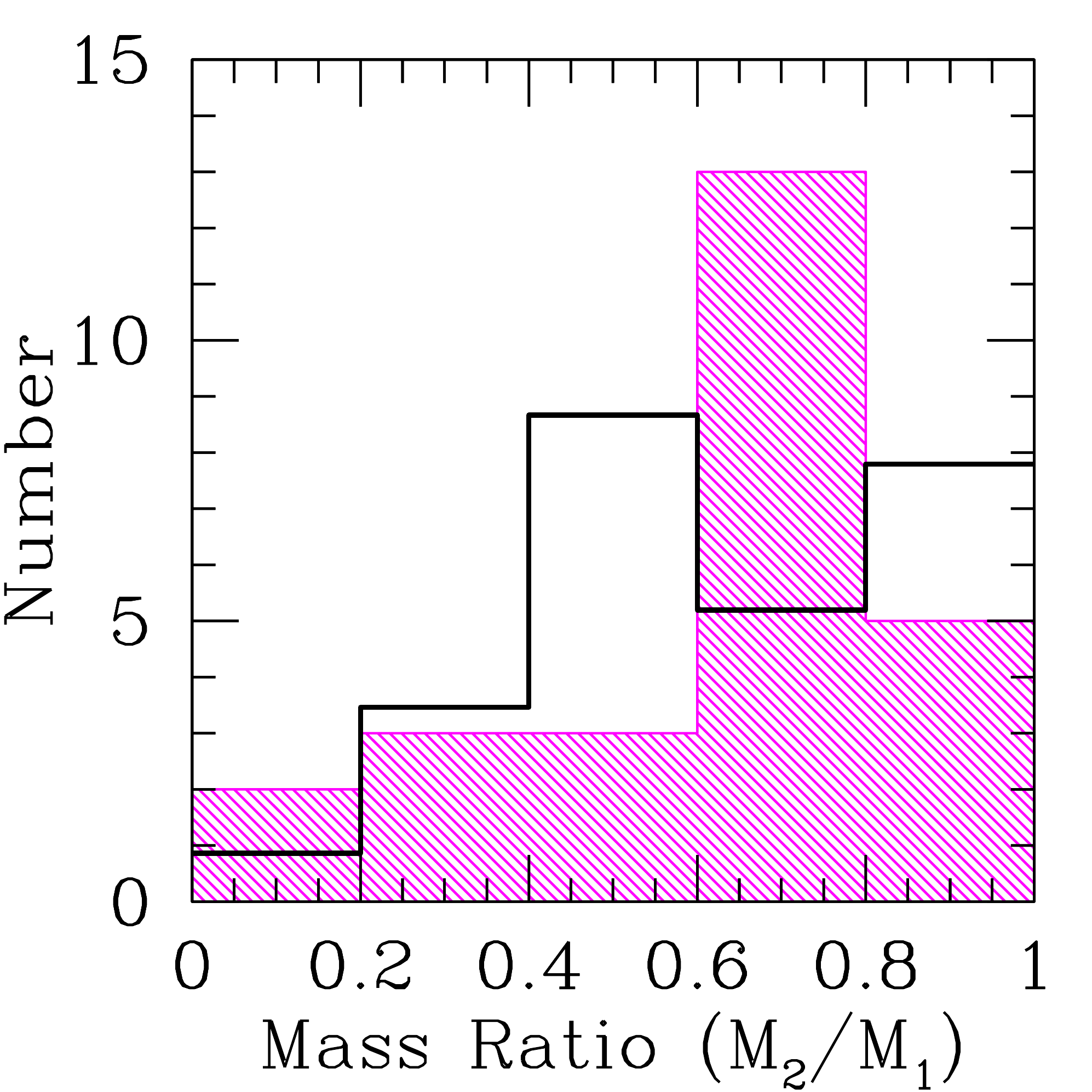}
    \includegraphics[width=4cm]{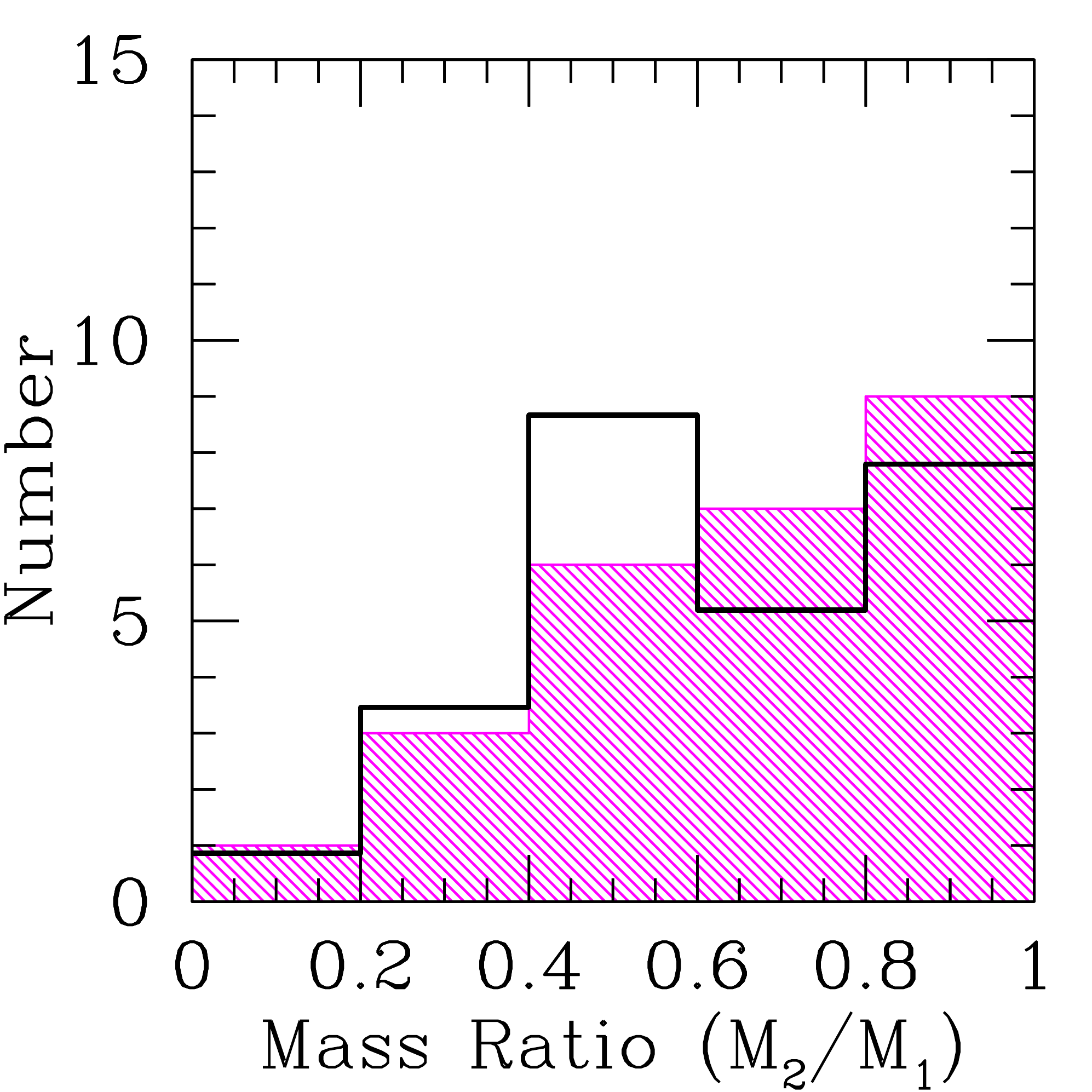}
    \caption{The mass ratio distributions of bound pairs with stellar primaries in the mass ranges $0.5<M_1<1.2$ M$_\odot$ (top row) and $0.1<M_1<0.5$ M$_\odot$ (bottom row) produced by the calculations with different metallicities (left to right).  The solid black lines give the observed mass ratio distributions of \citet{Raghavanetal2010} for binaries with solar-type primaries (top row) and \citet{Jansonetal2012} for M-dwarfs (bottom row).  The observed mass ratio distributions have been scaled so that the areas under the distributions match those from the simulation results.  There is no obvious dependence of the mass ratio distributions on opacity.}
\label{massratios}
\end{figure*}

In Fig.~\ref{separation_dist}, we present the separation (semi-major axis) distributions of the stellar ($M_1> 0.1$~M$_\odot$) multiples.  The distributions are compared with the lognormal distributions from the surveys of M-dwarfs by \cite{Jansonetal2012} and solar-type stars by \cite{Raghavanetal2010}.  As expected due to the absence of small-scale dissipation, there are few systems in the numerical simulations that have separations smaller than 1~AU.

The binned histograms indicate that there may be a trend of increasing peak separation with increasing metallicity.  In Fig.~\ref{cumsep_comp}, we provide the cumulative separation distributions for each of the calculations.  The median separations of the metal-poor calculations are smaller than those obtained for the calculations with solar and super-solar metallicities.  Performing Kolmogorov-Smirnov tests on each pair of cumulative distributions shows that they are statistically indistinguishable, except for the $Z=3~{\rm Z}_\odot$ distribution.  The most metal-rich calculation has a significant deficit of multiple systems with separations $\lsim 20$~AU.  We will discuss this result further in Section \ref{sec:closebinaries}.

Each calculation produces five or fewer VLM multiple systems (see Table \ref{tablemult}), so it is not possible to determine how the separations of VLM systems may depend on metallicity.  In Fig.~\ref{cumsep_comp}, we plot a single cumulative distribution that includes all 18 VLM multiple systems from all four calculations.  Unlike observed systems, the simulated VLM systems do not tend to have smaller separations than the stellar systems.  Either the simulations are missing some formation mechanism that preferentially produces close VLM systems, or VLM systems evolve to tighter separations on longer timescales \citep[e.g., via dynamical processing;][]{Bate2009a}.  This absence of tight VLM systems has been a problem for all similar calculations since radiative transfer was introduced \citep{Bate2012,Bate2014}.  Only the earlier barotropic calculations of \cite{Bate2009a} displayed significantly tighter VLM systems than stellar systems.  In the barotropic calculations, disc fragmentation was much more prevalent, so this may be an indication that there is insufficient disc fragmentation when radiative transfer is included.  This insufficient disc fragmentation could be numerical in nature, due to excess viscous heating and/or insufficient resolution \citep[e.g.][]{MerBat2012}. Alternately, the wider separations may be due to a lack of orbital decay through interaction with gas on small scales \citep[e.g., the loss of angular momentum to a circumbinary disc;][]{Artymowiczetal1991}.  For low-mass systems, any discs are expected to be low-mass and poorly numerically resolved or, indeed, completely unresolved \citep{Bate2018}.  This problem warrants future investigation.

\begin{figure}
\centering
    \includegraphics[width=8cm]{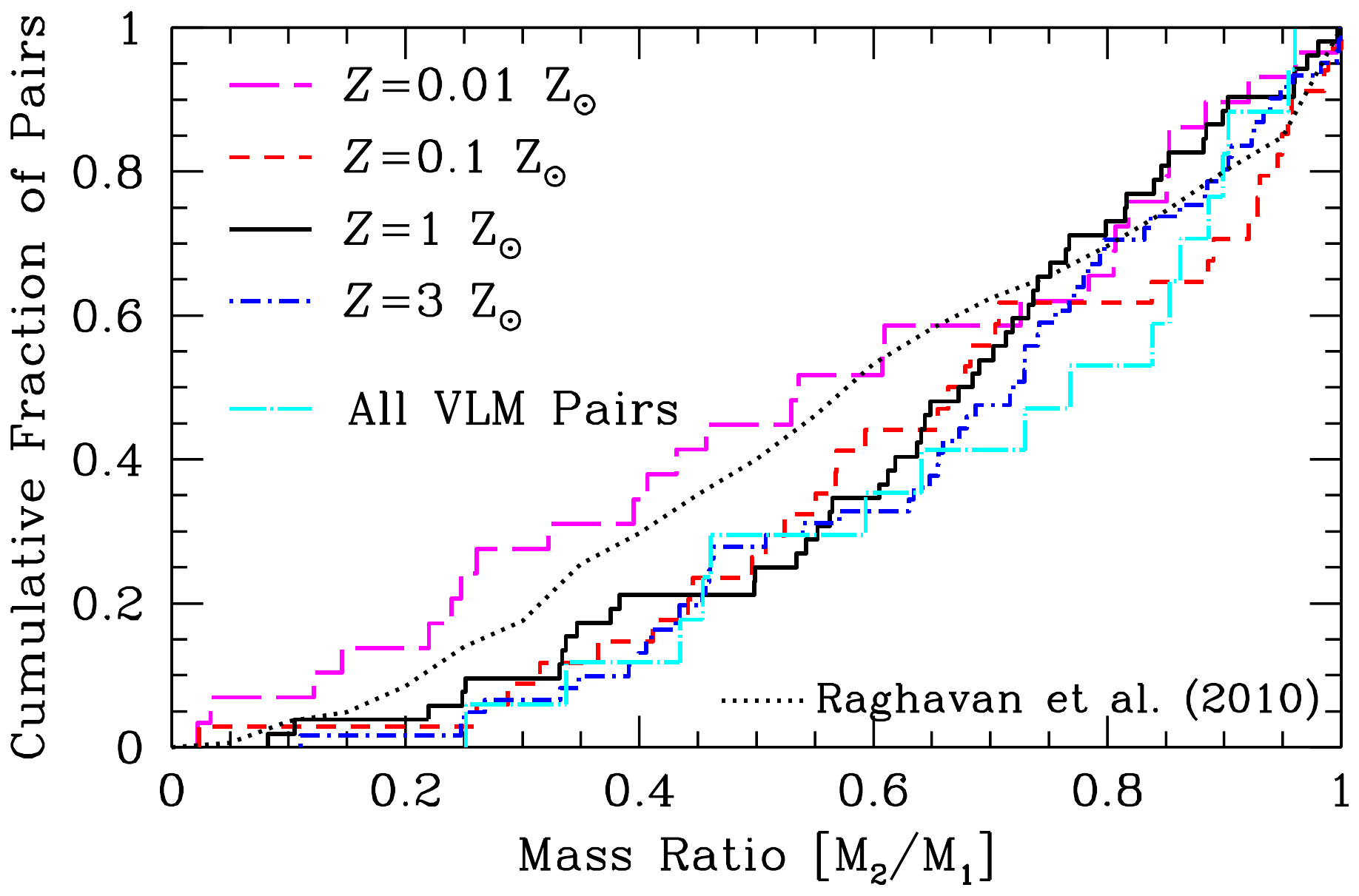}
\caption{The cumulative mass ratio distributions of pairs of objects produced by the four calculations with different metallicities.  Pairs include both binaries and bound pairs that may be components of triple or quadruple systems.  We also plot the mass ratio distribution of the 17 VLM pairs from all four calculations, and the observed mass ratio distribution of solar-type stars from \citet{Raghavanetal2010}. Performing Kolmogorov-Smirnov tests on the simulated distributions shows that they are statistically indistinguishable, despite the apparent excess of low mass-ratio systems for the lowest metallicity.  Each of the simulated distributions is also formally consistent with being randomly drawn from the \citet{Raghavanetal2010}, except for the $Z=3~{\rm Z}_\odot$ distribution for which the Kolmogorov-Smirnov probability is $9.7\times 10^{-3}$, but this distribution is in good agreement with the observed distribution for M-dwarf binaries (see the bottom-right panel of Fig.~\ref{massratios}).}
\label{cumq}
\end{figure}

\subsubsection{Mass ratio distributions of pairs}
\label{sec:q}

In Fig.~\ref{massratios}, we give the mass ratio distributions of bound pairs with primary masses $M_1 \ge 0.5$~M$_\odot$ (top row) and M-dwarf pairs with masses $0.1 \le M_1 < 0.5$~M$_\odot$ (bottom row).  We do not plot the distributions of VLM pairs because there are so few of them.  These distributions include binary systems, and pairs that are the inner components of triple and quadruple systems.  A triple system composed of a binary with a wider companion contributes the mass ratio from the closest pair, as does a quadruple composed of a triple with a wider companion. A quadruple composed of two pairs orbiting each other contributes two mass ratios -- one from each of the pairs.  We compare the M-dwarf distribution to the observed mass ratios from \cite{Jansonetal2012}, and the higher-mass stars to the mass ratio distribution of binaries with solar-type primaries from \cite{Raghavanetal2010}.  

The distributions from the calculations are in reasonable agreement with the observed distributions, and there is no strong evidence of a dependence on metallicity.  In Fig.~\ref{cumq} we plot the cumulative mass ratio distributions for each calculation (each covering all primary masses).  We also plot the mass ratio distribution of the 17 VLM pairs produced by all the calculations together.  There is a preference for equal-mass VLM pairs, but it is not as strong as is observed \citep{Closeetal2003, Siegleretal2005, Reidetal2006} and the VLM distribution is not significantly different from the overall mass ratio distributions.  Most of the distributions are biased toward equal-masses when compared to the mass ratio distribution of the solar-type pairs as observed by \cite{Raghavanetal2010}.  This is to be expected if  mass ratios become more biased toward equal masses as the primary mass decreases (consistent with the M-dwarf survey of \citealt{Jansonetal2012}) because most of the pairs in the simulations have M-dwarf primaries (Table \ref{tablemult}).  On the other hand, \citet{RegMey2013} argue that the observed mass ratio distributions of M-dwarf and solar-type binaries are currently indistinguishable. The lowest metallicity calculation has a greater fraction of low-mass ratio systems, but Kolmogorov-Smirnov tests comparing the $Z=0.01~{\rm Z}_\odot$ distribution with the distributions for higher metallicities do not find this difference to be significant, at least when including binaries of all separations.  However, we re-examine the mass ratio distribution of close binaries in Section \ref{sec:closebinaries}.

\begin{table}
\begin{tabular}{lccccc}\hline
Mass Range ~ [M$_\odot$]& Single & Binary  & Triple & Quadruple  \\ \hline
\multicolumn{5}{c}{Metallicity $Z=0.01~{\rm Z}_\odot$}  \\ \hline
\hspace{0.83cm}$M<0.03$       &      16     &     0     &      0      &     0    \\
$0.03\leq M<0.07$      &    22     &   0    &       0     &      1   \\
$0.07\leq M<0.10$      &      7      &    0      &     0     &      0   \\
$0.10\leq M<0.20$      &      8     &     2    &       1     &      0   \\
$0.20\leq M<0.50$      &      10     &     6    &       2     &     1   \\
$0.50\leq M<0.80$      &     4      &     3      &     1     &      0   \\
$0.80\leq M<1.2$        &       1      &     3      &     1     &      1   \\
\hspace{0.83cm}$M>1.2$        &       0       &    1      &     3     &      2   \\ \hline
\multicolumn{5}{c}{Metallicity $Z=0.1~{\rm Z}_\odot$}  \\ \hline
\hspace{0.83cm}$M<0.03$       &      12     &     0     &      0      &     0    \\
$0.03\leq M<0.07$      &    29     &   1   &       0     &      0   \\
$0.07\leq M<0.10$      &      8      &    2      &     1     &      0   \\
$0.10\leq M<0.20$      &      27     &     3    &      2     &      0   \\
$0.20\leq M<0.50$      &      13     &     4    &       2     &     1   \\
$0.50\leq M<0.80$      &     3      &     1      &     2     &      0   \\
$0.80\leq M<1.2$        &       0      &     2      &     0     &      3   \\
\hspace{0.83cm}$M>1.2$        &       1       &    3      &     0     &      3   \\ \hline
\multicolumn{5}{c}{Metallicity $Z={\rm Z}_\odot$}  \\ \hline
\hspace{0.83cm}$M<0.03$       &     20     &     0     &      0      &     0    \\
$0.03\leq M<0.07$      &    39     &   1    &       1     &      0   \\
$0.07\leq M<0.10$      &     20      &    3      &     0     &      0   \\
$0.10\leq M<0.20$      &      24     &     3    &       2     &      1   \\
$0.20\leq M<0.50$      &      16     &     12    &       4     &     1   \\
$0.50\leq M<0.80$      &     4      &     4      &     1     &      3   \\
$0.80\leq M<1.2$        &       2      &     1      &     4     &      0   \\
\hspace{0.83cm}$M>1.2$        &       3       &    2      &     1     &      4   \\ \hline
\multicolumn{5}{c}{Metallicity $Z=3~{\rm Z}_\odot$}  \\ \hline
\hspace{0.83cm}$M<0.03$       &      13     &     0     &      0      &     0    \\
$0.03\leq M<0.07$      &    28     &   2    &       0     &      0   \\
$0.07\leq M<0.10$      &      15      &    3      &     0     &      0   \\
$0.10\leq M<0.20$      &      23    &     9    &       4     &      1   \\
$0.20\leq M<0.50$      &      18     &     11    &       2    &     5   \\
$0.50\leq M<0.80$      &     7      &     3      &     1     &      2   \\
$0.80\leq M<1.2$        &       4      &     1      &     1     &      2   \\
\hspace{0.83cm}$M>1.2$        &       2       &    1      &     4     &      3   \\ \hline
All masses, 4  calculations                    &   399   &     87     &     40   &      34           \\ \hline
\end{tabular}
\caption{\label{tablemult} The numbers of single and multiple systems for different primary mass ranges at the end of the four radiation hydrodynamical calculations with different metallicities. }
\end{table}

\begin{table}
\begin{center}
\begin{tabular}{lccccl}\hline
Object Number & Mass & $t_{\rm form}$ & Accretion Rate\\
& [M$_\odot$] & [$t_{\rm ff}$] & [M$_\odot$~yr$^{-1}$] \\ \hline
  1 & 1.2586 & 0.6061  & $4.05\times10^{-5}$ \\
  2 & 0.2313 & 0.7067  & 0 \\
  3 & 0.5299 & 0.7100  & 0 \\
  4 & 1.0497 & 0.7515  & $3.20\times10^{-5}$ \\
  5 & 0.1456 & 0.7615  & $9.06\times10^{-7}$ \\
\hline
\end{tabular}
\end{center}
\caption{\label{tablestars} For each of the four calculations, we provide online tables of the stars and brown dwarfs that were formed, numbered by their order of formation, listing the mass of the object at the end of the calculation, the time (in units of the initial cloud free-fall time) at which it began to form (i.e. when a sink particle was inserted), and the accretion rate of the object at the end of the calculation (precision  $\approx 10^{-7}$~M$_\odot$~yr$^{-1}$).  The first five lines of the table for the solar metallicity calculation are provided above.}
\end{table}

\begin{table*}
\begin{tabular}{lcccccccccccccl}\hline
Object Numbers & No. of &  No. in & $M_{\rm max}$ & $M_{\rm min}$  & $M_1$ & $M_2$  & $q$ & $a$  & P & $e$  & Relative Spin  & Spin$_1$ & Spin$_2$ \\
& Objects & System & & & & & & & & & or Orbit  & -Orbit & -Orbit \\
&  &  & & & & & & & & & Angle & Angle & Angle\\
     & &    & [M$_\odot$] & [M$_\odot$] & [M$_\odot$] & [M$_\odot$] &  & [AU] & [yr] & & [deg] & [deg] & [deg] \\ \hline

 55,  62              &   2  &  3 &1.241 & 0.856 & 1.241 & 0.856 & 0.690  &   0.69   & 0.40 & 0.172  & 25 & 138 & 123 \\
171, 170            &     2  &  3 & 0.431 & 0.351 & 0.431 & 0.351 & 0.815 &   0.93  &  1.01 & 0.505 &  3  & 45 & 47 \\
(171, 170), 197     &     3  &  3 & 0.431 & 0.351 & 0.782 & 0.375 & 0.480  &  8.22  &  21.91 & 0.387 & 29 &   -- &  --     \\
(141, 204), (212, 215)\hspace{-0.5cm}  & 4  &  4 & 0.614 & 0.092 & 0.960 & 0.197 & 0.205 &   93.77 &  843.87 & 0.244  &  --  &  -- &   -- \\    
(( 21,  19),  25),  15  & 4   & 4 & 4.847 & 2.173 & 10.11 & 2.457 & 0.243  & 153.85  & 538.12 & 0.362  & --  & --  & --     \\
\hline
\end{tabular}
\caption{\label{tablemultprop} For each of the four calculations, we provide online tables of the properties of the multiple systems at the end of each calculation.  The structure of each system is described using a binary hierarchy.  For each `binary' we give the masses of the most massive star $M_{\rm max}$ in the system, the least massive star $M_{\rm min}$ in the system, the masses of the primary $M_1$ and secondary $M_2$, the mass ratio $q=M_2/M_1$, the semi-major axis $a$, the period $P$, the eccentricity $e$.  For binaries, we also give the relative spin angle, and the angles between orbit and each of the primary's and secondary's spins.  For triples, we give the relative angle between the inner and outer orbital planes. For binaries, $M_{\rm max}=M_1$ and $M_{\rm min}=M_2$.  However, for higher-order systems $M_1$ gives the combined mass of the most massive sub-system (which may be a star, binary, or a triple) and $M_2$ gives the combined mass of the least massive sub-system (which also may be a star, a binary, or a triple).  Multiple systems of the same order are listed in order of increasing semi-major axis.  As examples, we provide selected lines from the table from the solar metallicity calculation.}
\end{table*}

\section{Discussion}
\label{sec:discussion}

We have performed radiation hydrodynamical calculations of star formation in molecular clouds whose metallicity is varied by up to a factor of 300.  Despite very different temperature distributions within the clouds, the resulting stellar mass distributions, stellar multiplicity, and mass ratio distributions of bound pairs are statistically indistinguishable.  The only statistically-significant difference between the properties of the stellar systems produce by the four calculations is that there is a deficit of close multiple systems with super-solar metallicities.  In the following sections, we compare these results to those from past observational and theoretical studies to understand why metallicity is generally so unimportant and why close multiple system differ with different metallicities.

\subsection{Binary frequencies and separation distributions}
\label{sec:closebinaries}

Three papers have recently claimed that the close binary fractions for solar-type stars are anti-correlated with metallicity (\citealt{Badenes_etal2018}; \citealt*{MoeKraBad2018, ElBRix2018}).  Such an anti-correlation had been claimed before \citep{GreLin2007,Raghavanetal2010}, but the significance had always been limited by small numbers of stars and potential observational biases \citep[e.g.][]{Stryker_etal1985}, while other surveys found no significant difference \citep[e.g.][]{AbtWil1987,Latham_etal2002}.   \cite{Badenes_etal2018} considered radial velocity variable stars from the APOGEE survey (both main sequence and evolved stars) and found that metal-poor ($Z<0.3~{\rm Z}_\odot$) stars have a multiplicity fraction a factor of 2--3 higher than metal-rich ($Z>{\rm Z}_\odot$) stars.  \cite{MoeKraBad2018} examined spectroscopic binary fractions using five different datasets (spectroscopic binaries, radial velocity variables, and eclipsing binaries), finding that there was a consistent anti-correlation between the frequencies of close binaries (periods$<10^4$~days; separations $<10$~AU) across the datasets. \cite{ElBRix2018}  analysed the wide binary fraction using Gaia DR2 data and found that the wide binary fraction is independent of metallicity at separations $\gsim 250$~AU, but rapidly becomes anti-correlated at smaller separations (particularly $<100$~AU).

As seen in Section \ref{sec:separations}, we see evidence for such an anti-correlation from the numerical simulations.  In this section, we first compare our results to those of the above observational papers.  We then examine the origin of this anti-correlation.

\begin{figure}
\centering
    \includegraphics[width=7.5cm]{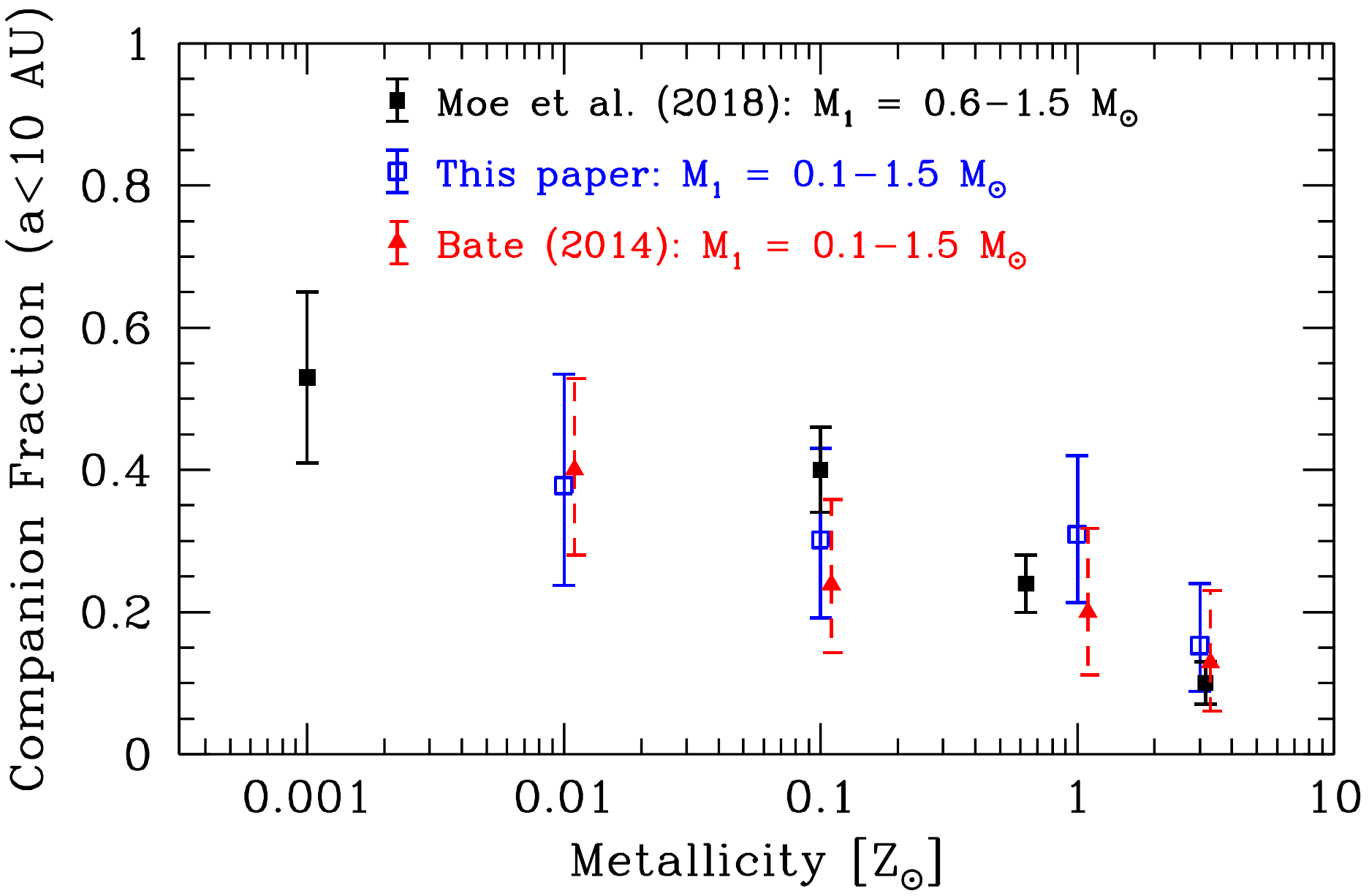} \vspace{-0.1cm}
\caption{ The frequencies of close companions (semi-major axes $a<10$~AU) for low-mass stars with masses $M_*=0.1-1.5$~M$_\odot$ that are produced by the four calculations with different metallicities (blue open squares and errorbars).  We compare the results with the equivalent frequencies from the numerical simulations of \citet{Bate2014} (red triangles and dashed errorbars), and with the observed values for stellar masses $M_*=0.6-1.5$~M$_\odot$ from \citet{MoeKraBad2018} (black filled squares and errorbars). The error bars for the numerical simulations give 95 percent confidence intervals.  The values of the metallicity have been slightly offset for the \citet{Bate2014} results for clarity. In all cases, there is an anti-correlation between the frequency of close companions and metallicity, and the numerical values are in reasonable agreement with the observed values.}
\label{fig:closebinaries}
\end{figure}

\begin{figure*}
\centering
    \includegraphics[width=6.5cm]{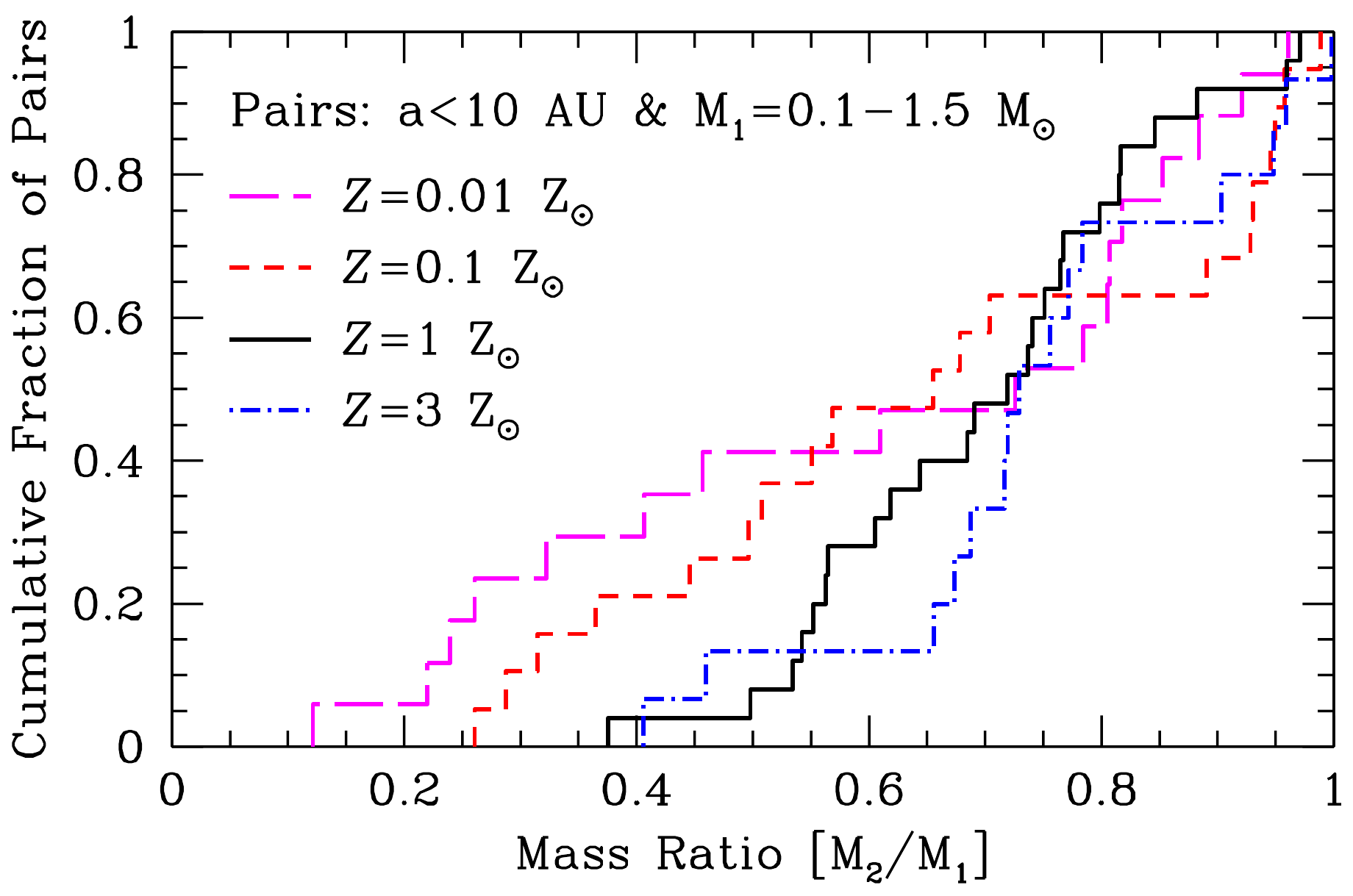} \hspace{1cm}
    \includegraphics[width=6.5cm]{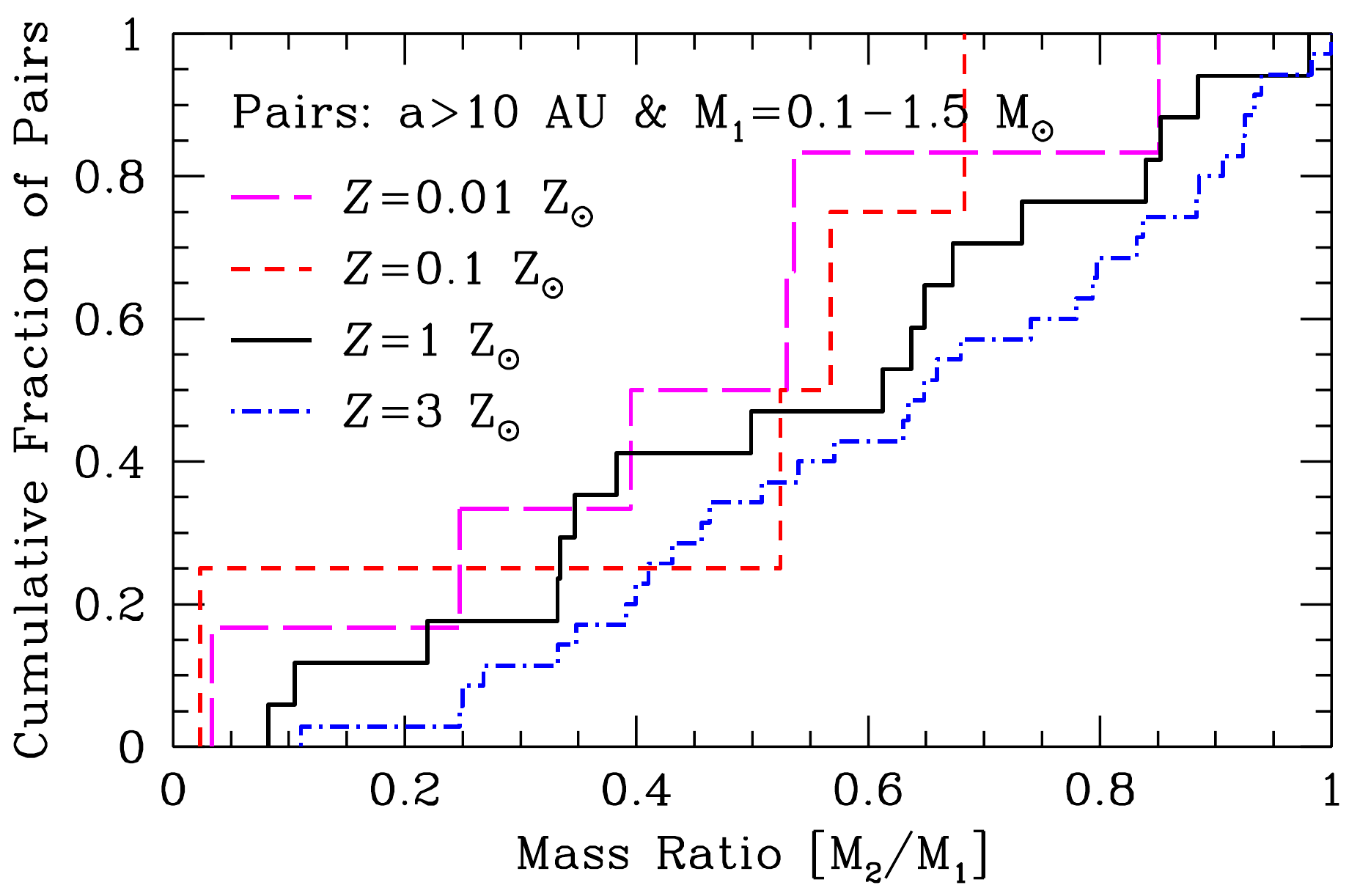} \vspace{-0.2cm}
\caption{ The mass ratio distributions of stellar pairs with primary masses $M_1=0.1-1.5$~M$_\odot$.  In the left panel, we give the distributions for pairs with semi-major axes $a<10$~AU, while in the right panel we give the distributions for pairs with $a>10$~AU. Due to the small numbers of objects, the probability of any two pairs of distributions being drawn from the same underlying distribution never falls below 9 percent in the left plot.  However, the close pairs (left panel) do display a consistent trend of more low mass ratio systems as the metallicity is reduced.  There is no such trend for the wider pairs (right panel).}
\label{fig:CBq}
\end{figure*}

\subsubsection{The close binary fraction and mass ratio distributions}
\label{sec:closeBF}

In Section \ref{sec:separations}, we found that the distribution of orbital separations from the highest metallicity calculation was statistically different from those with lower metallicities, but that the variations between the distributions at lower metallicities were consistent with those arising simply due to the small numbers of systems. However, these separation distributions included binary, triple, and quadruple systems, and systems of all primary masses.  This is not what the above observational papers considered.  \cite{Badenes_etal2018} and \cite{MoeKraBad2018} considered only the fractions of spectroscopic binaries (i.e., close binaries), while  \cite{ElBRix2018} only considered separations $\gsim 50$~AU.  Moreover, \cite{MoeKraBad2018} and \cite{ElBRix2018}  limited their studies to systems with solar-type primaries ($M_1\approx 0.6-1.5$~M$_\odot$ and $M_1 \approx 0.45-1.5$~M$_\odot$, respectively).  In Section \ref{sec:separations}, by including all systems in the analysis, variations in a limited fraction of the parameter space (i.e., close systems) may be hidden by statistical variations.  This is particularly important here because the number of systems produced by the numerical calculations is up to two orders of magnitude smaller than those in the above observational studies.

To better compare the simulations with the observations, we try to replicate as closely as possible the sample of  \cite{MoeKraBad2018}.  The problem is that we have relatively few systems with primary masses $M_1=0.6-1.5$~M$_\odot$ (see Table \ref{tablemult}).  Most systems are M-dwarf systems (as expected for a standard IMF).  To improve the statistical significant, we therefore expand our sample to consider $M_1 = 0.1-1.5$~M$_\odot$. We note that, observationally, M-dwarf binaries tend to be closer than more massive binaries with typical separations of $\approx 10$~AU \citep{Jansonetal2012}, and there have been few studies of whether the binary fraction of M-dwarfs depends on metallicity.  \cite*{RiaGizSam2008} and \cite*{LodZapMar2009} find that the binary fraction at separations $\gsim 5$~AU appears to be lower for metal-poor M-dwarfs systems, indicating that either the overall binary frequencies are lower, or that metal-poor systems are preferentially closer.  It is possible that the low binary fractions obtained by these metal-poor M-dwarf surveys was due to selection effects \citep[see the discussion in][]{MoeKraBad2018}.  But if both M-dwarf and solar-type binaries were preferentially closer at lower metallicity, this may help to explain both the M-dwarf results and the observed anti-correlation of close solar-type binaries with metallicity.

In Fig.~\ref{fig:closebinaries}, we present the close binary frequencies (semi-major axes $<10$~AU) from each of our simulations for systems with primary masses $M_1 = 0.1-1.5$~M$_\odot$.  We compare these to the equivalent fractions that were obtained by \cite{Bate2014}, and also to the observational results of \cite{MoeKraBad2018}.  The numerical results clearly slow the same anti-correlation of close binary fraction with metallicity as seen in the observed systems, albeit with larger error bars because of the much smaller sample sizes.  Interestingly, the same effect is seen in the \cite{Bate2014} dataset.   \citeauthor{Bate2014} did note that there was an apparent shifting of the peak of the separation distribution of multiple systems to smaller separations at lower metallicities, but concluded that this was not statistically significant given the small numbers of objects.  However, this result considered the populations as a whole; when restricting the dataset to consider only the close binary frequency of low-mass stars, the same trend is recovered. This implies that the physical mechanism that leads to the close binary metallicity dependence was already captured by the older calculations --- in other words, it originates from the metallicity dependence of the opacity used in the radiative transfer, rather than from the separate treatment of gas and dust temperatures or the model of the diffuse interstellar medium that are employed in the new calculations presented in this paper.

In Section \ref{sec:q} we noted that binary mass ratios may be preferentially lower in the $Z=0.01~{\rm Z}_\odot$ calculation.  Since binary frequencies seem to depend on whether close or wide systems are considered, we also investigate whether the mass ratios distributions of bound pairs differ between close and wide systems and whether they have a metallicity dependence.  In Fig.~\ref{fig:CBq}, we give the mass ratio distributions of pairs with low-mass stellar primaries ($M_1=0.1-1.5$~M$_\odot$) for close pairs (semi-major axes $a<10$~AU; left panel) and wider pairs (right panel).  For the close pairs, there is a consistent trend with metallicity such that lower metallicities produce more low-mass ratio systems.  It would be interesting to investigate whether real stellar systems display such a trend.  The wider pairs display no such trend.  We do note, however, that because of the relatively small numbers of systems produced by the calculations, even the most different of the close-pair mass-ratio distributions differ at less than the 2-$\sigma$ level of confidence.

\begin{table*}
\begin{center}
\begin{tabular}{ccccccc}\hline
Metallicity & \multicolumn{5}{c}{Numbers of Close Binaries Formed by Various Formation Mechanisms} & Total  \\
$[{\rm Z}_\odot]$ & Separate-SD & Filament-SD & Exchange & Filament-Network & Disc-Frag & \\ \hline
  0.01 &  3* & 7  & 4* & 0 & 5* &   17 \\
  0.1 &  10 & 4  & 3* & 0 & 3* &  19 \\
  1 &  2* & 12  & 1* & 6 & 5 &  25 \\
  3 &  7 & 2  & 7 & 0 & 0 &  16 \\
\hline
\end{tabular}
\end{center}
\caption{\label{CBform} For each of the four calculations, we provide a break down of the formation mechanisms involved in producing the close binaries (separations $<10$~AU) with primary masses $M_1=0.1-1.5$~M$_\odot$. We list the numbers of close binaries that involved stars forming in separate cores or filaments and becoming a close binary through a star-disc interaction, stars forming from the fragmentation of a single filament and becoming a close binary through a star-disc interaction, exchange interactions in multiple systems, fragmentation of a network of filaments, and disc fragmentation.  Some close binaries involved two mechanisms, so these are counted twice and the numbers are denoted with an asterisk. The last column gives the total number of close low-mass stellar binaries that were produced by that calculation.}
\end{table*}

\subsubsection{Wide companions}

Despite the dependence of close binary fraction on metallicity, studies of visual and proper-motion companions have found that the fraction of wide companions is relatively independent of metallicity \citep{Carney1983,ChaGou2004, ZapMar2004,ZinKohJah2004}, or perhaps lower for metal-poor systems \citep{Rastegaevetal2008,Jaoetal2009}. \cite{MoeKraBad2018} also showed that the frequency of wide companions (orbital periods longer than $10^6$ days, or semi-major axes $\gsim 200$~AU) in the \cite{Raghavanetal2010} sample does not display any significant metallicity dependence.  Similarly, \cite{ElBRix2018} find that at separations $\gsim 250$~AU there is no evidence that the companion fraction depends on metallicity.

\cite{MoeKraBad2018} don't find any evidence that the functional form of the separation distribution of close binaries varies with metallicity.  They point out that if the separation distribution has the same form, but a different normalisation, for separations $\lsim 10$AU, and the same distribution for separations $\gsim 200$~AU, then there must be a transition region at $\approx 10-200$~AU.  Furthermore, it is likely that peak of the separation distribution lies within this transition range of separations and the peak moves to greater separations at higher metallicities.  \cite{MoeKraBad2018} provide an illustrative figure (their Fig. 19) of this change in the separation distribution.

The results from the simulations are in qualitative agreement with this picture.  From Figs.~\ref{separation_dist} and \ref{cumsep_comp} it is clear that at $Z=0.01~{\rm Z}_\odot$ the peak of the separation distribution is at $\approx 10$~AU, while at $Z=3~{\rm Z}_\odot$ the peak increases to $\approx 30$~AU.  This is less of a shift than that proposed by \cite{MoeKraBad2018}, but the shift of the peak is in the correct sense and in the appropriate transition region.

There is some tension between the observations and the numerical simulations in terms of the overall multiplicity.  \cite{MoeKraBad2018} claim that the wide companion fraction is $0.21\pm0.03$ {\em independent of metallicity} over the separation range $\approx 200-20,000$~AU, but that the close companion fraction ($\lsim 10$~AU) ranges from $0.53\pm 0.12$ at $Z=0.001~{\rm Z}_\odot$ to $0.10\pm 0.03$ at $Z=3~{\rm Z}_\odot$.  Thus, unless the companion fraction is {\it positively correlated} with metallicity for separations $\approx 10-200$~AU, the overall companion fraction must be anti-correlated with metallicity, albeit less strongly than that for the close binaries alone.

However, in the numerical simulations, the overall multiplicity fraction for primary masses $M_1 = 0.1-1.5$~M$_\odot$ does not show any evidence for a metallicity dependence.  The multiplicity fractions are 22/45 = 0.49 ($Z=0.01~{\rm Z}_\odot$),  19/63 = 0.30  ($Z=0.1~{\rm Z}_\odot$),  38/84 = 0.46  ($Z={\rm Z}_\odot$), and 46/98 = 0.47 ($Z=3~{\rm Z}_\odot$).  There is no detectable metallicity dependence when considering either the multiplicity fraction (equation \ref{eq:mf}) or the companion fraction (defined as $(B+2T+3Q)/(S+B+T+Q)$). There are two main reasons for this.  First, in the simulations, there is a slightly higher fraction of wide systems ($100-10^4$~AU) at high metallicities than at low metallicities (see Fig.~\ref{separation_dist}, being careful to account for the fact that there are more stars overall in the higher metallicity calculations).  Second, although there is a significant deficit of close systems with semi-major axes $<10$~AU at $Z=3~{\rm Z}_\odot$, this is partially offset by a lot of systems with semi-major axes of $15-30$~AU (Fig.~\ref{cumsep_comp}).

It is possible that a weak anti-correlation of the overall multiplicity on metallicity is simply hidden by variations due to the relatively small numbers of stars produced by the simulations.  However, we note that the separation range $\approx 10-50$~AU is not well probed by observations.  Thus, it is also possible that the deficit of $\gsim 10$~AU companions at super-solar metallicities is largely made up for by an excess of companions at separations of $\approx 10-50$~AU.  This is an intriguing possibility that it would be fascinating to test with future observations.

Finally, we note that it is the wide population in these calculations that is most likely to undergo further dynamical evolution and, thus, least likely to be comparable to observed field stars.  The high stellar densities mean that it is difficult to form wide bound systems due to gravitational interactions with other protostars.  Moreover, \cite{MoeBat2010} and \cite{Kouwenhovenetal2010a} and  showed that wide binaries can actually be produced during the dissolution of a cluster and the dispersal of its stars into the field.

%
%
%

\vspace{-12pt}

\subsubsection{The origin of the close binary fraction anti-correlation with metallicity}
\label{sec:origin}

Both \cite{MoeKraBad2018} and \cite{ElBRix2018} propose that the higher close binary frequency at lower metallicities results from an increased propensity for disc fragmentation due to the lower opacities and, thus, enhanced disc cooling rates.  This is not unreasonable, but the advantage we have here is that we can actually go back and look at how each binary was formed to see whether this is the case, or whether the anti-correlation arises for some other reason.  

Forming close binaries by direct fragmentation on scales smaller than $\approx 5$~AU is not thought to be possible \citep{Bate1998,BatBonBro2002b} because collapsing molecular clouds are believed to produce an intermediate first hydrostatic core (FHSC) that has a typical radius of $\approx 5$~AU \citep{Larson1969} before a stellar core is formed.  However, by examining the formation of close binaries in the first hydrodynamical simulation to model the formation of a group of stars and resolve discs and binaries down to scales of a few AU, \cite{BatBonBro2002b} showed that, together, three mechanisms could produce realistic fractions of close binaries.  These mechanisms were dynamical encounters within and between multiple systems, the dissipative interaction of binaries and multiple systems circumbinary and circum-multiple discs, and the accretion by a binary of gas with low-specific angular momentum.  These mechanisms are involved in producing the close binary systems in the new calculations presented here.

Table \ref{CBform} gives the numbers of times that various mechanisms were involved in producing one of the close binaries ($M_1=0.1-1.5$~M$_\odot$).  This table was constructed by examining animations of how each close binary was formed (see the mosaic animations provided in the online Additional Supporting Information). We find that there are various ways in which the close binaries form.  Some protostars form in separate molecular condensations and subsequently encounter each other in a dissipative star-disc interaction to form a binary (`Separate-SD' in the table).  Others form as fragments along a single filament that fall towards each other and form a binary via a star-disc interaction (`Filament-SD' in the table).  Exchange interactions are sometimes involved, where a binary and a single star, or a binary and another multiple system, form separately and undergo a dynamical encounter during which a close binary system is formed consisting of one object from each of the original systems (`Exchange' in the table).  Some close binaries are formed via the fragmentation of a network of filaments and subsequent dissipative dynamical interactions (`Filament-Network' in the table).  Finally, some involve disc fragmentation (`Disc-Frag' in the table).  In some cases, more than one of the above mechanisms is involved. Asterisks are used in the table to denote that the numbers include such cases (and, thus, the numbers in that row sum to a greater number than the total number of close binaries that were formed in that calculation).

In each calculation, the majority of the close binaries were formed through dissipative star-disc interactions between two objects that formed separately (either in separate condensations or via filament fragmentation).  Disc fragmentation does play a role in the formation of some close binaries, but it is not the leading mechanism in any of the calculations and, apart from the highest metallicity calculation, the fraction of close binaries involving disc fragmentation does not appear to depend strongly on metallicity.

In examining how each close binary was formed, several differences were found between the high and low metallicity calculations.  First, in the super-solar metallicity calculation, there are many filaments that fragment into two FHSCs which fall together and merge into a single object before they have a chance to form stellar cores.  By contrast, in the lower metallicity calculations, such systems often produce close binaries because the FHSCs collapse to form stellar cores before they have time to merge \citep[FHSCs have typical diameters of $\approx 10$~AU, around 500 times larger than a stellar core;][]{Larson1969}. The reason for this difference is that with higher metallicity a FHSC has a higher opacity and a low cooling rate, and therefore takes longer to contract and trigger the second collapse to produce a stellar core.  This is one of the main reasons the close binary fraction is lowest in the $Z=3~{\rm Z}_\odot$ calculation.  

Second, in addition to the longer FHSC lifetimes, the $Z=3~{\rm Z}_\odot$ calculation does have much less disc fragmentation.  Only one disc fragments in the super-solar metallicity calculation (but does not form a close binary), compared to many discs that fragment in each of the other calculations.  However, it is important to note how the cases involving disc fragmentation produce close binaries.  In the solar-metallicity calculation, all five close binaries that involve disc fragmentation are produced from `classical' disc fragmentation, where the disc around an existing protostar (or in one case, a binary) fragments into one or more objects and one of these forms the close binary with the original protostar.  However, in the two lowest metallicity calculations, although the mechanisms for producing 8 close binaries involve disc fragmentation (see Table \ref{CBform}), they do so in a variety of different ways (only one involves `classical' disc fragmentation). In one case, the two objects that end up in a close binary begin as separate protostars. The disc around one of these fragments to produce two companions. Later the separate protostar exchanges into the multiple system and the close binary is formed from the two original protostars (neither of which was formed via disc fragmentation).  In two other similar cases, a passing star exchanges into a multiple system in which all but the original object formed by disc fragmentation, but in these cases the close binary is formed by pairing the passing protostar and one of the protostars that formed via disc fragmentation.  In another case, the fragmentation of a disc surrounding a binary is triggered by an encounter with another protostar and the fragment and the encountering protostar produce the close binary.  In another case, a protostar forms as one of more than 10 objects in a fragmenting disc and is ejected.  It later passes through a structured collapsing dense core that forms three protostars almost simultaneously and after a complex dynamical interaction ends up in a close binary with one of the three.  In the final example, two close binaries are formed when two objects that form via disc fragmentation pair up to form a close binary that is later ejected. Thus, in all of these cases disc fragmentation is involved in producing close binaries, but not usually in the simple manner that is often envisaged.

\begin{figure}
\centering
    \includegraphics[width=8cm]{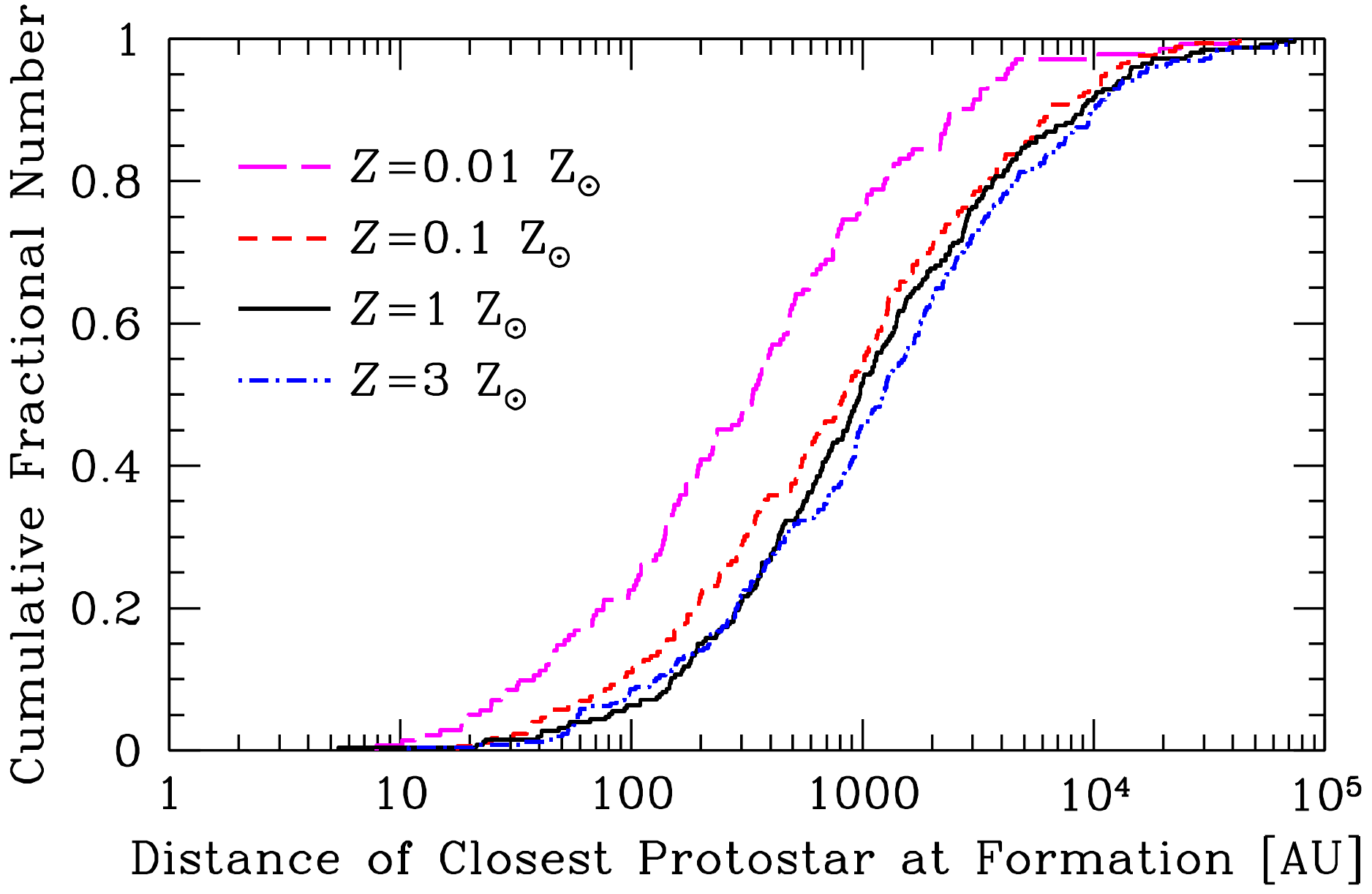}
\caption{The cumulative distributions of the distance between the closest existing protostar and a newly formed protostar from each of the four calculations with different metallicities.  Protostars tend to form closer to each other in the low metallicity calculations than at high metallicities. }
\label{cumradiuscreate}
\end{figure}

As mentioned in Section \ref{sec:closeBF}, the fact that the anti-correlation of the close binary fraction with metallicity appears in both the calculations of \cite{Bate2014} and the new calculations presented here means that it is the metallicity dependence of the opacity that is the cause.  Fundamentally, the opacity decreases with decreasing metallicity.  At high densities where gas and dust are thermally well coupled and the optical depth is high, this results in an increased cooling rate at low metallicities which is likely to produce enhanced fragmentation on small scales. This will apply to the collapse of structure molecular cloud cores, filaments, and discs.  To test this assertion, when each protostar is formed (i.e., a sink particle is inserted) we record the distance to the closest already existing protostar.  In Fig. \ref{cumradiuscreate} we plot the cumulative distributions of these separations for all the protostars that form in each calculation.  It can be clearly seen that at the lowest metallicity ($Z=0.01~{\rm Z}_\odot$), the protostars tend to form closer to each other than at higher metallicities, and the distribution for the $Z=0.1~{\rm Z}_\odot$ calculation is also  shifted slightly towards smaller scales compared to the distributions for the two highest metallicities.  This provides further evidence that it is the reduced opacity leading to enhanced small-scale fragmentation that enhances the close binary fraction at low metallicities.  

In summary, in both the calculations discussed in this paper and in \cite{Bate2014} there is an anti-correlation between the close binary fraction and metallicity that is similar to that which is observed.  However, enhanced disc fragmentation is not the main reason for the higher fractions of close binaries at lower metallicities.  Rather, it is the lower opacity and higher cooling rate of dense gas in general which results both in more small-scale fragmentation (whether it be core, filament, or disc fragmentation) and shorter lifetimes of the first hydrostatic core phase which reduces the propensity for FHSCs to merge before producing stellar cores.

\vspace{-6pt}

\subsection{Comparison with previous theoretical results}

With the exception of the close binary properties and the abundance of stellar mergers, we find no evidence for a strong dependence of stellar properties when varying the metallicity from 0.01 to 3 times the solar value.  The initial mass function seems to be quite insensitive, although there is a hint that the fraction of brown dwarfs may increase with lower metallicity.  The overall stellar multiplicities and the properties of wide binaries also seem to be relatively insensitive.  This invariance is somewhat surprising, given that the temperature and density structure of the star-forming clouds themselves do vary greatly and the onset of the star formation is greatly delayed at low metallicities due to the increased temperatures and the associated pressure support (Section \ref{sec:clouds}).

\cite{Myersetal2011} and \cite{Bate2014} performed two similar prior studies. They performed radiation hydrodynamical calculations of star cluster formation in which the opacity of the gas was varied (over a factor of 20 for the former study, and a factor of 300 for the latter study).  Both studies found that the IMF did not vary significantly with such changes to the opacity.  The calculations of \cite{Myersetal2011} only produced a few dozen stars each and employed sink particles with radii of 28~AU so they could not investigate the dependence of multiple systems on metallicity.  

The calculations presented here are identical in scale and resolution of those of \citet{Bate2014}, but they substantially improve the thermal modelling of low-density gas compared to the previous studies.  Not only are the heating and cooling processes that dominate the low-density (number densities $\lsim 10^4$~cm$^{-3}$) ISM included, but gas and dust temperatures are treated separately.  This has an enormous effect on the gas temperatures at low densities, which are typically much greater than in the calculations of \cite{Myersetal2011} and \cite{Bate2014}.  Furthermore, at the highest metallicity, the temperature of high-density gas is often lower ($T_{\rm gas}<10$~K) due to extinction of the interstellar radiation field.  The different temperatures substantially alter the rate at which stars are produced by the clouds.  In \cite{Bate2014}, the clouds all produced stars at similar rates regardless of the opacity.  In the new calculations, the star formation is substantially delayed at sub-solar metallicities (Fig.~\ref{massnumber}).  However, the changes in the large-scale temperature and density structure and the delayed onset of star formation seem to be the only significant effects of the improved thermal modelling.  The properties of the stellar systems are very similar to those obtained by \cite{Bate2014}.

\cite{Myersetal2011} and \cite{Bate2014} both explained the relative invariance of the IMF as being due to the effects of radiative heating from protostellars on the fragmentation of nearby gas, as first proposed by \cite{Bate2009b}.  The idea is that the characteristic (median) mass of the IMF does not depend strongly on the large-scale density and temperature  of a molecular cloud (i.e., the global Jeans mass) because as protostars form they heat the surrounding cloud, inhibiting fragmentation, and increasing the cloud's effective Jeans mass.  The key is that in the simple case of a dust opacity that depends on wavelength, $\lambda$, as $\kappa \propto \lambda^{-\beta}$ then the temperature at distance, $r$, from a protostar of luminosity, $L_*$, will approximately follow $T \propto L_*^{1/4}r^{-2/(4+\beta)}$ (at intermediate distances where the radiation is optically thin, but the temperatures are not dominated by the ISM) which is {\em independent of the magnitude of the opacity}.  Thus, the temperature structure at intermediate distances from an accreting protostar is expected to be largely independent of the metallicity, and it is the temperature that determines whether a given gas structure will fragment or not.  If gas does not fragment, much of it will be able to accreted by the existing protostar rather than fragment and produce new protostars.  This leads to thermal self-regulation of the IMF.  

We note that there are many complications to this simplified argument. For example, for a deeply embedded protostar with a given luminosity the spectral energy distribution of the radiation will depend somewhat on the metallicity (it will be redder for higher metallicity) and, therefore, a small effect on the surrounding temperatures is to be expected.  There is also considerable uncertainty regarding the luminosities of young protostars.  Accretion luminosity is believed to overwhelm the intrinsic luminosity of low-mass protostars at high accretion rates \citep[e.g.,][]{Offneretal2009}.  Therefore, the above argument also assumes that the protostellar radius does not vary significantly with metallicity. In the calculations presented here, protostellar luminosities are typically underestimated as the sink particles themselves do not emit radiation and they use (metallicity-independent) accretion radii of 0.5~au.  This treatment should be improved upon in future studies. In-depth discussion regarding protostellar luminosities and other issues can be found in \citet{Bate2009b,Bate2012} and \citet{Krumholz2011}.
 
Of course, the effect of radiative heating discussed above requires existing nearby protostars.  In regions of collapsing gas away from protostars that formed earlier, the reduced opacities at lower metallicity allow high-density gas to cool more quickly than at high metallicities, increasing the likelihood of fragmentation.  Past hydrodynamical calculations of \cite{Machida2008} and \cite{Machidaetal2009} showed that the fragmentation of unstable Bonnor-Ebert spheres increases with decreasing metallicity, although their calculations were performed using a barotropic equation of state rather than employing radiative transfer and a realistic equation of state.  \cite{Bate2014} performed simple spherically-symmetric calculations of collapsing \mbox{1-M$_\odot$} Bonnor-Ebert spheres with radiative transfer and varyied the opacities, showing that  gas temperatures reduce by factors of 5--7 at densities ranging from $10^{-13}$ to $10^{-9}$~g~cm$^{-3}$ when the opacity is reduced from 3 to 0.01 times the opacity of solar-metallicity gas \citep[see Fig. 23 of][]{Bate2014}.  As argued in Section \ref{sec:origin}, it is these enhanced cooling rates with lower metallicity that produces more small-scale fragmentation, shorter FHSC lifetimes, and the increased close binary frequencies (and protostellar mergers).

\vspace{-12pt}

\section{Conclusions}
\label{conclusions}

We have presented results from four radiation hydrodynamical simulations of star cluster formation that have identical initial conditions except for their metallicity which ranges from 1/100 to 3 times the solar value.  The calculations resolve the opacity limit for fragmentation, protoplanetary discs (radii $\gsim 1$~AU), and multiple stellar systems.  Unlike previous similar calculations, gas and dust temperatures are treated separately and a thermochemical model of the diffuse ISM is used to provide more accurate temperatures, particularly at low densities.

We draw the following conclusions:
\begin{enumerate}
\item  Lower metallicity generally results in higher gas and dust temperatures at low densities due to reduced rates of cooling. The associated increase in thermal pressure produces smoother gas distributions and delays the onset of star formation, particularly for sub-solar metallicities. However, as the simulations progress, the star formation rates eventually become similar.

\item Most stellar properties do not display a strong dependence on metallicity.  The stellar mass functions produced by the calculations are statistically indistinguishable, although there is a hint that the ratio of brown dwarfs to stars may increase slightly at low metallicity. All of the calculations produce IMFs that are similar to the parametrisation of the observed IMF by \citet{Chabrier2005}, but with marginally lower median masses.

\item The stellar multiplicity strongly increases with primary mass, similar to observed systems.  But we do not detect a significant dependence of the overall multiplicity on metallicity.

\item The separation distributions of multiple stellar systems are found to be metallicity dependent. Metal-rich systems ($Z=3~{\rm Z}_\odot$) have a statistically-significant deficit of close binaries relative to the systems produced in the lower metallicity calculations (Fig.~\ref{cumsep_comp}).  There is an apparent trend for multiple systems to be preferentially tighter at lower metallicities, but the significance of this trend is limited by the small numbers of systems produced by the calculations. Examining the frequencies of close (semi-major axes $a<10$~AU) bound protostellar pairs with primary masses $M_1=0.1-1.5~{\rm M}_\odot$, we find an anti-correlation between the close binary fraction and metallicity that is similar to that which has recently been observed (Fig.~\ref{fig:closebinaries}).  We do not find any evidence for a dependence of wide binaries $a\gsim 100$~AU) on metallicity.

\item Considering bound stellar pairs (binaries, or bound pairs in higher-order systems) of all masses and separations, we find no evidence for a strong dependence of mass ratio on metallicity.  However, if we restrict our samples to close ($a<10$~AU), low-mass bound stellar pairs we uncover a consistent preference for more unequal-mass pairs at lower metallicities.  The statistical significance of this result is at about the 2-$\sigma$ level.

\item We investigate the origin of the close ($a<10$~AU) binary systems and the anti-correlation between their frequency and metallicity.  We find that as the metallicity is decreased, the lower opacities and higher cooling rates of dense gas results in more small-scale fragmentation and in the protostars being formed closer together (Fig.~\ref{cumradiuscreate}).  All types of fragmentation are enhanced, including the fragmentation of collapsing dense cores, filament fragmentation, and disc fragmentation.  In addition, the lifetimes of the first hydrostatic core phase are longer at higher metallicities (due to the higher optical depths and reduced cooling rates), which means that two metal-rich FHSCs are more likely to merge and produce a single stellar core than two metal-poor FHSCs which may collapse to form two (tightly bound) stellar cores before the systems have a chance to merge.
The close binaries are produced through a variety of different mechanisms, including dissipative star-disc encounters and interactions with circumbinary discs, dynamical exchange encounters between multiple stellar systems, and disc fragmentation.  Disc fragmentation alone is not responsible for the anti-correlation between close binary fraction and metallicity in the simulations.
\end{enumerate}

In general, variations in metallicity between 1/100 and 3 times the solar value do not have a large effect on the resulting stellar properties, with the exception of close multiple systems.  Simulations that produce much larger numbers of stellar systems $\gsim 1000$ will be required to confirm some of the above trends and to seek out other weak dependencies of stellar properties on metallicity.

\vspace{-12pt}

\section*{Acknowledgements}


MRB thanks Kareem El-Badry for comments that helped improve the paper. 
This work was supported by the European Research Council under the European Commission's Seventh Framework Programme (FP7/2007-2013 Grant Agreement No. 339248).  The calculations discussed in this paper were performed on the University of Exeter Supercomputer, Isca, and on the DiRAC Complexity system, operated by the University of Leicester IT Services, which forms part of the STFC DiRAC HPC Facility (www.dirac.ac.uk). The latter equipment is funded by BIS National E-Infrastructure capital grant ST/K000373/1 and STFC DiRAC Operations grant ST/K0003259/1. DiRAC is part of the National e-Infrastructure.
Some of the figures were produced using SPLASH \citep{Price2007}, an SPH visualization tool publicly available at http://users.monash.edu.au/$\sim$dprice/splash.

\vspace{-12pt}

\section*{Supporting Information}

Additional Supporting Information may be found in the online version of this article:

{\bf Data files for protostars.} We provide text files of Tables \ref{tablestars} and \ref{tablemultprop} that give the properties of the protostars and multiple systems for each of the four calculations.  These files contain the data necessary to construct Figs.~\ref{fig:IMF} to \ref{fig:CBq}.  Their file names are of the format {\tt Table3\_MetalX.txt} and  {\tt Table4\_MetalX.txt} where `X' gives the metallicity (`001' for 0.01, `01' for 0.1, `1', or `3').  

{\bf Animations.} We provide animations of the large-scale evolution of the column density, gas temperature, and dust temperature for each of the four calculations (i.e., 12 animations).  Sink particles are represented by white circles.  These animations support the snapshots provided in Figs.~\ref{fig:DTZ001} to \ref{fig:DTZ3}.   In addition, we provide one `mosaic' animation for each of the calculations, which displays a region with dimensions of $400\times 400$ au centred on each protostar (sink particle) that is produced during the simulation.  For these animations, the colour scale shows the logarithm of column density, ranging from 1 to $10^4$~g~cm$^{-2}$.  The protostars appear in the order in which they form in the calculation (excluding those that are lost in mergers).  These mosaic animations allow the evolution of each protostar and its disc(s) to be followed.  The file names are of the format {\tt Bate2019\_MetalX\_Y.txt} where `X' gives the metallicity and `Y' gives the variable (`Density', `Tgas', `Tdust', or `Discs'). 

{\bf SPH output files.} Finally, the data set consisting of the output from the four calculations that are discussed in this paper is available from the University of Exeter's Open Research Exeter (ORE) repository at: https://ore.exeter.ac.uk/repository/.

\vspace{-12pt}

\bibliography{mbate}

\end{document}